# Photo-Rechargeable Li Ion Batteries using $TiS_2$ Cathode

*Amar Kumar,*[1] *Raheel Hammad,*[1] *Mansi Pahuja,*[2] *Raul Arenal,*[3,4,5] *Kaushik Ghosh,*[2] *Soumya Ghosh,* [1*] *and Tharangattu N. Narayanan*[1*]

[1]Tata Institute of Fundamental Research-Hyderabad, Sy No. 36/P Serilingampally Mandal, Hyderabad-500046, India.

[2]Institute of Nano Science & Technology, 140306 Mohali, Punjab, India.

[3]Instituto de Nanociencia y Materiales de Aragon (INMA), Universidad de Zaragoza, 50009 Zaragoza, Spain.

[4]Laboratorio de MicroscopiasAvanzadas (LMA), Universidad de Zaragoza, 50018 Zaragoza, Spain.

[5]Fundación ARAID, 50018 Zaragoza, Spain.

(* Corresponding Authors: tnn@tifrh.res.in (T.N.N.) and soumya.ghosh@tifrh.res.in (S.G.)

## Abstract

Photo-rechargeable (solar) battery can be considered as an energy harvesting *cum* storage system, where it can charge the conventional metal-ion battery using light instead of electricity, without having other parasitic reactions. Here we demonstrate a two-electrode lithium ion solar battery with multifaceted $TiS_2$-$TiO_2$ hybrid sheets as cathode. Choice of $TiS_2$-$TiO_2$ electrode ensures the formation of a type II semiconductor heterostructure while the lateral heterostructure geometry ensures high mass/charge transfer and light interactions with the electrode. $TiS_2$ has a higher lithium binding energy (1.6 eV) than $TiO_2$ (1.03 eV), ensuring the possibilities of higher amount of Li ion insertion to $TiS_2$ and hence the maximum recovery with the photocharging, as further confirmed by the experiments. Apart from the demonstration of solar solid-state batteries, the charging of lithium ion *full cell* with light indicates the formation of lithium intercalated graphite compounds, ensuring the charging of the battery without any other parasitic reactions at the electrolyte or electrode-electrolyte interfaces. Possible mechanisms proposed here for the charging and discharging processes of solar batteries, based on our experimental and theoretical results, indicate the potential of such systems in forthcoming era of renewable energies.

**TOC**:

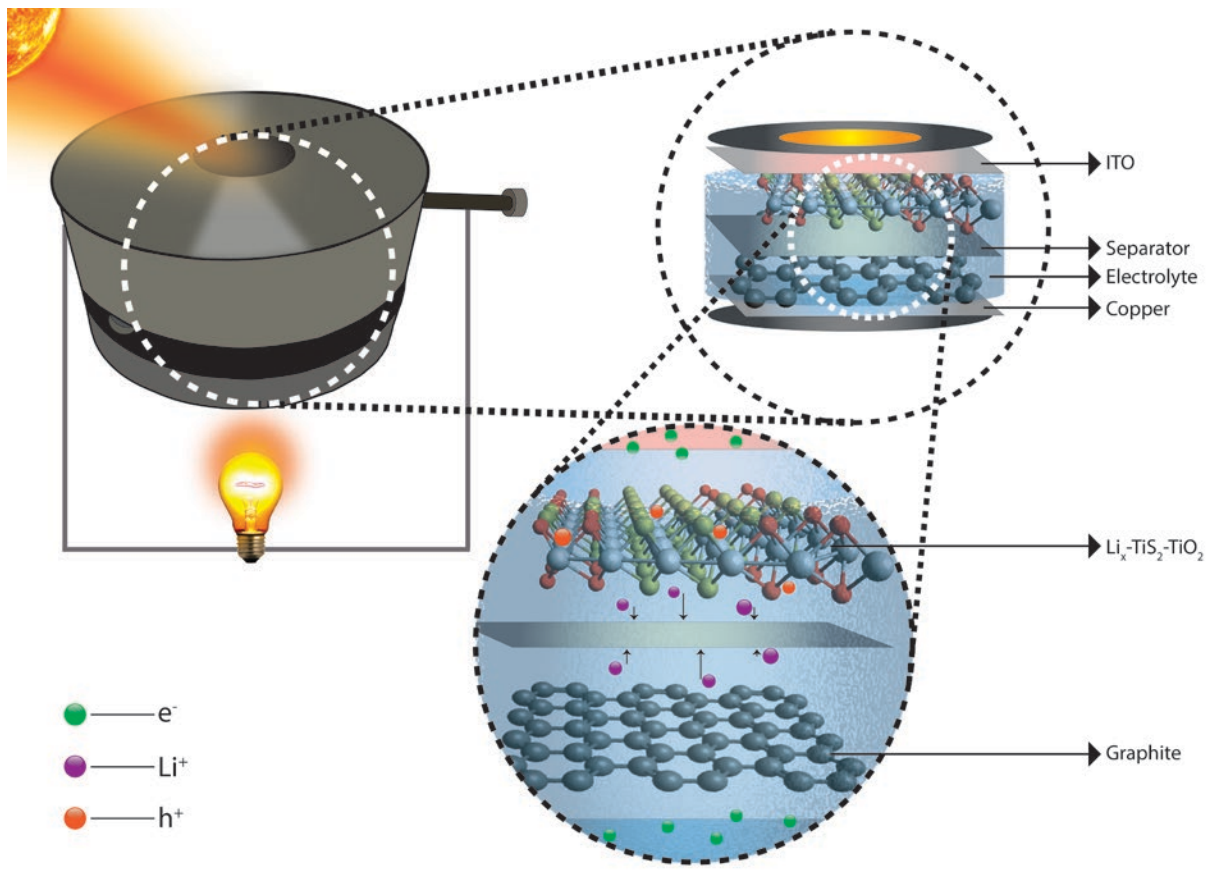

Development of a photo-rechargeable lithium ion battery *full cell* using multifaceted layered cathode and graphite anode, and demonstration of its charging and discharging processes both theoretically and experimentally. The cathode consists a lateral heterostructure of $TiS_2$-$TiO_2$ domains, ensuring the formation of type II heterostructure with high lithium binding energy for $TiS_2$, indicating the possibilities of batteries with high energy efficiency.

## Introduction:

Exploring new ways of harvesting solar energy is receiving tremendous attention due to its potential in relation to tomorrow's green energy economy.[1] Storing solar energy directly inside an electrochemical system is one such viable method that has been recently explored.[2] Along similar lines, directly photo-rechargeable battery or solar battery has recently been highlighted. In the past, two-electrode photo-batteries (solar batteries) were demonstrated mostly with metal-based electrodes. For example, lithium[3–6] or zinc[7,8] has been used as one of the electrodes. In these cells, excitons are generated in the 'cathode' upon exposure to light, and the 'charging' process of the battery/cell is triggered by increasing the lifetime of the holes in the cathode due to efficient electron-hole separation that relies on the band alignment of the different constituents in the electrode.[9] Apart from the metal based '*half cells*' where the metal, say lithium, is directly used as the anode, there is no report on photo-rechargeable Li ion '*full cells*', where a metal-ion containing electrode is employed as the cathode and a layered electrode (say graphite) serves as the anode. Paolella *et al.* showed that a two-electrode system with a hybrid mixture of cathode material (lithium iron phosphate (LFP) and N719 dye) can function as a photo-rechargeable cathode, where dye-sensitization upon light illumination helps delithiation, presumably by the transfer of the generated holes to the Li intercalated LFP.[3] One of the major issues of such organic dye based systems is its cyclability due to the instability of the dye while interacting with the conventional organic electrolytes used in a battery.[3] This idea has been extended to perovskites[4] as well as organic molecule based [5] solar batteries, but the cyclability of such cells and their full-charging to the maximum voltage were found to be limited.

Electrodes based on inorganic frameworks can be alternatives to address the stability issue. In one such demonstration, the authors could show that the $MoS_2/MoO_x$ based layered type-II semiconductor is a viable option for cyclable, high photo-efficient solar (*half cell*) battery development.[9]More recently, Hu *et al.* extended the concept to a $SnO_2/TiO_2$ heterojunction nano arrays based electrode for battery applications. In this set-up, $Li_xTiO_2$ is shown to transfer electron to $SnO_2$ upon light adsorption, resulting in an increase in reversible $Li^+$ intercalation in $SnO_2$.[1] Having demonstrated the potential of such inorganic structures, the key is to develop a layered cathode material, since most of the demonstrated photoelectrodes are battery 'anode' materials.[1] A photoactive (in the visible range) cathode material based cell can have enhanced energy density with high energy efficiency (since it operates at higher potentials), which is highly required for the successful implementation of the light-chargeable two electrode batteries.

.

A type II semiconductor heterojunction-based electrode system can intercalate metal ions (e.g., Li ions) in both the materials simultaneously during discharge.[9] In such a case, photo-rechargeable (recoverable) Li ions (hence charging capacity) is limited to one electrode and the rest of the Li ions from the other material need to be recovered by electrically charging the battery.[9] Hence, the desired combination of materials for cathode should have *(i)* substantial difference in Li ion (or metal ion of concern) binding capabilities while *(ii)* their bands should be aligned in such a way that hole accumulation occurs in the material with higher Li ion binding capability. These attributes will ensure high Li ion recovery with light and hence similar charging and discharging behaviour.

With these concepts in mind, we propose that the combination of $TiS_2/TiO_2$is a potential photocathode material for Li ion *full cells*. Furthermore, theoretical investigations demonstrate that $TiS_2$ has a higher Li ion adsorption energy (1.6 eV, details in the later part) as compared to $TiO_2$ (1.03 eV) along with higher valence band and conduction band edges when forming a staggered band structure with $TiO_2$, where that will ensure the hole accumulation in $TiS_2$ uponvisible light illumination.

$TiS_2$ is known to be a cathode material for Li ion battery, as established byStanley Whittingham *et al*. during their early research.[10] $TiS_2$ possesses high electrical conductivity as well ashigh $Li^+$diffusion rates[10] – the two required properties of a battery cathode material. Moreover, it is the lightest and cheapest of all the group IVB and VB layered dichalcogenides.[11] Hence, $TiS_2$ based cathode electrode can be used in both liquid and solid-state batteries.[12] Furthermore, $TiS_2$-$TiO_2$ heterostructure constitutes type II heterojunction band alignment which has not only a wide range of solar energy coverage (~80%) with near-infrared absorption edge, but also possesses fast electron transfer rate.[13]Along with the ability to absorb a wide range of solar energies, $TiS_2$-$TiO_2$ type II heterostructure also possesses high specific capacity for metal ion battery.[14] Herein, we prepared $TiS_2$-$TiO_2$ type II heterostructure nanosheets as the photocathode of a two electrode photo-rechargeable battery, developed *via* simple but scalable chemical method. The *half cell* and *full cell* (with graphite anode) solar batteries show better photo-efficiency and capacitance in comparison to the systems that have been previously reported in the literature (supporting information (SI), **Table S1**).

Apart from the band structure and $Li^+$ binding capabilities, the photoelectrode structure is also highly crucial in the performance of a solar battery. The electrode design demands high interactions with light along with an efficient mass transfer to the electrolyte being used in the battery. Two-dimensional (2D) sheets or nanosheets in general can offer high surface area, which is found to be helpful for metal ion batteries such as Li and sodium.[14] Here the type II heterostructure consists of $TiO_2$ nanosheets along with partially sulfurized regions of $TiS_2$. This arrangement ensures an electrode structure with high light-matter as well as electrode-electrolyte interactions. Moreover, $TiO_2$ being a wide bandgap semiconductor (~3.47 eV),[15] the exciton formation with solar radiation (1 sun, 100 mW/cm$^2$ power density) can be mainly restricted in $TiS_2$ (~1.10 eV,[16] as proven in the later part) and hence an effective electron-hole separation in the heterostructure can be expected.

Electrolyte degradation is a common issue in liquid electrolyte based Li ion batteries and it is particularly a concern when implemented in solar batteries. This issue can be addressed by solid-state electrolytes in conjunction with a good Li ion diffusing electrode. $TiS_2$ has high Li ion diffusion coefficient (9.75 X $10^{-12}$ cm$^2$s$^{-1}$) [10,12] and hence, we further demonstrate the applicability of $TiS_2$-$TiO_2$ indirect light chargeable batteries in conjunction with solid-state electrolyte system.

**Results and discussion:**

Initially, multi-layered$TiO_2$ nanosheets ($Ti_{0.87}O_2$ NSs, **Figure S1**; stoichiometry was confirmed in the previous report[14]) were synthesized *via* a chemical exfoliation process.[14] The $Ti_{0.87}O_2$ NSs were further annealed in air to obtain anatase phase of $TiO_2$ NSs (named as annealed $TiO_2$ NS, details in SI and later part). Partially and fully sulfurized regions containing $TiO_2$ NSs were then developed (**Figure 1(a)**) by a chemical vapour transport deposition (CVD) technique (details in supporting information method section).[14] Morphology of the nanosheets and the elemental constituents of the samples (annealed $TiO_2$ NSs, $TiS_2$-$TiO_2$ NSs, and $TiS_2$ NSs) were confirmed by scanning electron microscopy (SEM) imaging and energy dispersive spectroscopy (EDS) based mapping (**Figure S2 and S3**). All the samples possess nanosheet geometry (**Figure S2**). Furthermore, the EDS based elemental mapping shows uniform distribution of titanium and oxygen in the annealed $TiO_2$ NSs, titanium and sulphur in $TiS_2$ NSs, and titanium, oxygen, and sulphur in $TiS_2$-$TiO_2$ NSs (**Figure S3**).

Furthermore, Raman spectroscopy analysis indicates the formation of annealed $TiO_2$ (**Figure 1 (b)**), $TiS_2$-$TiO_2$ and $TiS_2$ (details discussion in section 1, **Figure S4**). Chemical states of all the atoms in each NSs are identified using X-Ray Photoelectron spectroscopy (XPS) analysis. The XPS survey spectra (**Figure 1 (c)**) of the annealed $TiO_2$ samples confirm the presence of Ti (2p) and O (1s), while in $TiS_2$-$TiO_2$NSs (**Figure 1 (c)**),S (2p) is present along with Ti (2p) and O (1s). A detailed discussion on the high resolution XPS spectra of samples is given in the SI (section 1, **Figure S5**). The XPS, Raman and X-ray diffraction (XRD) analyses (discussed in the following section) indicate the formation of $TiO_2$ and $TiS_2$-$TiO_2$, along with the surface (fully) sulfurized $TiS_2$ sheets.[17]

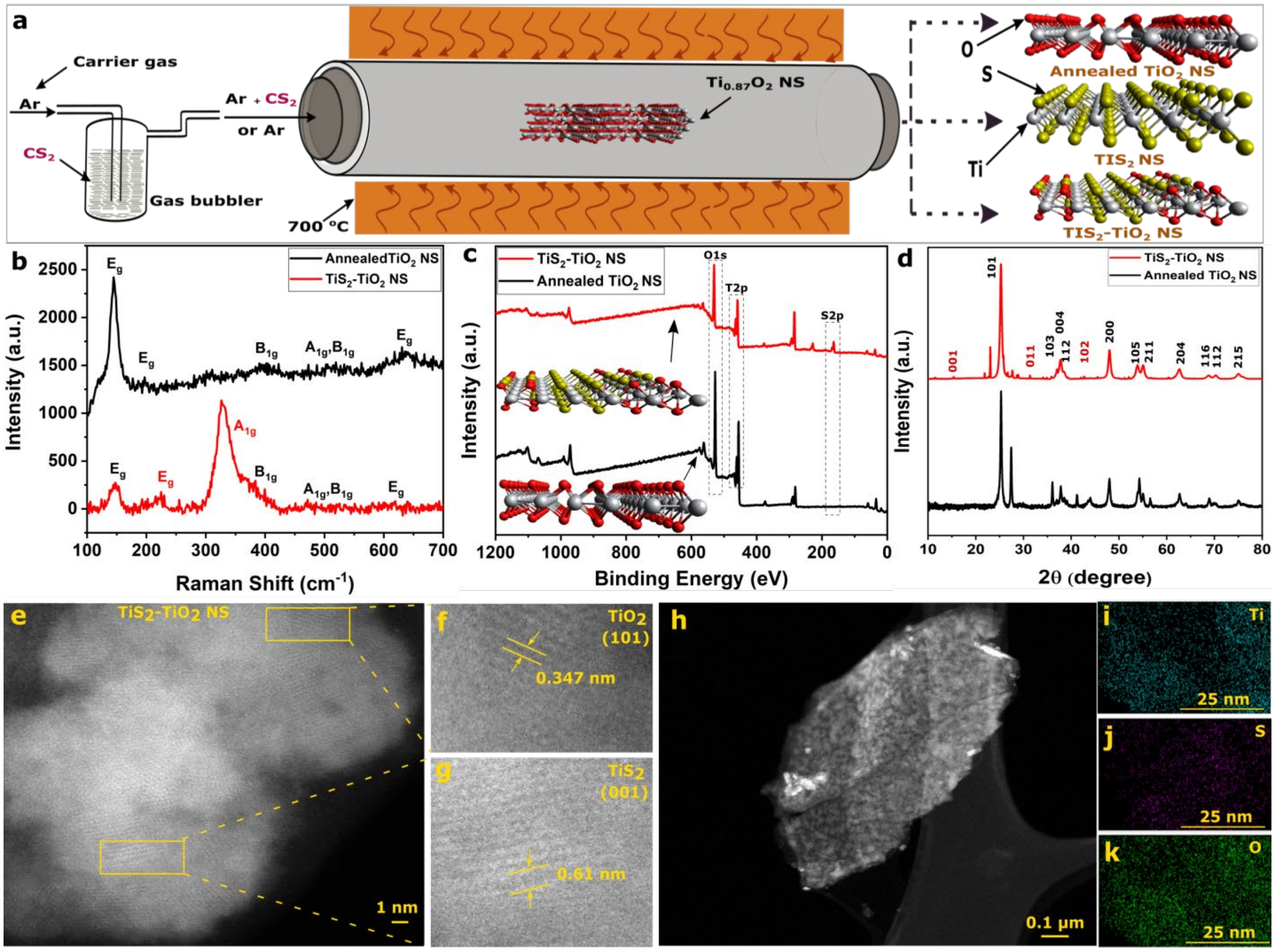

**Figure 1.** (a) Schematic of the reaction procedure for annealed $TiO_2$ NSs, heterostructured (regions containing $TiS_2$ and $TiO_2$) $TiS_2$-$TiO_2$ NSs, and $TiS_2$ NSs*via* the CVD set up. The characterisation of annealed $TiO_2$ NSs and heterostructured$TiS_2$-$TiO_2$ NSs using: (b) Raman spectroscopy, (c) XPS survey scan (inset shows the respective structures), and (d) XRD, indicating their characteristic features as explained in the main text. (e-g) HAADF-(HR) STEM images of $TiS_2$-$TiO_2$ NSs, ((f) area corresponding to $TiO_2$ and (g) the region where $TiS_2$ can be observed, respectively). (h) Another HAADF-STEM image recorded on another flake. An EDS spectrum image has been recorded in this $TiS_2$-$TiO_2$ NSs. (i-k) STEM-EDS elemental mapping maps of Ti, S, and O, respectively have been extracted from the analysis of the EDS spectrum

image recorded in (h). Those images (e-h) and chemical maps (i-k) show the crystallinity and the elemental distribution of the nanosheets.

The XRD diffraction pattern (as seen in **Figure 1(d)**) of annealed $TiO_2$ NS (**Figure 1 (d),** black) indicates the formation of a pure anatase phase of $TiO_2$. The diffraction planes of annealed $TiO_2$ NS (**Figure 1 (d),** black) can be indexed as (101), (103), (004), (112), (200), (105), (211), (204), (116), (112), and (215).[14]However, in the $TiS_2$-$TiO_2$ NSs along with all the $TiO_2$ planes, the crystallographic planes of 2H-$TiS_2$ are also present, marked as (001), (011), and (102).[18] This indicates the existence of crystalline phases of both $TiS_2$ and $TiO_2$ in $TiS_2$-$TiO_2$ NSs. To verify the atomic level structure of the nanosheets and also to check the elemental chemical distribution of $TiS_2$ and $TiO_2$ phases on $TiS_2$-$TiO_2$ NSs, transmission electron microscopy (TEM) analysis have also been carried out on the heterostructured $TiS_2$-$TiO_2$ NSs. In the high-resolution scanning TEM (HRSTEM) images, $TiS_2$ and $TiO_2$ phase were observed on the surface of $TiS_2$-$TiO_2$ NSs (**Figure 1 (e-g)**). $TiO_2$ is found to be in anatase phase,in tune with the XRD analysis. The HRSTEM images show an inter-planar distance of 0.61 nm,[14]that belongs to the crystalline $TiS_2$. The 0.347 nm crystalline planes belong to the anatase $TiO_2$.[14]Furthermore, STEM-EDS elemental mapping of $TiS_2$-$TiO_2$ NSs shows homogenous distribution of titanium, sulphur and oxygen on $TiS_2$-$TiO_2$ NSs(**Figure 1 (h-k))**. Hence, in the heterostructure $TiS_2$-$TiO_2$ NSs, $TiS_2$ nano domains are distributed over the $TiO_2$ NSs. The sheets (flakes) have a lateral width of ~ 600 nm.

The work functions, valance band edges, and band gaps that are experimentally estimated for annealed $TiO_2$ and $TiS_2$ NSs are given in SI (section 2, **Figure S6**). UV-Visible spectroscopy analyses indicate that the absorption edges of $TiO_2$ annealed and $TiS_2$ NSs are at ~ 330 nm and

~830 nm (**Figure S6 (a-c)**),[19,20] indicating the large band gap of $TiO_2$ (~3.36 eV) and full visible light absorption in $TiS_2$ (~1.10 eV) and $TiS_2$-$TiO_2$ NSs. The ultraviolet photoelectron spectroscopy (UPS) analyses were conducted to estimate the valence band edges and work functions. Based on these analyses, the valance band (VB) edge and conduction band (CB) edge of the annealed $TiO_2$, and $TiS_2$ NSs are found to be: VB ($TiO_2$) 7.24 eV, CB ($TiO_2$) 3.88 eV, and VB ($TiS_2$) 4.64 eV, CB ($TiS_2$) 3.54 eV,[16,20] respectively and it indicates the possibilities of the formation of a type II heterostructure in $TiS_2$-$TiO_2$, as shown in the **Figure S6 (f)**. This is further verified computationally using density functional theory (DFT) based calculations. The computed electronic band structures of $TiS_2$ and $TiO_2$ are provided in **Figure S21** and **Figure S23**, respectively. The computed VB and CB edges of the $TiO_2$ and $TiS_2$ monolayers are found to be in line with the experimentally observed trend:VB ($TiO_2$) 7.62 eV, CB ($TiO_2$) 4.28 eV, and VB ($TiS_2$) 4.60 eV, CB ($TiS_2$) 3.86 eV, respectively, (**Figure 2(a)**).The computed work functions for monolayer $TiS_2$ and $TiO_2$ are found to be 5.758 eV and 6.025 eV, respectively (further computational details are provided in the SI). A high photo response from the $TiS_2$-$TiO_2$ NSs in comparison to others indicates effective charge separation happening in the hybrid nanosheet, as expected from the band structure analyses, both experimentally and theoretically. The results of the photo response (photoconductivity) of the different nanosheets are provided in SI (section 2, **Figure S7**).[9]

To check the efficacy of theTiS2-TiO2 system as Li insertion electrode, electrochemical Li ion insertion and de-insertion cyclic voltammetry (CV) experiments were conducted using $TiS_2$-$TiO_2$ NS electrode (details of the cell construction is provided in the SI) and compared with that of annealed $TiO_2$ and $TiS_2$ NSs. The CV was recorded in the range of 1.0-2.9V *vs* $Li/Li^+$at a scan rate of 0.1mV/s (**Figure 2(b)**). In the cathodic region of annealed $TiO_2$ NS, the peak at 1.7

V is attributed to $Li^+$ insertion in $TiO_2$ NS(**Figure 2(b),** black).[21] In $TiS_2$ NS, $Li^+$ insertion peak appears at 2.34 V (**Figure 2(b),** blue).[22] In $TiS_2$-$TiO_2$NS, two different $Li^+$ insertion peaks appear at 2.31 V and 1.71 V indicating that the insertion is happening in both $TiO_2$ and $TiS_2$ (**Figure 2(b),** red).[21,22] This further confirms that both $TiO_2$ and $TiS_2$ regions are exposed on the surface of $TiS_2$-$TiO_2$ NS. Furthermore, annealed $TiO_2$ NS,$TiS_2$ NS, and $TiS_2$-$TiO_2$NS show peaks at 2.1V for $TiO_2$ NS, broad peaks at 2.13 V for $TiS_2$ NS and a sharp and a broad peakat 2.04 V for $TiS_2$-$TiO_2$ NS (**Figure 2(b)**).[21,22] These potentials are related to the de-lithiation from $Li_xTiO_2$, $Li_xTiS_2$, and $Li_xTiS_2$-$TiO_2$ electrodes, respectively. The discharge curves of these cells (coin cells) are shown in **Figures S8** and the specific capacities of the Li ion coin cells are found to be as follows (annealed $TiO_2$): 69 $mAhg^{-1}$, ($TiS_2$-$TiO_2$) 208 $mAhg^{-1}$,and ($TiS_2$) 300 $mAhg^{-1}$.[23] Moreover, the discharge plateaus of $TiO_2$, $TiS_2$, and $TiS_2$-$TiO_2$ based *half cells* at the current density of 21 $mAg^{-1}$ are found to be at 1.67 V, 2.40 and 2.40 V, respectively, indicating the potential of $TiS_2$-$TiO_2$ electrodes as cathodes. Electrochemical cyclability of these cells were also tested (**Figure S9**) and it can be seen that a high capacity of 100 $mAhg^{-1}$ is achievable with $TiS_2$-$TiO_2$ electrode even after 100 cycles. The voltage stabilisation of the cell during discharge will increase the open circuit voltage of the cell (as seen in **Figure S10**) due to the re-arrangement of metal ions within the electrode,[24] but upon further discharge, a sudden drop in the voltage can be seen (**Figure S10** inset).

The photocharging capability of the $TiS_2$-$TiO_2$NS was measured by potentiostatic and galvanostatic charge–discharge measurements. Before the potentiostatic measurement, the $TiS_2$-$TiO_2$ NS cell with an optical window (details in SI) was first discharged in the dark (without light) to 1.0 V *vs* $Li/Li^+$ to lithiate the $TiS_2$-$TiO_2$ NS. After the discharge process, a constant voltage of 2.5 V *vs* $Li/Li^+$ was applied in the presence and absence of white light irradiation (white light (70mWcm$^{-2}$)), and the charging current was recorded.[5] As shown in **Figure 2c**, the current is found to be increased under white light irradiations compared to that in dark by 0.29 mAcm$^{-2}$. The light response is further studied using repeated light on/off experiments.[5] The increase in the charging current in the presence of light indicates the possibilities of photoelectron generation during the charging process, where the current density generated is shown to be sufficient for battery charging, as shown in the subsequent sections.

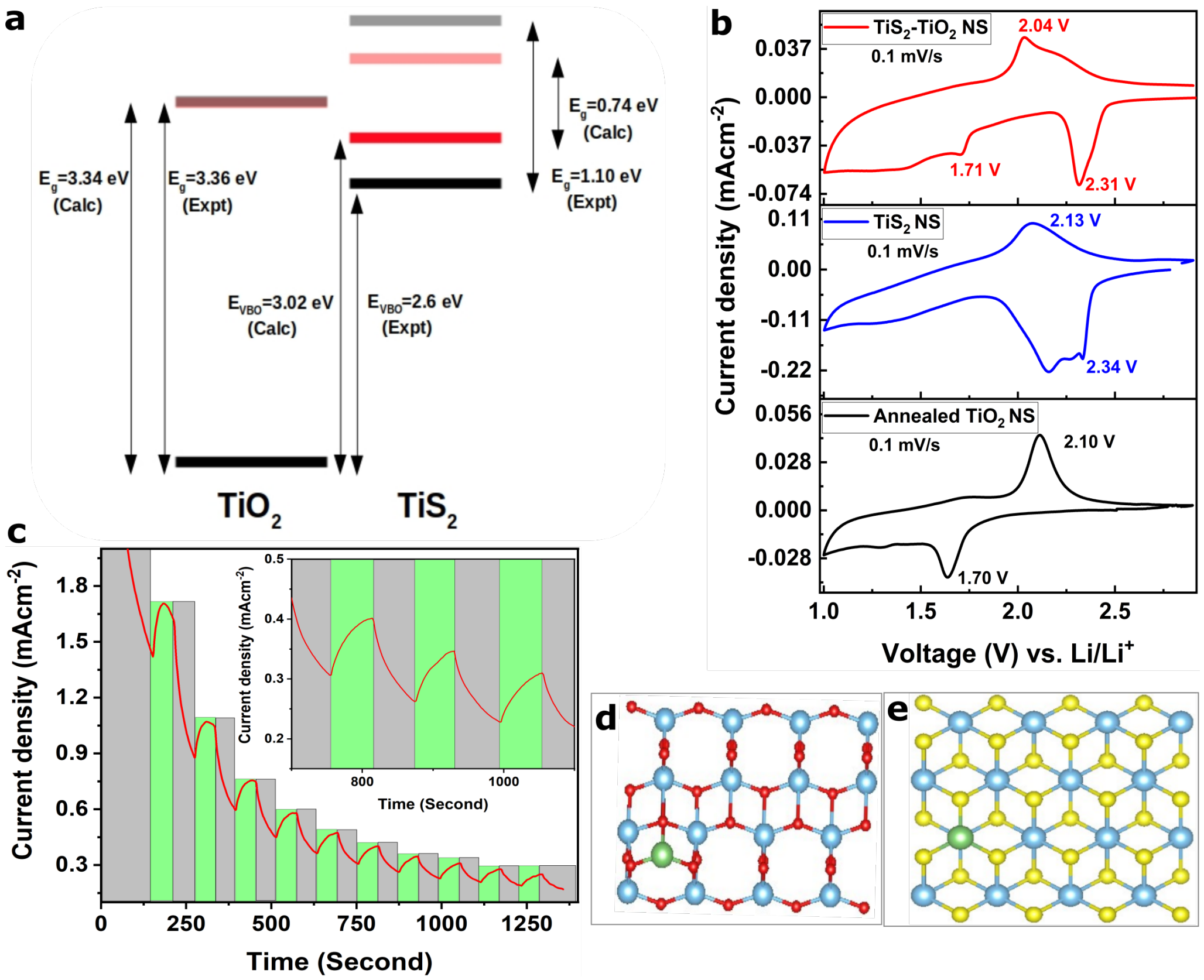

**Figure 2.** Band offset, surface structure, charge transfer mechanism and optical response: (a) band-offset for $TiS_2$ and $TiO_2$ . Black and Red colors represent experimental and computational valence bands respectively while lighter shades of black and red represent the corresponding conduction bands. (b) CV of (with lithium metal as counter electrode) annealed $TiO_2$ NSs (black), $TiS_2$ NSs (blue), and $TiS_2$-$TiO_2$ NSs (red) at a scan rate of 0.1mV/s (c) Potentiostatic

measurement where the potential was held at 2.5 V *vs* $Li/Li^+$ and the charging current were measured in the presence (green) and absence (grey) of white light (70mWcm$^{-2}$), Before the measurement, the cell was discharged to 1.0V*vs* $Li/Li^+$ to lithiate the $TiS_2$-$TiO_2$ NSs.(d) and (e) Indicate the energy stabilised structures, as obtained from DFT calculations (details in supporting information), of Li ion intercalated $TiO_2$ and $TiS_2$ nanosheets.

The solar battery discharge process is also tested with a discharge current density of 21 mAg$^{-1}$ (coin cell (CR2032) having liquid (EC/EMC with $LiPF_6$) electrolyte with a light entering window (quartz), The designand details of the set-up are provided in SI.[9] The open circuit potentials of the constructed batteries (*half cells*) are found to be ~ 3.1 V *vs* $Li/Li^+$. Upon discharge of the cell up to 1V,the potential of the cell was allowed to stabilise. After the voltage stabilisation, irradiation light (white light, 70mWcm$^{-2}$) through the quartz window of the cell is found to increase the cell voltage, as shown in **Figure S11**, indicating the efficacy of the photoelectrons to charge the cell back to the original potential.

To further confirm the effect of light on the Li-insertion and de-insertion in $TiS_2$-$TiO_2$ NS electrode, CV experiment was performed with Li ion *half cells* with and without the exposure of red (100 mWcm$^{-2}$) and UV light (100 mWcm$^{-2}$). The CV was recorded in the range of 1.0–2.9 V *vs* $Li/Li^+$ at a scan rate of 0.1 mV $s^{-1}$ (**Figures 3(a)** and **3 (b)**). In the cathodic region, the peaks around 1.55 V in dark are attributed to $Li^+$ insertion in $TiS_2$-$TiO_2$ NSs. This peak in the 3$^{rd}$ cycle is different from the first cycle as shown in **Figure 2a**, because stable solid electrolyte interface formation can occur after the first discharge.[22,25] Similar effects with light are observed in CVs under both red and UV light illuminations. The faradic current enhancement and the slight shift in the voltage to higher value in the presence of light are indications of the favourable Li insertion process augmented due to the efficient charge transfer mechanism, as recently explained in $SnO_2/TiO_2$ heterostructure based solar battery.[1] Also, in the anodic region, de-insertion peaks in the dark shifted are from 2.15 V to 2.06 V in red light and 2.03 V in UV light because the generated holes help in the de-intercalation of lithium ion from $TiS_2$-$TiO_2$ NSs.

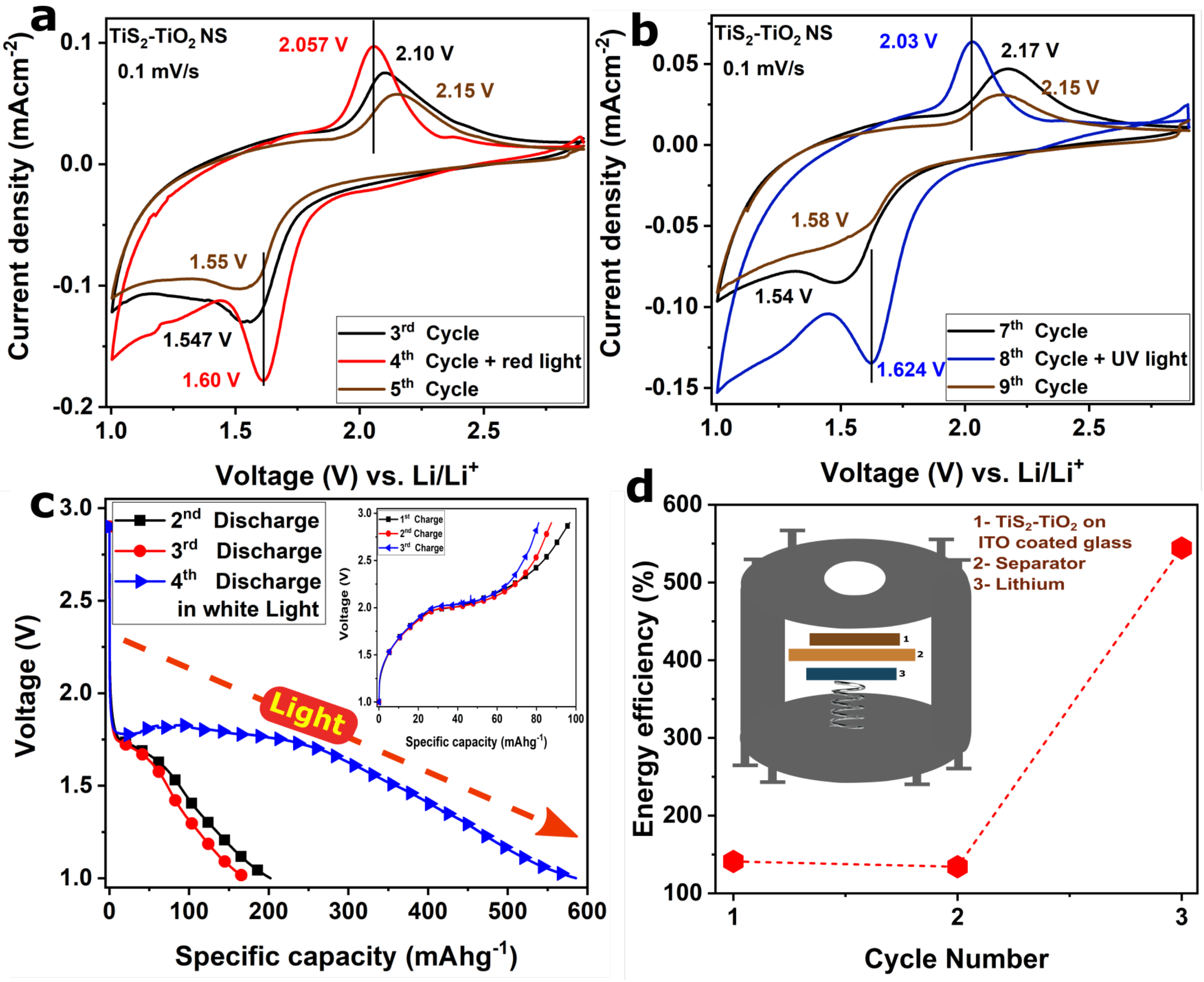

**Figure 3.** The CV Comparisons in $TiS_2$-$TiO_2$ NSs electrodes (with lithium metal as the other electrode) in dark and light. (a) Red LED, (b) UV light at a scan rate of 0.1mV/s. (c) Competitive photocharging during the discharge of the battery with $TiS_2$-$TiO_2$ photoelectrode (inset- charging capacity plots). (d) Energy efficiency (%) for discharge in dark at constant current load for two

cycles and discharge in white light during the third cycle (inset schematic of photo-battery model with a light window).

To further test the effect of light on the charging and discharging processes of these cells, the cells were first electrochemically charged and discharged three times (in dark) till 1 V (**Figure 3 (c)**) and charged back to 2.9 V by a constant currentdensity of ~ 21 mAg$^{-1}$ (**Figure 3 (c)**). During this electrochemical charging, the capacity values are found to be ~ 97 mAhg$^{-1}$ (1$^{st}$ charge), ~ 87 mAhg$^{-1}$ (2$^{nd}$ charge), and ~81 mAhg$^{-1}$ (3$^{rd}$ charge), respectively in dark (**Figure 3 (c),** inset). Discharge capacity values in dark are found to be ~201 mAhg$^{-1}$ (2$^{nd}$ dark) and ~ 169 mAhg$^{-1}$ (3$^{rd}$dark) (**Figure 3 (c)**).The fourth discharge of the same cell in presence of white light (< 1 sun white light) delivers a significantly higher capacity value of ~ 585 mAhg$^{-1}$. This capacity is ~ 245% higher than that obtained in the 3$^{rd}$ cycle. A comparison of the specific discharge capacity in presence of light in $TiS_2$-$TiO_2$ NSs based photoelectrode with other reported photo-rechargeable two systems is given in **Table S1**. This comparison clearly demonstrates that the present cell delivers the highest capacity enhancement in light with reasonably good specific capacity. The energy efficiency (details in SI section 5) enhancement in lightis also calculated by taking the ratio of discharging and charging profiles for discharge in dark and light (see, SI section 5 for details). This calculation shows that the energy efficiency in dark is ~141% in 1$^{st}$ cycle and ~134% in 2$^{nd}$ cycle (**Figure 3 (d)**). This value is found to be enhanced to ~544% when discharged in white light. This result confirms that competitive charging is happening upon light exposure during discharge (**Figure 3 (c)**).

In addition to the competitive charging process as demonstrated above, direct charging of the cell using light is also investigated. To check the light assisted de-lithiation (or photocharging) process in the $TiS_2$-$TiO_2$ NSs electrode, the lithium *half cell* was constructed that shows an initial open circuit voltage (OCV) of 3.1 V. The battery was discharged in dark with ~21 mA $g^{-1}$ discharge current density (**Figure 4a**, black area). After the discharge of cell up to 1.0 V, the cell voltage was allowed to equilibrate (**Figure 4a**, pink area). After the voltage stabilization (at ≈1.95 V),[24] a white light having a power density of 70 $mWcm^{-2}$ is turned on. The voltage enhancement is visible during the exposure of light as shown in **Figure 4a** (white area, enhanced up to 2.89 V within 14.6 h). This rise in the voltage is due to the light assisted de-lithiation or photocharging of lithiated $TiS_2$-$TiO_2$ NSs electrode. After charging, the cell is allowed to discharge with current density of ~ 21 $mAg^{-1}$. The ensuring discharge capacity of ~ 278 $mAhg^{-1}$, is close to the initial discharge capacity 283 $mAhg^{-1}$ (**Figure 4b**),indicating photocharging. In another set of control experiments, the same cell is allowed to discharge until 1 V as discussed earlier (**Figure S10**) with a current density of ~21 $mAg^{-1}$. After the voltage stabilization in dark, the cell was further stabilised at 1.97 V (similar to the previous value). Following this, equilibration procedure, the cell is again discharged with the same current load of ~21 $mAg^{-1}$. It can be seen that the voltage has suddenly dropped without giving any significant capacity (inset of **Figure S10**). Further photocharging was conducted in this cell and then it was discharged with a higher current load (compared to previous) of ~ 46 $mAg^{-1}$, to check its efficacy in delivering higher power density. Further light charging and discharging capacities of the cell are shown in **Figure 4c**. In this case, consistent charging and discharging capacities were obtained as shown in **Figure 4c:**,1$^{st}$ photo charge:~254 $mAhg^{-1}$, 1$^{st}$ discharge:~237 $mAhg^{-1}$, 2$^{nd}$ photo charge:~228 $mAhg^{-1}$, 2$^{nd}$ discharge:~217 $mAhg^{-1}$.Similar photocharging and discharging capacities indicate that the maximum lithium intercalation is happening in $TiS_2$, in tune with the DFT based binding energy calculations (discussed later).

Photoefficiency was calculated for $TiS_2$-$TiO_2$ NSs electrode and compared with a previously reported single material photo-battery electrode material (**Figure S12**). The figure indicates a high photoefficiency (0.23% at high current density load of 0.46 $mAg^{-1}$, the corresponding calculation is shown in section 5 and **Figure S13**) even in comparison to $MoS_2/MoO_x$ solar battery [9] or other reported organic or inorganic photo-batteries (**Table S2**).

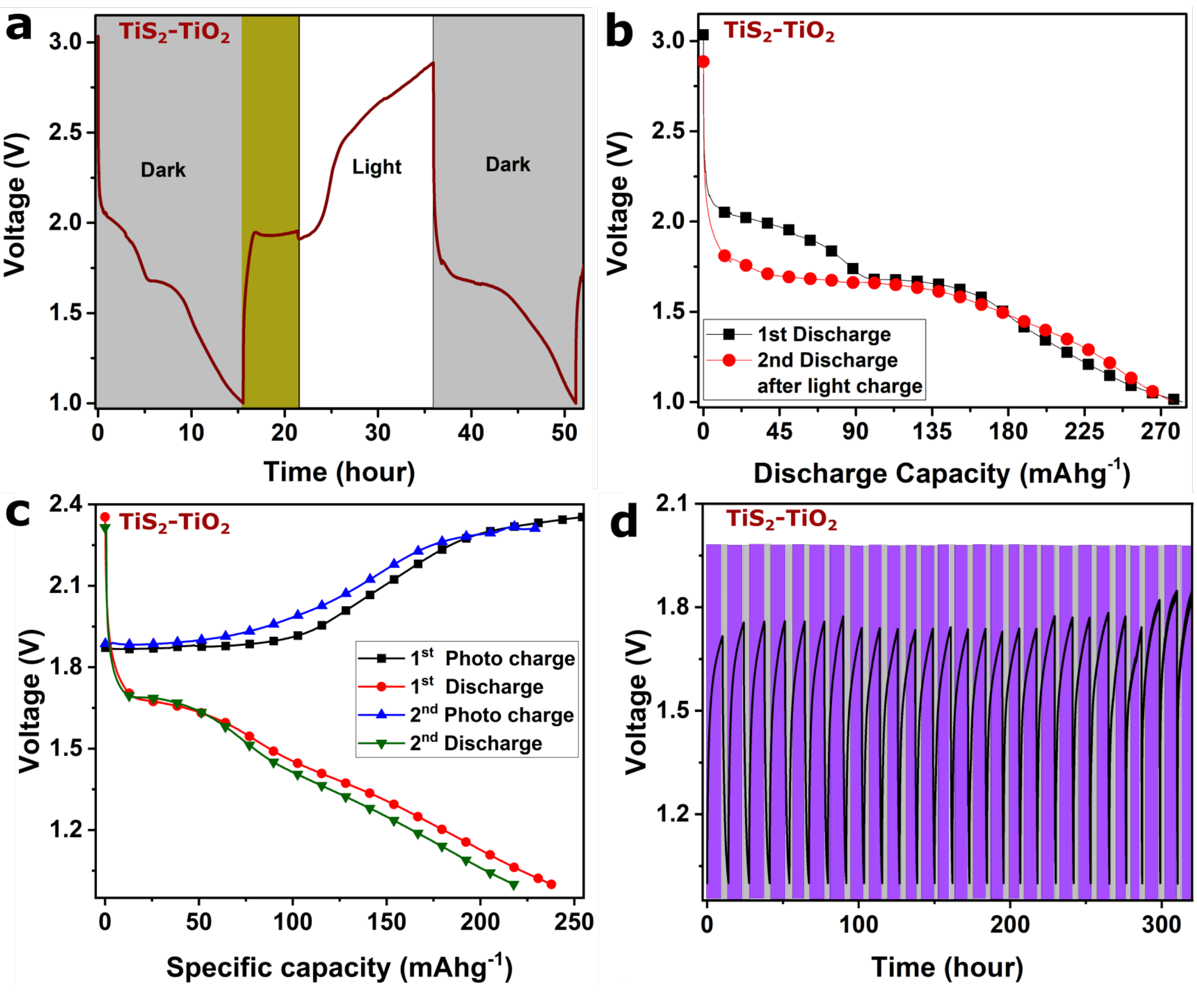

**Figure 4.** Direct photo charge performance of solar battery of $TiS_2$-$TiO_2$ NSs electrodes (with Lithium metal as counter electrode: (a) Discharging in dark by constant current load (light grey area) followed by voltage stabilization (dark yellow area), photocharging (white area) and discharge after photo charge (black area). (b) Comparison of specific capacity of 1$^{st}$ discharge and 2$^{nd}$ discharge after photo charge of $TiS_2$-$TiO_2$ NSs electrodes (inset capacity comparison of conventional and photo Li ion battery). (c) Photo charge capacity and corresponding discharge capacity comparison(d) Long-term photo-battery (cell) cycling done at constant current for discharge (in light grey) and photo charge by red light (LED red light , 100 $mWcm^{-2}$) shown in light violet colour.

The long-term cyclability of photo-rechargeable batteries is another concern among the reported systems.[3] In **Figure 4 (d)** we demonstrate the cyclability of the current set-up. The photocharging time was kept fixed at 10 h and the cell was discharged with a constant current density of ~ 21 $mAg^{-1}$. The mechanism of photocharging of the $TiS_2$-$TiO_2$ NSs based photoelectrode for Li ion battery isassumed to be similar to the $MoS_2/MoO_x$ based photoelectrode (SI, section 5).[9]

In addition to the liquid electrolyte containing Li ion batteries discussed above, where they bring major safety as well as battery lifespan concerns due to the flammability of the electrolyte,[26,27] solid-state batteries have particular interest in light assisted charging process as the decomposition of the electrolyte can be avoided. Due to the high Li ion diffusion capabilities of $TiS_2$, it has been shown to be an efficient electrode in solid-state batteries.[26] Using our previously reported method,[28] we developed a solid electrolyte for Li ions and constructed a solid-state solar battery(details in SI). Initially, the electrochemical charging and discharging of the solid-state battery, with $TiS_2$-$TiO_2$ NSs electrode and Li metal having solid electrolyte (20% $LiClO_4$ in Polyethylene oxide) are tested, **Figure S14a**. The initial capacity of ~ 214 $mAhg^{-1}$ is obtained for this cell and further electrochemical charging and discharging gave a capacity values of ~ 196 $mAhg^{-1}$ in both the second and third cycles (**Figure S14b**). The photocharging of such a solid-state cell is shown in **Figure S14c**. It is found that a capacity of ~ 58 $mAhg^{-1}$(**Figure S14d**)is obtained after the first discharging (after photocharging the cell) of the cell thatis photo charged to a voltage of 2.60 V from 1.88 V. This indicates the possibilities of solid-state solar batteries, which addresses safety issues and electrolyte degradation issues while ensuring a high discharge capacity with light (**Figure S14**).

Furthermore, the mechanism of intercalation and de-intercalation of Li ions in the $TiS_2$-$TiO_2$ electrode is studied using *in situ* Raman spectroscopy experiments. Details of the experiment and the results are shown in SI (section 7 and **Figure S15-16**). In a nutshell, it can be seen that Li ion intercalation happens in both $TiS_2$ and $TiO_2$ (to a lesser extent) and the photocharging brings the Li ions from the $TiS_2$, indicating the veracity of the above-mentioned charging mechanism. The preferential binding of Li to $TiS_2$ as compared to $TiO_2$ is also supported by the calculated binding energies of $Li^+$ to $TiS_2$ and $TiO_2$, -1.6 eV and -1.0 eV, respectively.

Control experiments were also conducted where photo irradiation with and without external circuit (Schematic representation in **Figure S17**) was performed. During this experiment, in the absence of any external electron transport path, voltage did not enhance even after 5 h of light exposure, **Figure S17 (a)**. But the voltage of the same cell enhanced to 2.65 V from 1.88 V within 5 h, once connected with an external circuit, **Figure S17 (b)**. This observation indicates that light assisted degradation (oxidation) of the electrolyte is not the cause for the voltage enhancement. Paolella *et al.* reported that the photoelectrons oxidise the electrolyte only (parasitic process) but cannot reduce Li ion on the other electrode.[3] Such a process cannot ensure the proper long term cycling of the solar battery and hence the process happening at the other electrode needs to be studied by constructing a *full cell*.

*Full cell* solar battery is also constructed using pre-lithiated $TiS_2$-$TiO_2$ electrode (cathode, details of preparation in SI) and graphite anode (schematics of the experiment and the cell are shown in **Figure S15 (c)** and **Figure S19**, respectively). The light assisted charging of such a cell is studied using *in situ* Raman spectroscopy, where the Raman signal is collected from the graphite side while doing the photo exposure (white light, $70mWcm^{-2}$) from the other side (section 7 and **Figure S15 (c-d)**). The changes in the 'D' (disorder) and 'G' ($E_{2g}$) Raman modes of graphite are visible with charging process (**Figure S15d**) and the details can be seen in SI. Powder XRD of the graphite is also taken after the charging process (**Figure S15e**) and it is compared with that of electrochemically charged graphite electrode made out of a similar full cell (details in SI, section 7). It can be seen from the XRD comparisons that the formation of $LiC_6$, $LiC_{12}$ and/or similar lithium intercalated graphite compositions are there on both the graphite electrodes, indicating the successful photoelectron transfer to the anode (detail mechanism discuss in SI, section 8). Theoretical possibilities of such a process to form lithium intercalated graphite is also further verified.

While the reduction of work function of graphene upon increasing the concentration of metal adatom is well documented,[29] current study reveals that the work function of the Li-intercalated graphene can be matched with the work function of the photoexcited lithiated $TiS_2$-$TiO_2$ heterostructure (detailed calculations in SI, **Figures S20-S30**). The relative positions of the Fermi level of $Li_1C_{12}$ as compared to $TiO_2$ and $TiS_2$ is shown in **Figure S31**. Relatively small difference between $LiC_{12}$ and $TiS_2$ energy levels indicate the possibilities of such reactions with the generated photoelectrons at the photocathode. Moreover, reversible changes in the diffraction planes of $TiS_2$ are also observed while conducting post-XRD analyses on lithiated and de-lithiated (photocharging) $TiS_2$-$TiO_2$ electrodes, while such a reversible process was not happening in $TiO_2$ electrode.

In conclusion, possibilities of lithium ion based solar battery full cells having conventional graphite anode and intercalated cathode material is tested here and it is shown that such a system can be realized. In this attempt, $TiS_2$-$TiO_2$ nanosheets were functioned as cathode material where $TiS_2$ nanosheet regions were formed on nanosheets of anatase $TiO_2$ nanosheets, which were prepared using a chemical route. This type II semiconductor lateral heterostructure ensures high photoelectron generation while having high binding energy of Li for $TiS_2$ (-1.6 eV), as confirmed using DFT based calculations. Band alignment calculations using both theoretical and experimental techniques are in line with each other indicating the formation of this type II structure. Apart from the photoelectron formation, the heterostructure based electrode is shown to work efficiently as photocathode. Both *half cell* and *full cell* studies are conducted, and a discharge energy efficiency of ~>500% is achieved with this electrode under white light illumination (<1 sun). Similar charging and discharging capacities (~250 $mAhg^{-1}$) were obtained under battery cycling in 'light' indicating the efficient shuttling of Li ions between the electrodes during the charge-discharge processes. The formation of intercalated graphite compounds (possibly $LiC_{12}$), shown with *in situ* and *ex situ* experiments and theoretical analyses, establishes the functioning of battery full cells with light assisted charging process. Hence this work establishes the potential of the concept of solar battery, where it can be fully functional as an electrochemical battery while harvesting solar energy.

## 1. Experimental Section

*Materials:* Titanium oxide ($TiO_2$, Sigma-Aldrich), Potassium carbonate ($K_2CO_3$, Sisco Research Laboratories Pvt. Ltd., (SRL)), Lithium carbonate ($Li_2CO_3$, SRL), Hydrochloric acid (Sigma Aldrich), tetramethylammonium hydroxide aqueous solution (SRL), Carbon disulfide ($CS_2$, Spectrochem), Lithium hexafluoro phoshphate in EC/EMC solvent (Sigma Aldrich), Ethylene carbonate (Sigma Aldrich), Polythelene carbonate (Sigma Aldrich), Lithium perchlorate ($LiClO_4$, ACS, >95%), ITO coated glass 25mm x 25 mm (Sigma Aldrich), Acetylene black (MTI Corporation), PVDF (Sigma Aldrich), N-Methyl-2-Pyrrolidone (NMP, Avra Synthesis Pvt. Ltd.), Whatman Glass microfiber filter paper (GE Healthcare Life sciences)

*Methods:* Synthesis of 2D $Ti_{0.87}O_2$ nanosheets: First of all we synthesised layered titanate ($K_{0.8}[Ti_{1.73}Li_{0.27}]O_4$) by calcined of solid-state mixture at 1273K for 20 hour, where 5 g of $Li_2CO_3$, $K_2CO_3$, and $TiO_2$ (solid-state mixture at molar ratio, 0.14:0.4:1.73) were placed in a crucible.[14] Further for proton exchange of layered titanate, the layered titanate precursor was immersed in acidic solution (0.5 mol $L^{-1}$ HCl). To completely obtain protonated titanate ($H_{1.07}Ti_{1.73}O_4 \cdot H_2O$), we stirred solution for 48 h, during course of this period the HCl solution were refreshed in every 24 h. After that, the mixture solution was centrifuged at 7500 rpm, and the solid material was washed with distilled water. Further solid compound were dried at room temperature under a vacuum for 48 h. To exfoliate the layered protonated titanate ($H_{1.07}Ti_{1.73}O_4 \cdot H_2O$) into the single-layer $Ti_{0.87}O_2$ nanosheets by exchanging the protons of protonated titanate. The protonated titanate were re-dissolved in the saturated tetramethylammonium hydroxide aqueous solution (TMAOH) (keeping 1:1 molar ratio), were shaken for 48 h. To completely remove water, $Ti_{0.87}O_2$ solution was placed in the freezer for 40 hand then freeze dried for 72 h in Lyophilizer.[14]

Synthesis of 2D Annealed $TiO_2$: 1g of 2D $Ti_{0.87}O_2$ was placed in a quartz container of the chemical vapour deposition (CVD) setup. The flow parameters were set at Ar at 700 °C for 2 h using a tube furnace. Black and white annealed $TiO_2$ nanosheets were obtained.

Synthesis of 2D $TiS_2$-$TiO_2$ hybrid and 2D $TiS_2$: 1g of 2D $Ti_{0.87}O_2$ was placed in a quartz container of the chemical vapour deposition (CVD) setup. The flow parameters were set at $CS_2$/Ar at 700 °C for 20 minute for $TiS_2$-$TiO_2$ hybrid and 1 hour for 2D $TiS_2$ using a tube furnace. Slight and deep Olive-colour $TiS_2$-$TiO_2$and $TiS_2$ 2D nanosheets were obtained.[14]

*Preparation of the photo electrodes*: For the photo electrode fabrication, the sample ($TiS_2$-$TiO_2$ NS or Annealed $TiO_2$ NS or $TiS_2$ NSs) was mixed with 10 weight % carbon (C) black and 10 weight % PVDF (polymer) binder.[9] They were mixed well in N-Methyl-2-Pyrrolidone (NMP) solvent and drop-casted on an ITO coated glass plate.[9] The coating was further dried at 80 °C for 4 hours. In order to prepare solid electrolyte of polyethylene oxide (PEO) and $LiClO_4$, 1 g of PEO and 0.2 g of $LiClO_4$ were dissolved in excess of acetonitrile at 60 °C till a homogeneous thick slurry is obtained.[28] The bubble free slurry was casted onto aluminium foil. The coating was first allowed to dry at room temperature then dried at 60 °C.[28] The thickness was optimized for 0.6 mm and is kept same for all the different membrane. The other electrode (anode) of the cell was a lithium chip. The coin cells were assembled inside the glove box as follows: a cathode disc was placed on the coin cell. The SE was then dipped in ethylene carbonate–propylene carbonate solution (0.2 g ethylene carbonate in 1 mL of propylene carbonate) and placed on cathode followed by lithium chip as anode. The coin cell was then closed and sealed through punching.[28]

*Characterisation:* To determine vibrational Raman modes of $TiS_2$-$TiO_2$ NS, Annealed $TiO_2$ NS, $TiS_2$ NSs and conduct in-situ Raman analysis, Renishaw in Via Raman microscope (532 nm excitation) was used. Further to know the surface topography and composition of the sample, FESEM, JEOL JSM-7200 Falong with EDS was used. To identify the elemental composition and chemical structure of NSs, PHI Quantera X-ray photoelectron spectrometer (Survey: pass energy, 140 eV, high-resolution spectra, 26 eV) was used. To determine valance band edge we used Ultraviolet photon spectroscopy (UPS) using He I excitation (21.2 eV) and recorded with a constant pass energy of 5 eV in the ultrahigh vaccum (UHV) chamber of the XPS instrument (Thermo Fisher Scientific). Transmission electron microscopy (TEM) was performed using a probe aberration corrected microscope, (Thermo Fisher Scientific Titan Low Base working at 300 kV). This microscope is equipped with a high-brightness field-emission gun (X-FEG) and with an EDS spectrometer (Oxford Instruments Ultim Max 100). Powder XRD data of the materials is acquired from RIGAKU Smartlab X-Ray diffractometer at operational power of 9kW with Hypix-3000 detector. Diffuse reflectance UV-visible (DR UV-vis) spectra have been recorded on a Shimadzu UV-2550 spectrophotometer. Photo conductivity measurements were conducted using a three-electrode system with Ag/AgCl as reference electrode and graphite as a counter. A potentiostat -Biologic SP-300 was used for the measurements. For such studies, the samples were coated on a glassy carbon electrode and applied 0.3 V external bias in 0.2 M $Na_2SO_4$. White, UV and red LED light (Holmarc Opto-Mechatronics) sources were used for the measurements. For freeze drying we used Lyophilizer from Christ. Furthermore, galvanostatic charge and discharge tests were conducted using Neware Battery testing system. During in-situ Raman analysis, we used potentiostat Biologic SP-300 for galvanostatic charge and discharge.

All the electrochemical experiments were carried out at room temperature, and all the experiments were repeated more than thrice under identical conditions to check the reproducibility.

**Acknowledgements**:

TNN and SG acknowledge the support from Infosys-TIFR "*Leading Edge*" Research Grant. Authors from TIFR also thank the support of Department of Atomic Energy, Government of India, under Project Identification No. RTI 4007. Authors thank Pallavi Thakur and Dr. Sai Smruti Samantaray for valuable suggestions on some experiment. R.A. acknowledges financial support from the Spanish Ministry of Science and Innovation MCIN and the Spanish Research Agency AEI (PID2019-104739GB-100/AEI/10.13039/501100011033), Gobierno de Aragon (DGA, E13-20R), and from the European Union H2020 programs "ESTEEM3" (grant agreement No 823717) and Graphene Flaghsip CORE3 (grant agreement No. 881603).TEM and XPS measurements were performed in the Laboratorio de MicroscopicasAvanzadas (LMA) at the Universidad de Zaragoza (Spain).Wethank G.Antorrea (LMA, U Zaragoza,Spain) for help with XPS acquisition.K.G. is grateful for getting the financial support from Ministry of Textiles [Grant 2/3/2021- NTTM(Pt.)] and Nanomission (Grant SR-/NM/NS-91/ 2016), Department of Science and Technology, Government of India for the execution of this work.

**Competing interests**

The authors declare no competing interests.

**Author contributions**

T.N.N. conceived the idea with S.G. and planned the work with A.K. A.K. performed most of the electrochemical experiments and photo-battery design. R.H. did most of the computational calculations. M.P. performed in UPS and DRS experiments. R.A. helped to conduct the XPS and TEM analyses. A.K., R.H., M.P., R.A., K.G., S.G. and T.N.N. co-wrote the paper, and all the authors discussed the manuscript.

**Data Availability Statement**

The data that support the findings of the study are available from the corresponding author upon reasonable request. Figure S20 and S22 shows the structural units of $TiS_2$ and $TiO_2$ respectively that are employed in band structure calculations, Figure S21 and S23. Figure S24 and S25 show the planar averaged Hartree potential in $TiS_2$ and $TiO_2$ monolayers for work function calculations whereas Figure S26 shows the non-monotonous convergence of work function with respect to number of layers. The simulation cells with multiple layers of $TiS_2$ and $TiO_2$ are shown in Figure S27 and S29 respectively. The corresponding plots for planar averaged Hartree potentials are provided in Figure S28 and S30. The simulation cell for $Li_1C_{12}$ is shown in Figure S32 whereas the corresponding computed planar averaged Hartree potential is shown in Figure S33.

**References:**

(1) Hu, C.; Chen, L.; Hu, Y.; Chen, A.; Chen, L.; Jiang, H.; Li, C. Light-Motivated

SnO2/TiO2 Heterojunctions Enabling the Breakthrough in Energy Density for Lithium-Ion Batteries. *Adv. Mater.* **2021**, *33* (49), 2103558. https://doi.org/10.1002/ADMA.202103558.

(2) Wang, R.; Liu, H.; Zhang, Y.; Sun, K.; Bao, W. Integrated Photovoltaic Charging and Energy Storage Systems: Mechanism, Optimization, and Future. *Small* **2022**, *18* (31), 2203014. https://doi.org/10.1002/SMLL.202203014.

(3) Paolella, A.; Faure, C.; Bertoni, G.; Marras, S.; Guerfi, A.; Darwiche, A.; Hovington, P.; Commarieu, B.; Wang, Z.; Prato, M.; Colombo, M.; Monaco, S.; Zhu, W.; Feng, Z.; Vijh, A.; George, C.; Demopoulos, G. P.; Armand, M.; Zaghib, K. Light-Assisted Delithiation of Lithium Iron Phosphate Nanocrystals towards Photo-Rechargeable Lithium Ion Batteries. *Nat. Commun. 2017 81* **2017**, *8* (1), 1–10. https://doi.org/10.1038/ncomms14643.

(4) Ahmad, S.; George, C.; Beesley, D. J.; Baumberg, J. J.; De Volder, M. Photo-Rechargeable Organo-Halide Perovskite Batteries. *Nano Lett.* **2018**, *18* (3), 1856–1862. https://doi.org/10.1021/ACS.NANOLETT.7B05153/SUPPL_FILE/NL7B05153_SI_001.PDF.

(5) Kato, K.; Puthirath, A. B.; Mojibpour, A.; Miroshnikov, M.; Satapathy, S.; Thangavel, N. K.; Mahankali, K.; Dong, L.; Arava, L. M. R.; John, G.; Bharadwaj, P.; Babu, G.; Ajayan, P. M. Light-Assisted Rechargeable Lithium Batteries: Organic Molecules for Simultaneous Energy Harvesting and Storage. *Nano Lett.* **2021**, *21* (2), 907–913. https://doi.org/10.1021/ACS.NANOLETT.0C03311/SUPPL_FILE/NL0C03311_SI_001.PDF.

(6) Boruah, B. D.; Wen, B.; De Volder, M. Light Rechargeable Lithium-Ion Batteries Using V2O5Cathodes. *Nano Lett.* **2021**, *21* (8), 3527–3532. https://doi.org/10.1021/ACS.NANOLETT.1C00298/ASSET/IMAGES/LARGE/NL1C00298_0005.JPEG.

(7) Boruah, B. D.; Mathieson, A.; Wen, B.; Feldmann, S.; Dose, W. M.; De Volder, M. Photo-Rechargeable Zinc-Ion Batteries. *Energy Environ. Sci.* **2020**, *13* (8), 2414–2421. https://doi.org/10.1039/D0EE01392G.

(8) Deka Boruah, B.; Mathieson, A.; Ki Park, S.; Zhang, X.; Wen, B.; Tan, L.; Boies, A.; De Volder, M.; Deka Boruah, B.; Mathieson, A.; Park, S. K.; Zhang, X.; Wen, B.; Tan, L.; Boies, A.; De Volder, M. Vanadium Dioxide Cathodes for High-Rate Photo-Rechargeable Zinc-Ion Batteries. *Adv. Energy Mater.* **2021**, *11* (13), 2100115. https://doi.org/10.1002/AENM.202100115.

(9) Kumar, A.; Thakur, P.; Sharma, R.; Puthirath, A. B.; Ajayan, P. M.; Narayanan, T. N. Photo Rechargeable Li-Ion Batteries Using Nanorod Heterostructure Electrodes. *Small* **2021**, *17* (51), 2105029. https://doi.org/10.1002/SMLL.202105029.

(10) Whittingham, M. S. Electrical Energy Storage and Intercalation Chemistry. *Science* **1976**, *192* (4244), 1126–1127. https://doi.org/10.1126/SCIENCE.192.4244.1126.

(11) Trevey, J. E.; Gilsdorf, J. R.; Stoldt, C. R.; Lee, S.-H.; Liu, P. Electrochemical Investigation of All-Solid-State Lithium Batteries with a High Capacity Sulfur-Based Electrode. *J. Electrochem. Soc.* **2012**, *159* (7), A1019–A1022. https://doi.org/10.1149/2.052207JES/XML.

(12) Kim, J. Y.; Park, J.; Kang, S. H.; Jung, S.; Shin, D. O.; Lee, M. J.; Oh, J.; Kim, K. M.; Zausch, J.; Lee, Y. G.; Lee, Y. M. Revisiting TiS2 as a Diffusion-Dependent Cathode with Promising Energy Density for All-Solid-State Lithium Secondary Batteries. *Energy Storage Mater.* **2021**, *41*, 289–296. https://doi.org/10.1016/J.ENSM.2021.06.005.

(13) Han, J. H.; Lee, S.; Yoo, D.; Lee, J. H.; Jeong, S.; Kim, J. G.; Cheon, J. Unveiling Chemical Reactivity and Structural Transformation of Two-Dimensional Layered Nanocrystals. *J. Am. Chem. Soc.* **2013**, *135* (10), 3736–3739. https://doi.org/10.1021/JA309744C/SUPPL_FILE/JA309744C_SI_001.PDF.

(14) Bayhan, Z.; Huang, G.; Yin, J.; Xu, X.; Lei, Y.; Liu, Z.; Alshareef, H. N. Two-Dimensional TiO2/TiS2Hybrid Nanosheet Anodes for High-Rate Sodium-Ion Batteries. *ACS Appl. Energy Mater.* **2021**, *4* (9), 8721–8727. https://doi.org/10.1021/ACSAEM.1C01818/SUPPL_FILE/AE1C01818_SI_001.PDF.

(15) Challagulla, S.; Tarafder, K.; Ganesan, R.; Roy, S. Structure Sensitive Photocatalytic Reduction of Nitroarenes over TiO 2. *Sci. Reports 2017 71* **2017**, *7* (1), 1–11. https://doi.org/10.1038/s41598-017-08599-2.

(16) Greenaway, D. L.; Nitsche, R. Preparation and Optical Properties of Group IV–VI2 Chalcogenides Having the CdI2 Structure. *J. Phys. Chem. Solids* **1965**, *26* (9), 1445–1458. https://doi.org/10.1016/0022-3697(65)90043-0.

(17) Etghani, S. A.; Ansari, E.; Mohajerzadeh, S. Evolution of Large Area TiS2-TiO2 Heterostructures and S-Doped TiO2 Nano-Sheets on Titanium Foils. *Sci. Reports 2019 91* **2019**, *9* (1), 1–14. https://doi.org/10.1038/s41598-019-53651-y.

(18) Plashnitsa, V. V.; Vietmeyer, F.; Petchsang, N.; Tongying, P.; Kosel, T. H.; Kuno, M. Synthetic Strategy and Structural and Optical Characterization of Thin Highly Crystalline Titanium Disulfide Nanosheets. *J. Phys. Chem. Lett.* **2012**, *3* (11), 1554–1558. https://doi.org/10.1021/JZ300487P/SUPPL_FILE/JZ300487P_SI_001.PDF.

(19) Zhu, L.; Lu, Q.; Lv, L.; Wang, Y.; Hu, Y.; Deng, Z.; Lou, Z.; Hou, Y.; Teng, F. Ligand-Free Rutile and Anatase TiO2 Nanocrystals as Electron Extraction Layers for High Performance Inverted Polymer Solar Cells. *RSC Adv.* **2017**, *7* (33), 20084–20092. https://doi.org/10.1039/C7RA00134G.

(20) Huckaba, A. J.; Gharibzadeh, S.; Ralaiarisoa, M.; Roldán-Carmona, C.; Mohammadian, N.; Grancini, G.; Lee, Y.; Amsalem, P.; Plichta, E. J.; Koch, N.; Moshaii, A.; Nazeeruddin, M. K. Low-Cost TiS2 as Hole-Transport Material for Perovskite Solar Cells. *Small Methods* **2017**, *1* (10), 1700250. https://doi.org/10.1002/SMTD.201700250.

(21) Zheng, P.; Liu, T.; Su, Y.; Zhang, L.; Guo, S. TiO2 Nanotubes Wrapped with Reduced Graphene Oxide as a High-Performance Anode Material for Lithium-Ion Batteries. *Sci. Reports 2016 61* **2016**, *6* (1), 1–8. https://doi.org/10.1038/srep36580.

(22) Fleischmann, S.; Shao, H.; Taberna, P. L.; Rozier, P.; Simon, P. Electrochemically Induced Deformation Determines the Rate of Lithium Intercalation in Bulk TiS2. *ACS Energy Lett.* **2021**, *6* (12), 4173–4178. https://doi.org/10.1021/ACSENERGYLETT.1C01934/ASSET/IMAGES/LARGE/NZ1C01934_0004.JPEG.

(23) Zhang, L.; Sun, D.; Kang, J.; Wang, H. T.; Hsieh, S. H.; Pong, W. F.; Bechtel, H. A.; Feng, J.; Wang, L. W.; Cairns, E. J.; Guo, J. Tracking the Chemical and Structural

Evolution of the TiS2 Electrode in the Lithium-Ion Cell Using Operando X-Ray Absorption Spectroscopy. *Nano Lett.* **2018**, *18* (7), 4506–4515. https://doi.org/10.1021/ACS.NANOLETT.8B01680/SUPPL_FILE/NL8B01680_SI_001.PDF.

(24) Hu, Y.; Iwata, G. Z.; Mohammadi, M.; Silletta, E. V.; Wickenbrock, A.; Blanchard, J. W.; Budker, D.; Jerschow, A. Sensitive Magnetometry Reveals Inhomogeneities in Charge Storage and Weak Transient Internal Currents in Li-Ion Cells. *Proc. Natl. Acad. Sci.* **2020**, *117* (20), 10667–10672. https://doi.org/10.1073/PNAS.1917172117.

(25) Fleischmann, S.; Mitchell, J. B.; Wang, R.; Zhan, C.; Jiang, D. E.; Presser, V.; Augustyn, V. Pseudocapacitance: From Fundamental Understanding to High Power Energy Storage Materials. *Chem. Rev.* **2020**, *120* (14), 6738–6782. https://doi.org/10.1021/ACS.CHEMREV.0C00170/ASSET/IMAGES/MEDIUM/CR0C00170_0037.GIF.

(26) Trevey, J. E.; Stoldt, C. R.; Lee, S.-H. High Power Nanocomposite TiS2 Cathodes for All-Solid-State Lithium Batteries. *J. Electrochem. Soc.* **2011**, *158* (12), A1282. https://doi.org/10.1149/2.017112JES/XML.

(27) Mizuno, F.; Hayashi, A.; Tadanaga, K.; Minami, T.; Tatsumisago, M. All-Solid-State Lithium Secondary Batteries Using a Layer-Structured LiNi0.5Mn0.5O2 Cathode Material. *J. Power Sources* **2003**, *124* (1), 170–173. https://doi.org/10.1016/S0378-7753(03)00610-4.

(28) Patra, S.; Thakur, P.; Soman, B.; Puthirath, A. B.; Ajayan, P. M.; Mogurampelly, S.; Karthik Chethan, V.; Narayanan, T. N. Mechanistic Insight into the Improved Li Ion

Conductivity of Solid Polymer Electrolytes. *RSC Adv.* **2019**, *9* (66), 38646–38657. https://doi.org/10.1039/C9RA08003A.

(29) Legesse, M.; Mellouhi, F. El; Bentria, E. T.; Madjet, M. E.; Fisher, T. S.; Kais, S.; Alharbi, F. H. Reduced Work Function of Graphene by Metal Adatoms. *Appl. Surf. Sci.* **2017**, *394*, 98–107. https://doi.org/10.1016/J.APSUSC.2016.10.097.

**Supplementary Materials**

# Photo-Rechargeable Li Ion Batteries using $TiS_2$ Cathode

*Amar Kumar,*[1] *Raheel Hammad,*[1] *Mansi Pahuja,*[2]*Raul Arenal,*[3,4,5]*Kaushik Ghosh,*[2] *Soumya Ghosh,* [1*] *and Tharangattu N. Narayanan*[1*]

[1]Tata Institute of Fundamental Research-Hyderabad, Sy No. 36/P Serilingampally Mandal, Hyderabad-500046, India.

[2]Institute of Nano Science & Technology, 140306 Mohali, Punjab, India.

[3]Instituto de Nanociencia y Materiales de Aragon (INMA), Universidad de Zaragoza, 50009 Zaragoza, Spain.

[4]Laboratorio de Microscopias Avanzadas (LMA), Universidad de Zaragoza, 50018 Zaragoza, Spain.

[5]Fundación ARAID, 50018 Zaragoza, Spain.

(* Corresponding Authors: tnn@tifrh.res.in (T.N.N.) and soumya.ghosh@tifrh.res.in (S.G.)

Section 1: SEM, Raman analysis, XPS, XRD, and photo-response of $Ti_{0.87}O_2$ NSs, annealed $TiO_2$ NSs, $TiS_2$-$TiO_2$ NSs, and $TiS_2$ NSs:

- Figure S1: Powder X-ray data of $Ti_{0.87}O_2$ NSs
- Figure S2 and S3: SEM and EDS mapping of annealed $TiO_2$ NSs, $TiS_2$-$TiO_2$ NSs, and $TiS_2$ NSs
- Figure S4: Comparison of the Raman spectrum of $TiS_2$-$TiO_2$ NSs and $TiS_2$ NSs
- Figure S5: High resolution XPS spectra of annealed $TiO_2$ NSs, $TiS_2$-$TiO_2$ NSs, and $TiS_2$ NSs

Section 2: Optical band gap, Tauc plot and Photo response comparison:

- Figure S6: UV-Vis absorption spectra of annealed $TiO_2$ NS (black) and $TiS_2$- $TiO_2$ NS (red), Optical band gap, Tauc plot and Experimental band alignment of Type II heterostructure $TiS_2$-$TiO_2$ NSs
- Figure S7: Photo response comparison of annealed $TiO_2$ NSs, $TiS_2$-$TiO_2$ NSs, and $TiS_2$ NSs with white light exposure

Section 3: Conventional Li ion battery performance of all NS compound and Discharge without photo charge Li-ion battery experiment:

- Figure S8: Conventional Li ion battery performance of annealed $TiO_2$ NSs, $TiS_2$-$TiO_2$ NSs, and $TiS_2$ NSs
- Figure S9: Long term cyclibility test of Conventional $Li^+$-ion battery performance of $TiS_2$-$TiO_2$ NSs
- Figure S10: Discharge without photo charge Li-ion battery experiment of TiS2-TiO2 NSs electrodes (with Lithium metal as counter electrode)
- Figure S11. Photo charging curve

Section 4:

- Energy efficiency calculation:
- Table S1. Comparison of enhancement in specific capacity in light of $TiS_2$-$TiO_2$ NSs based photo-electrode with other reported photo-rechargeable two electrode system

Section 5: Photo conversion calculation & comparison:

- Figure S12: Comparison of photoefficiency for $TiS_2$-$TiO_2$ NSs based electrode with reported single material photo battery electrode material
- Figure S13: After photo charge, during discharge potential-time integration calculation and photo charge duration for $TiS_2$-$TiO_2$ NSs
- Table S2: Comparison of discharge specific capacity after light charge and photoefficiency of $TiS_2$-$TiO_2$ NSs based photo-electrode with other reported photo-rechargeable two electrode system
- Photo charge mechanism

Section 6: Solid electrolyte based conventional and photo Li ion battery performance of $TiS_2$-$TiO_2$ NSs (with Lithium metal as counter electrode and 20% $LiClO_4$ in Polyethylene oxide) electrode

- Figure S14: Solid-state electrolyte (20% $LiClO_4$ in Polyethylene oxide) based conventional and photo battery performance

Section 7: *In situ* Raman analysis during lithiation and de-lithiation in $TiS_2$-$TiO_2$ NSs electrodes (with Lithium metal as counter electrode) by electrochemically and Schematic illustrations of control experiment of light assisted de-lithiation with and without external connection:

- Figure S15: Schematic illustrations of *in situ* Raman and *ex situ* powder x-ray diffraction characterisation and experiment of lithiation and delithiation in $TiS_2$-$TiO_2$ NSs electrodes (with Lithium metal as counter electrode) and Graphite electrodes (with $Li_xTiS_2$-$TiO_2$ NSs as counter electrode)

**Section 1:**

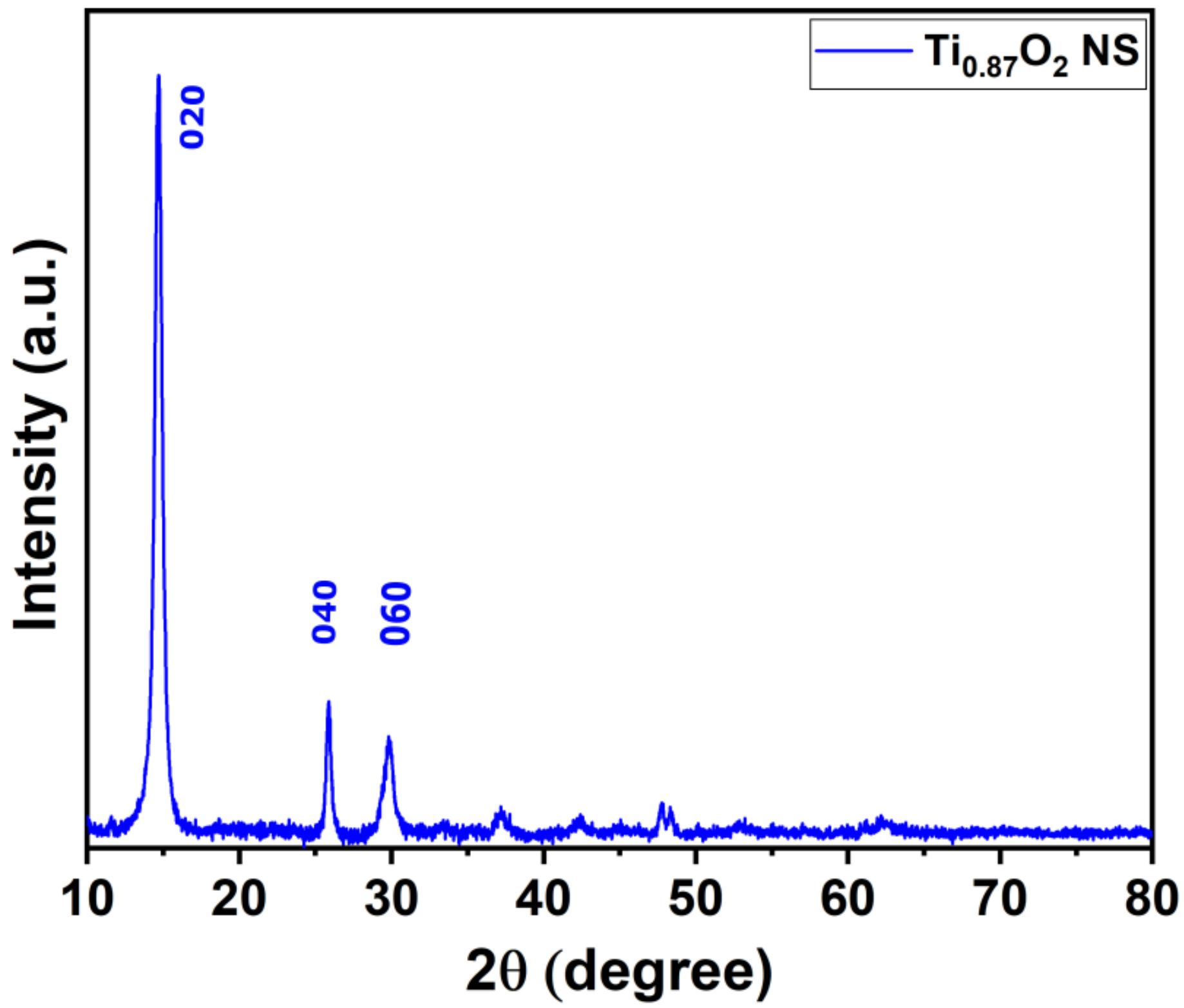

**Figure S1.** Powder X-ray data of $Ti_{0.87}O_2$ NSs

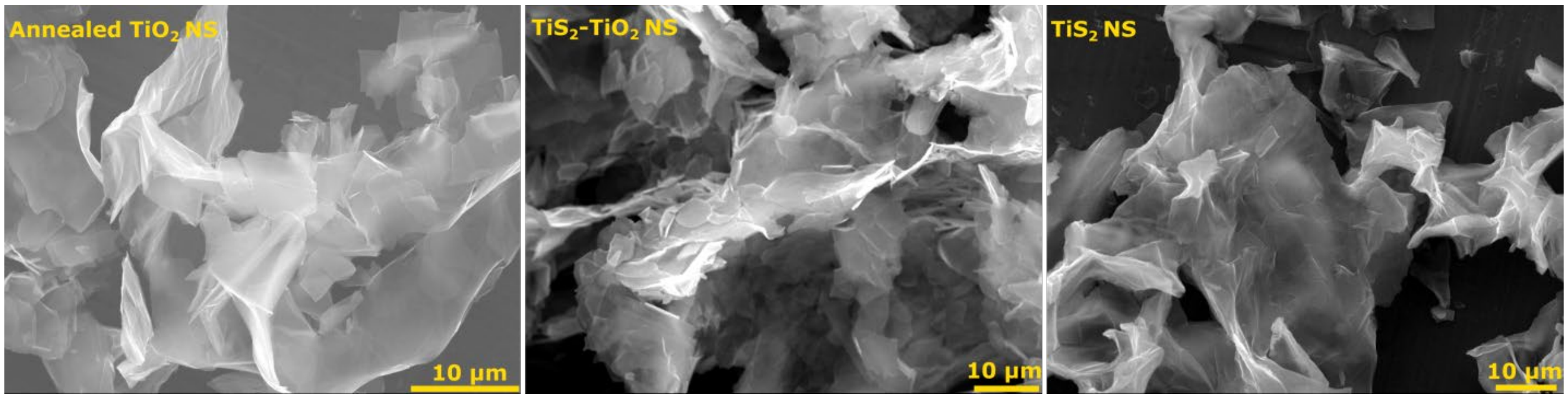

**Figure S2.** The SEM images of annealed $TiO_2$ NS (left), $TiS_2$-$TiO_2$ NS (middle), and $TiS_2$ NS (right)

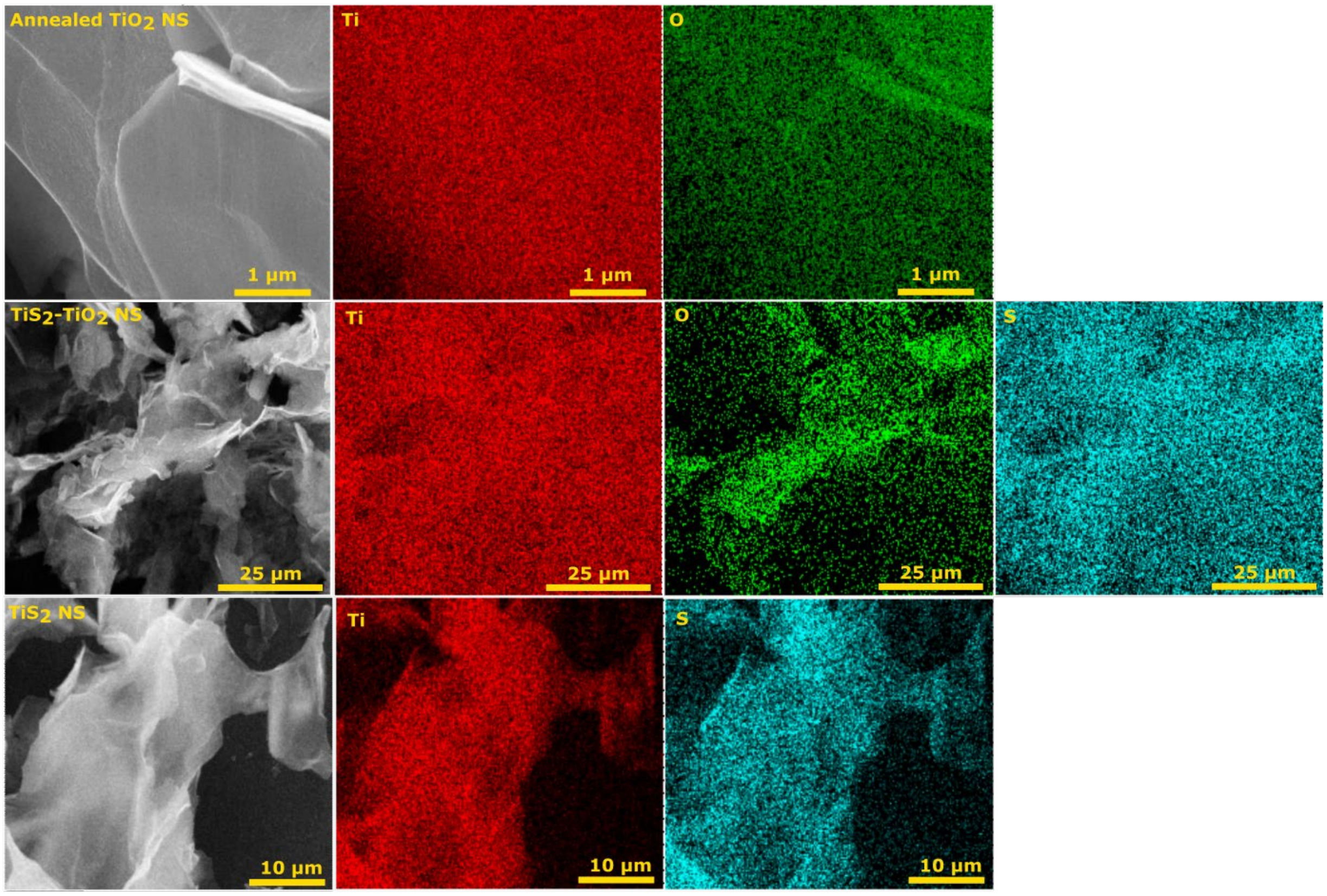

**Figure S3.** SEM images and EDS mapping of annealed $TiO_2$ NS (top), $TiS_2$-$TiO_2$ NS (middle), and $TiS_2$ NS (down)

*Raman analysis*:-

Raman spectroscopy analyses by 532 nm excitation of the annealed $TiO_2$ (**Figure 1 (b)**, black curve) show all the four modes of an anatase $TiO_2$ namely, $E_g$ (144.7 $cm^{-1}$), $E_g$ (197.5 $cm^{-1}$), $B_{1g}$ (397.4 $cm^{-1}$), $A_{1g}$+$B_{1g}$ (510.8 $cm^{-1}$), and $E_g$ (639.12 $cm^{-1}$).[4] The $TiS_2$-$TiO_2$ Raman spectrum has both $TiS_2$ and anatase $TiO_2$ Raman modes (**Figure 1 (b)**, red curve). In $TiS_2$-$TiO_2$, A1g (327.0 $cm^{-1}$) and Eg (223.0 $cm^{-1}$) modes correspond to $TiS_2$ and $E_g$ (147.9 $cm^{-1}$), $E_g$ (192.0 $cm^{-1}$), $B_{1g}$ (402.0 $cm^{-1}$), $A_{1g}$+$B_{1g}$ (473.0 $cm^{-1}$), $E_g$ (606.0 $cm^{-1}$) correspond to $TiO_2$ Raman mode are present.[1,4] The Raman spectrum of $TiS_2$ NS shows vibrational modes corresponding to that of $A_{1g}$ (335.0 $cm^{-1}$) and $E_g$ (223.0 $cm^{-1}$), **Figure S4**.[1] This shows the typical behaviour of bulk $TiS_2$.[5] Hence, it verifies the conversion of $TiO_2$ NS surface to $TiS_2$ NS upon long term sulfurization process.

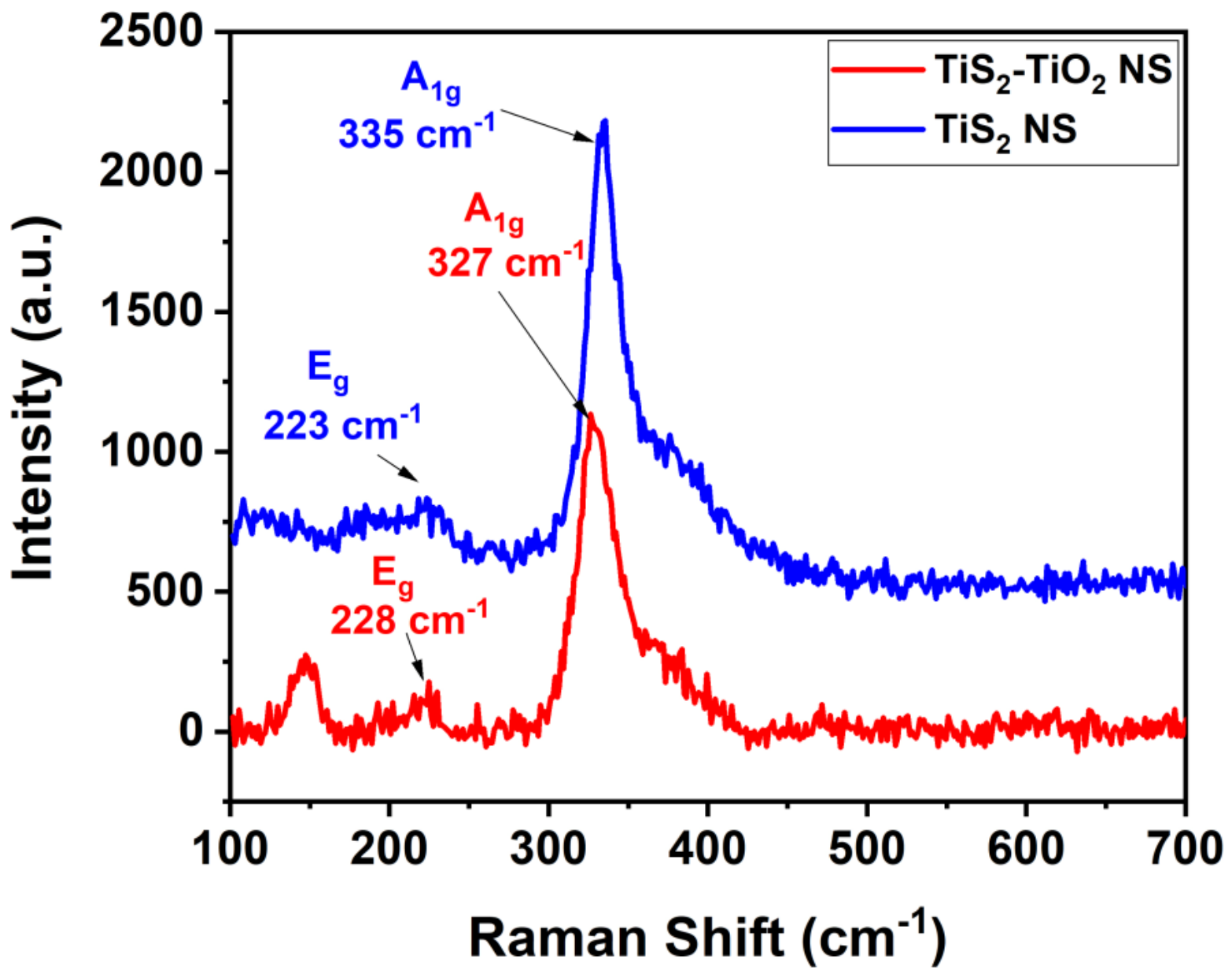

**Figure S4.** Comparison of the Raman spectra of $TiS_2$-$TiO_2$ NSs and $TiS_2$ NSs

*High resolution XPS analysis:*

For annealed $TiO_2$, Ti 2p shows two peaks positioned at 459.1 eV and 464.8 eV (**Figure S5 (a)**), arising due to spin-orbit splitting of Ti 2P3/2 and Ti 2P1/2, respectively.[6] This suggests the presence of $Ti^{4+}$ chemical state. The binding energy (BE) peak at 530.3 eV (**Figure S5 (b)**) is attributing to O 1s of annealed $TiO_2$. This further confirms the formation of Titanium oxide ($TiO_2$) NSs. Along with O1s peak two more oxide peak at 530.7 eV (O 1s a) and 531.5 eV (O 1s b) which shows hydroxyl OH is present in $TiO_2$.[6] In $TiS_2$-$TiO_2$ NSs (**Figure S5 (c)**), Ti 2p, positioned at 459.0 eV and 464.7 eV where they correspond to that of $Ti^{4+}$ chemical states.[7] Two peaks at the BEs of 530.4 eV and 531.9 eV correspond to O1s and hydroxyl OH (O 1s a) present in $TiS_2$-$TiO_2$ NSs (**Figure S5 (d)**).[7] While the other four at the BEs of 164.0 eV, 165.2 eV, 168.5 eV and 169.7 eV correspond to S2p and S2p of cationic $S^{4+}$ state (S II) chemical state (all the peaks are normalised by C 1s peak) (**Figure S5 (e)**).[7] This corroborates the formation of $TiS_2$-$TiO_2$ NSs heterostructure. The high resolution XPS spectra of Ti 2p for $TiS_2$ NSs (**Figure S5 (f)**) also shows two peaks. Here, the two peaks positioned at the BEs of 458.9 eV and 464.7 eV are indicating the presence of $Ti^{4+}$ chemical state as in $TiS_2$, Furthermore, O1s and hydroxyl OH shows two peaks at the BEs of 530.4 eV and 531.9 eV (O 1s a) in $TiS_2$ NSs (**Figure S5 (g)**). While the S2p and S2p of cationic $S^{4+}$ state (S II) has four peaks at the BEs of 163.8 eV, 164.9 eV, 168.8 eV and 170.0 eV of (**Figure S5 (h)**).[7]

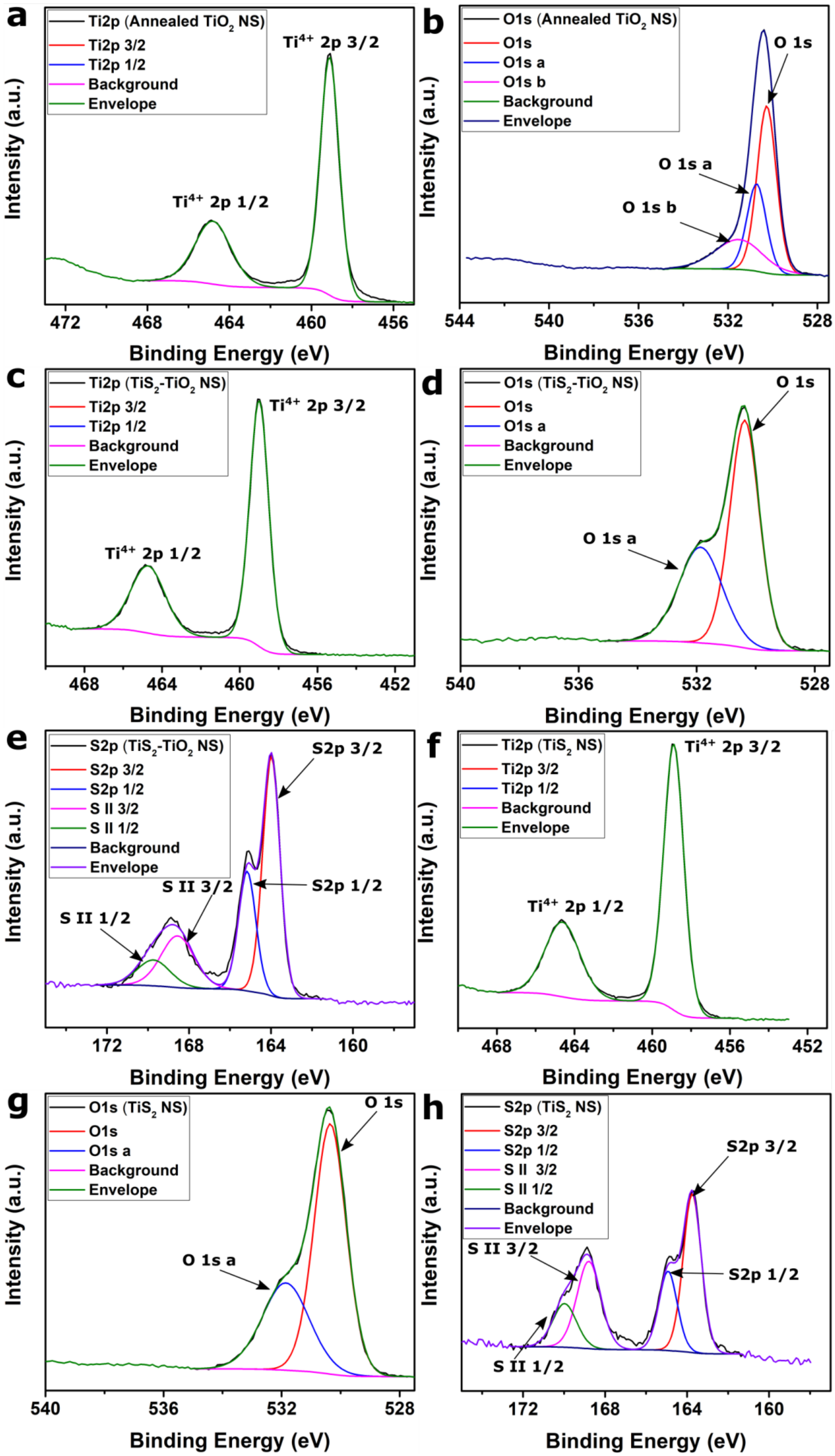

**Figure S5.** High resolution XPS spectra of: (a) Ti 2p of annealed $TiO_2$ NS, (b) O 1s of annealed $TiO_2$ NS, (c) Ti 2pof $TiS_2$-$TiO_2$ NS, (d) O 1s of $TiS_2$-$TiO_2$ NS, (e) S2p of $TiS_2$-$TiO_2$ NS, (f) Ti 2p of $TiS_2$ NS, (g) O 1s of $TiS_2$ NS, and (h) S2p of $TiS_2$ NS.

**Section 2:**

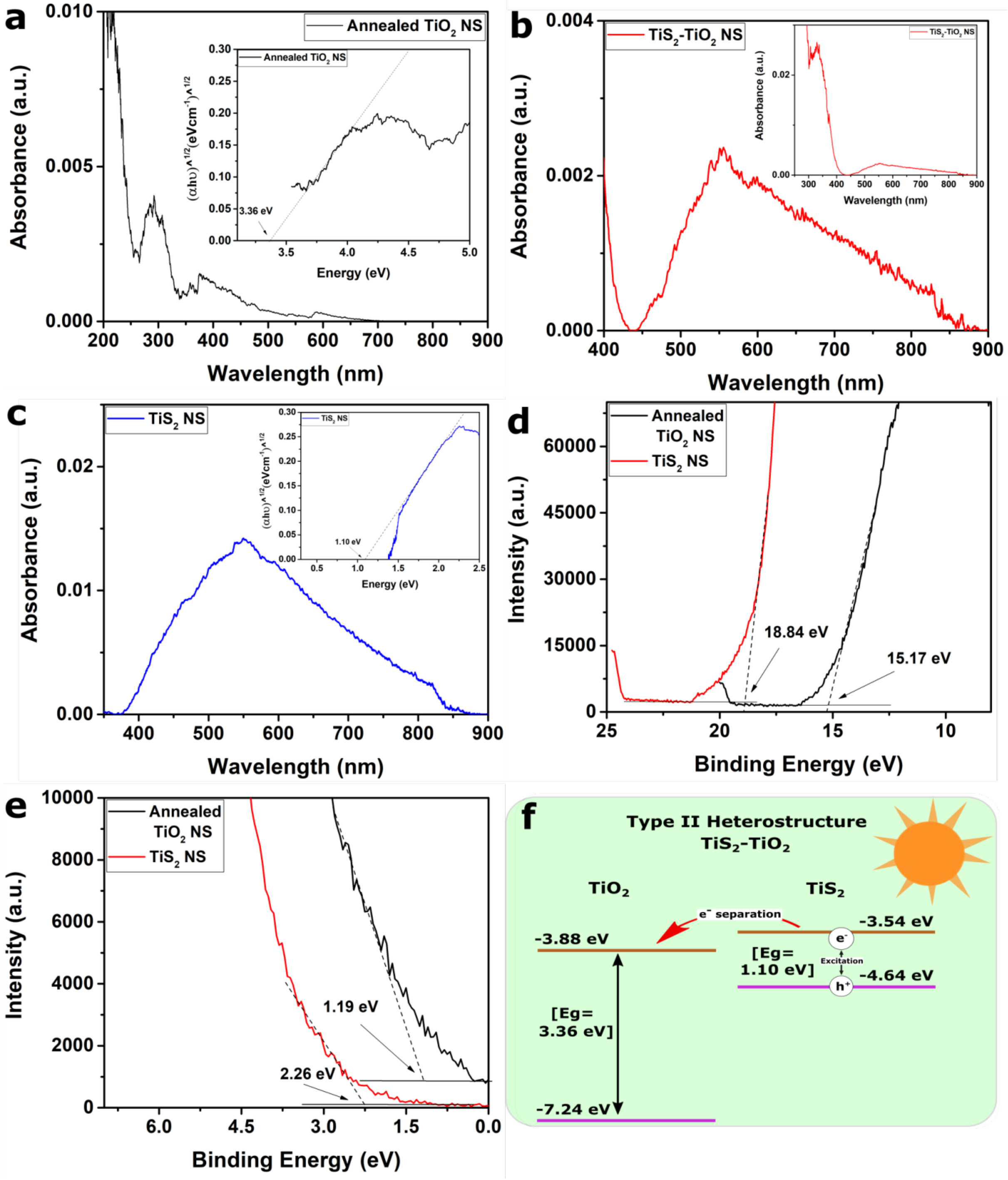

**Figure S6.** (a) UV-Vis DRS of annealed (a) $TiO_2$ NS (inset- Tauc plot of annealed $TiO_2$ NS) (b) $TiS_2$-$TiO_2$ NS (inset- Full range absorbance showing $TiS_2$ and $TiO_2$ absorption in visible and UV region) and (c) $TiS_2$ NS (inset- Tauc plot of $TiS_2$ NS from the light absorbance measurement). (d) Secondary electron cut off from Ultraviolet photoelectron spectroscopy (UPS) spectra of annealed $TiO_2$ NS (black) and $TiS_2$ NS (red) (e) Valence band region from UPS spectra of annealed $TiO_2$ NS (black) and $TiS_2$ NS (red) (f) Final Type II heterostructure band diagram schematic of annealed $TiO_2$ NS and $TiS_2$ NS calculated from the UPS results and Tauc plot.

Photo-response was also checked for the active material ($TiS_2$-$TiO_2$NS). For this, we have coated active material on a glassy carbon electrode and tested in 0.2 M $Na_2SO_4$ solution with an applied bias voltage of 0.2 V *vs* Ag/AgCl reference electrode.[2] The photo-response was tested with a white light source (using an LED of 70 $mW/cm^2$ power density). For comparison we also performed this experiment for Annealed $TiO_2$ NS and $TiS_2$ NS. As shown in **figure S7**, it is clear that because of easily separation of electron-hole pair. $TiS_2$-$TiO_2$ NS shows high photo current and fast decay in dark compare with $TiO_2$ NS and $TiS_2$ NS.[2,8]

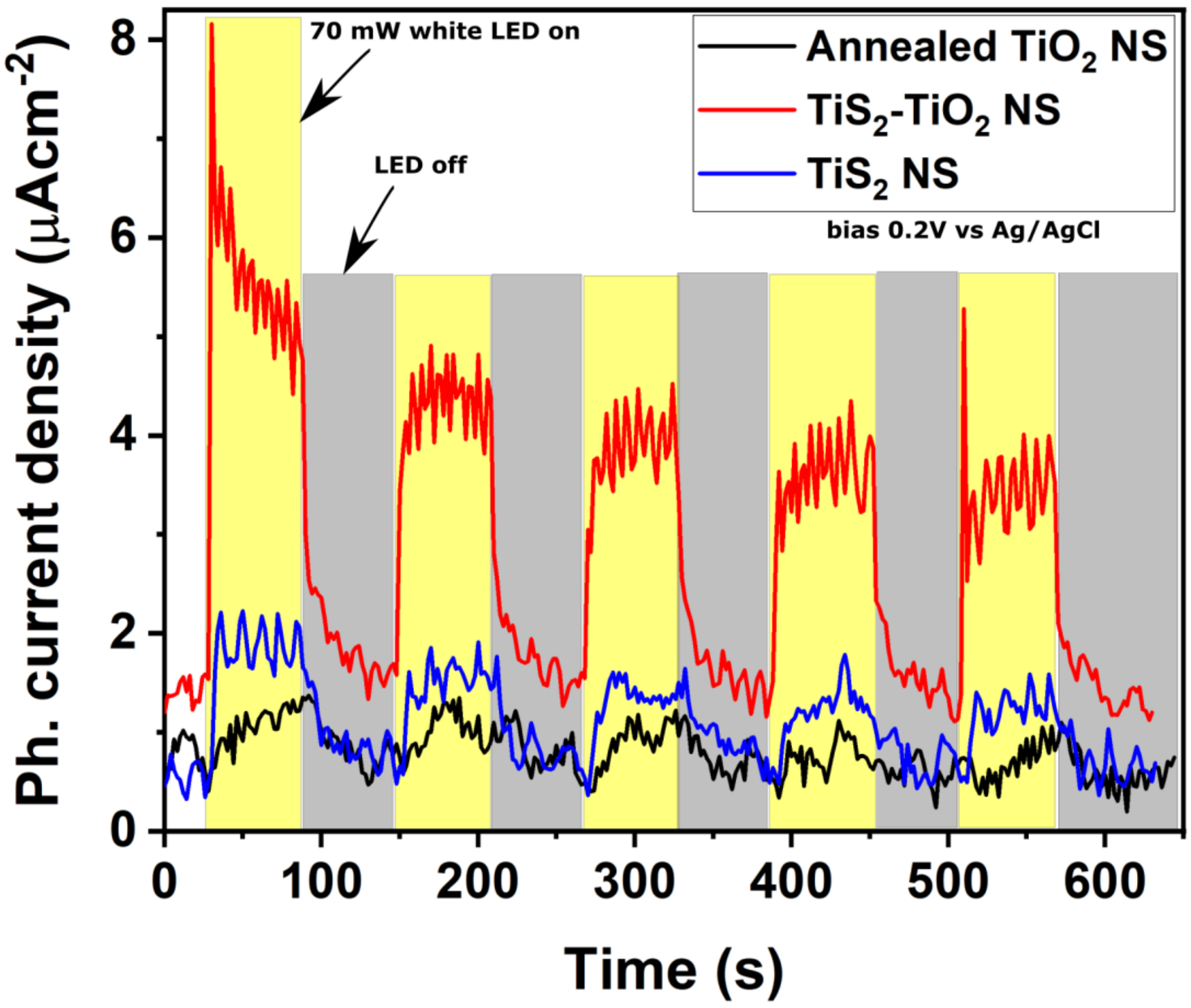

**Figure S7.** The photo-response of annealed $TiO_2$ NS (black), $TiS_2$-$TiO_2$ NS (red), and $TiS_2$ NS (blue)) in white LED light illuminations.

**Section 3:**

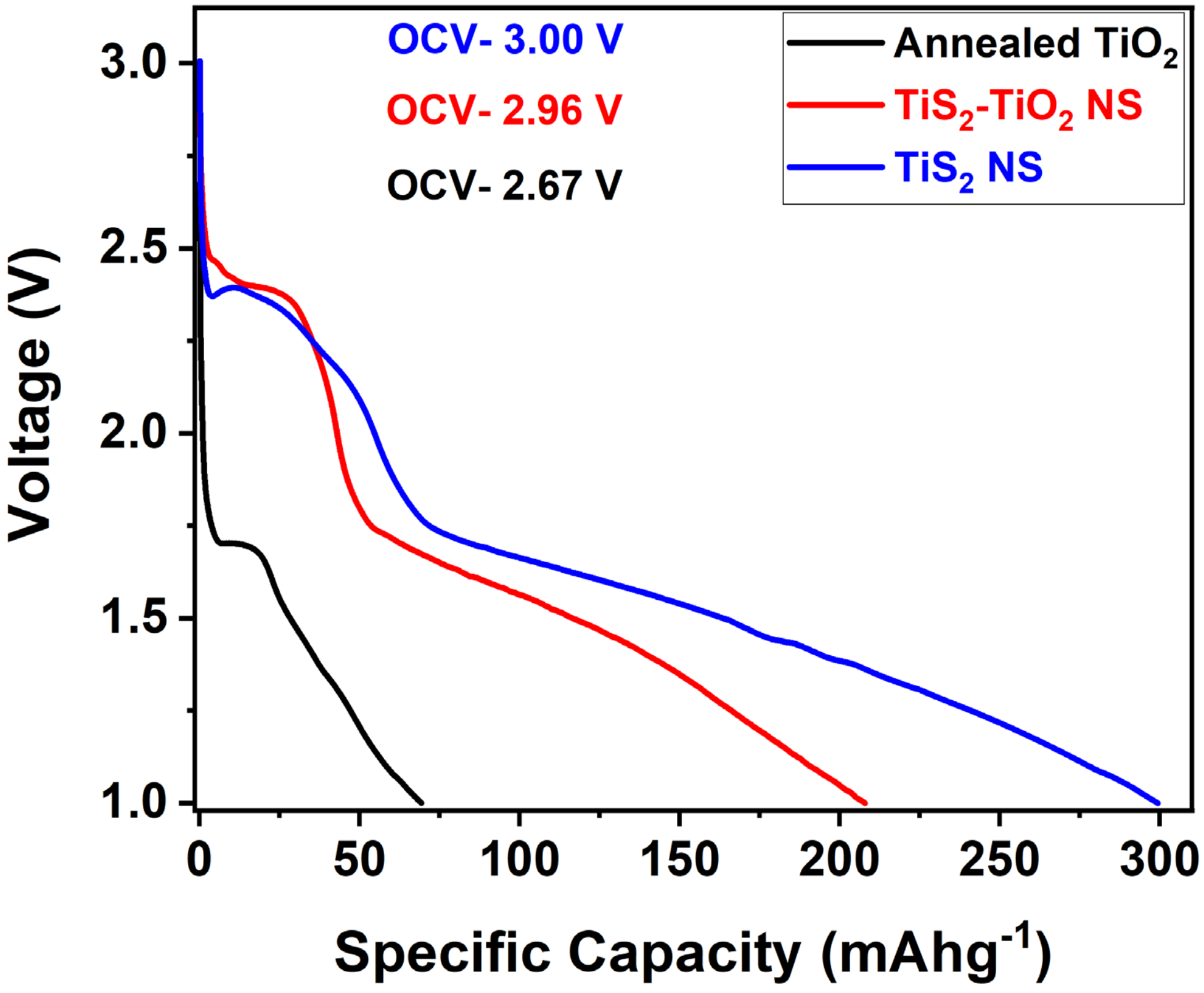

**Figure S8.** Li ion coin cell OCVs and specific capacity comparison of annealed $TiO_2$ NS (black), $TiS_2$-$TiO_2$ NS (red), and $TiS_2$ NS (blue)) with Li metal as other electrode

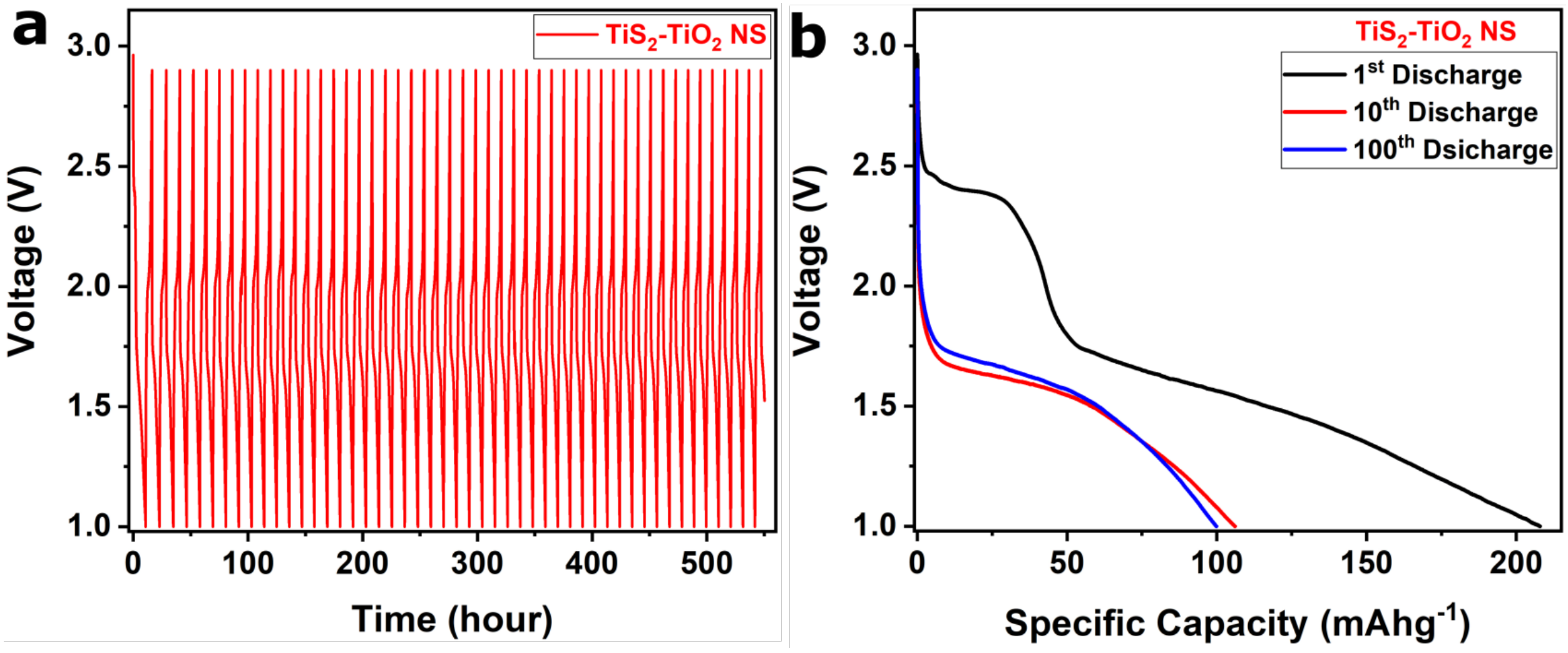

**Figure S9.** Conventional Li ion battery long cycles and specific capacity comparison after 100 cycles, of: (a,b) $TiS_2$-$TiO_2$ NS with Li metal as other electrode

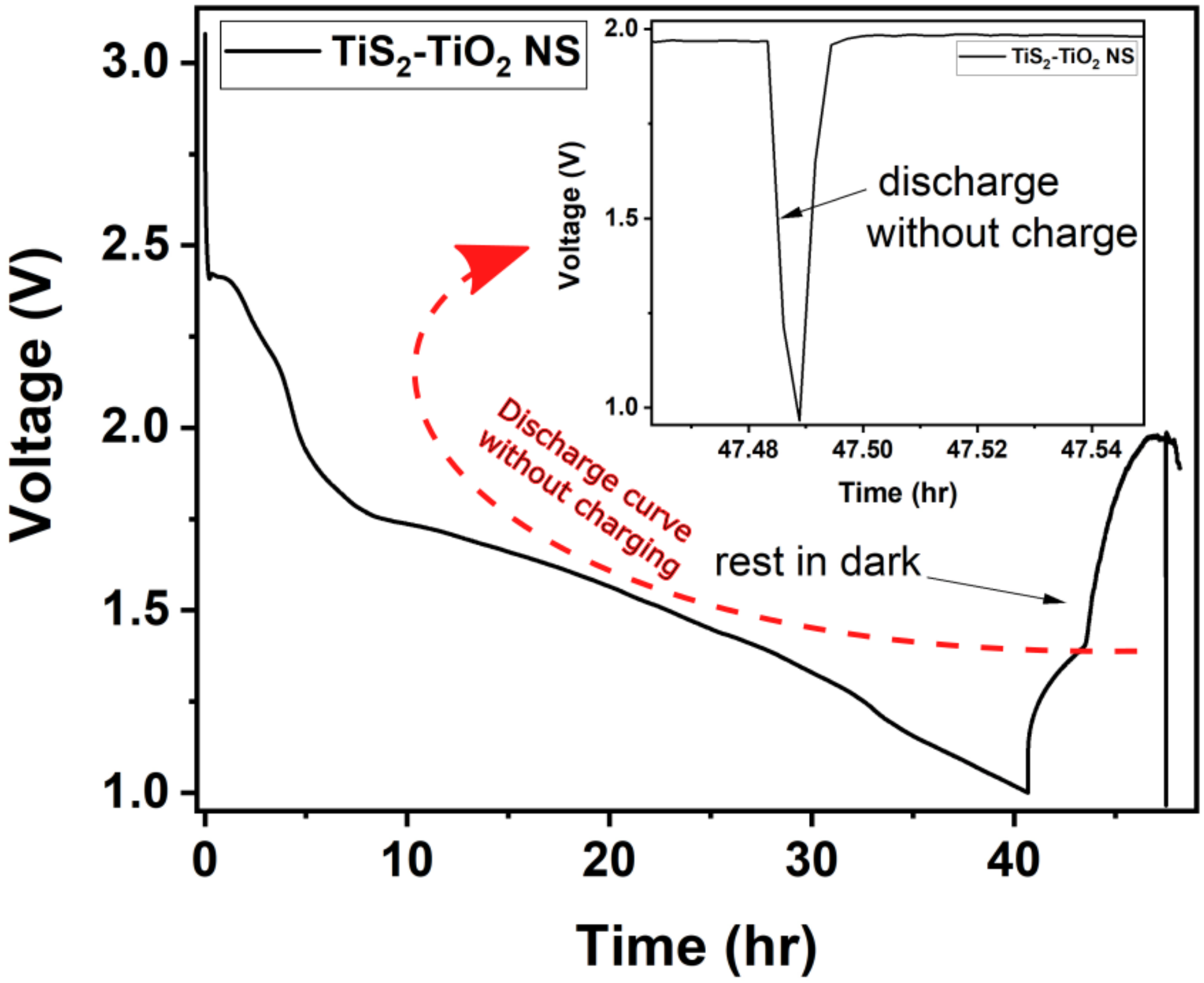

**Figure S10.** Discharge without photo charge Li-ion battery experiment of $TiS_2$-$TiO_2$ NSs electrodes (with Lithium metal as other electrode). After complete discharge by $21mAg^{-1}$ current load, we allowed for voltage stabilization. Further we discharge this battery by same $21mAg^{-1}$ current load without photo charge (inset Voltage-Time plot of sudden discharge part).

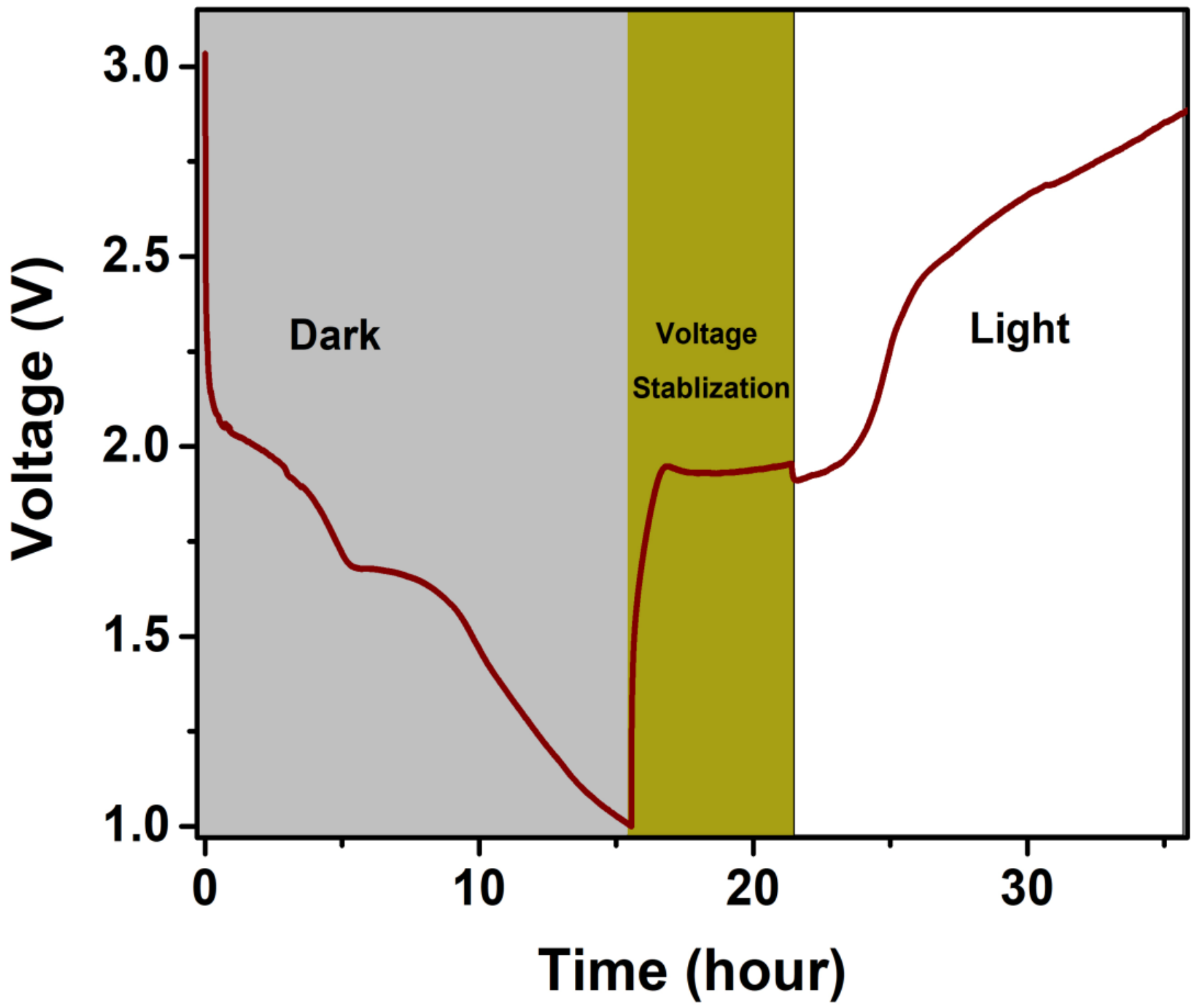

**Figure S11.** Discharging in dark by constant current load (light grey area) followed by voltage stabilization (dark yellow area), photo-charging (white area). (photo-battery of $TiS_2$-$TiO_2$ NSs based electrode with Lithium metal as other electrode)

**Section 4:**

Energy efficiency is calculated by taking ratio of discharge and charge profile curves. That is,

$$\text{Energy efficiency} = \frac{\text{Area under the discharge curve (Potential(V) vs. time (hrs))} * \text{const. current (mA)}}{\text{Area under the charge curve (Potential(V) vs. time (hrs))} * \text{const. current (mA)}}$$

**Table S1**. Comparison of enhancement in specific capacity in light of $TiS_2$-$TiO_2$ NSs based photoelectrode with other reported photo-rechargeable two electrode system

| Photo-active Electrode | Types of Metal ion | Specific Capacity in dark (mAhg$^{-1}$) | Specific Capacity in light (mAhg$^{-1}$) | Enhanced Specific Capacity in light (%) | References | Light type | power |
|---|---|---|---|---|---|---|---|
| **$V_2O_5$** | Li ion | 81 | 127 | 57[*)] | Nano Lett. 2021, 21, 3527–3532 | 455 nm | 12 $mWcm^{-2}$ |
| **Tetrakislawsone (TKL)** | Li ion | 296 | 332 | 12[#)] | Nano Lett. 2021, 21, 2, 907–913 | 405 nm , diode laser | 120 mW |
| **$MoS_2$-ZnO** | Zn ion | 245 | 340 | 38.77[#)] | ACS Nano 2021, 15, 16616–16624 | 455 nm | 12 $mWcm^{-2}$ |
| **$SnO_2/TiO_2$** | Li ion | 687 | 1308 | 90.39[a)] | Adv. Mater. 2021, 33, 2103558 | 365 nm Xenon lamp | 25 $mWcm^{-2}$ |
| **$TiS_2$-$TiO_2$** | Li ion | 170 | 586 | 245[#) c)] | Current work | White LED light | 70 mW $cm^{-2}$ |

[*)] Composite electrode; [#)] Single component electrode, [a)] anode material, [c)] cathode material

**Section 5:**

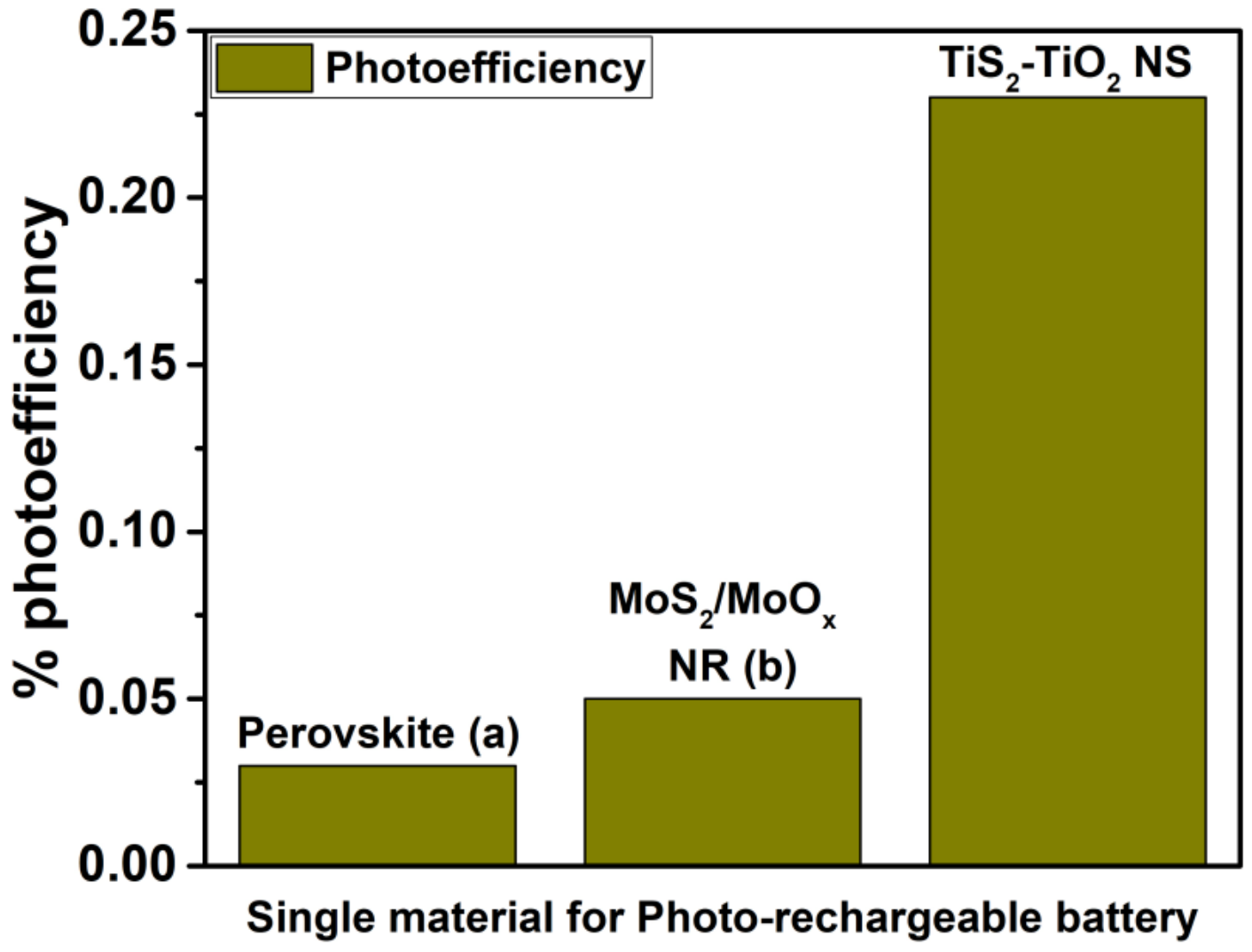

**Figure S12.** Comparison of photoefficiency for $TiS_2$-$TiO_2$ NSs based electrode with reported single material photo-battery electrode material. (a- [9], b- [2])

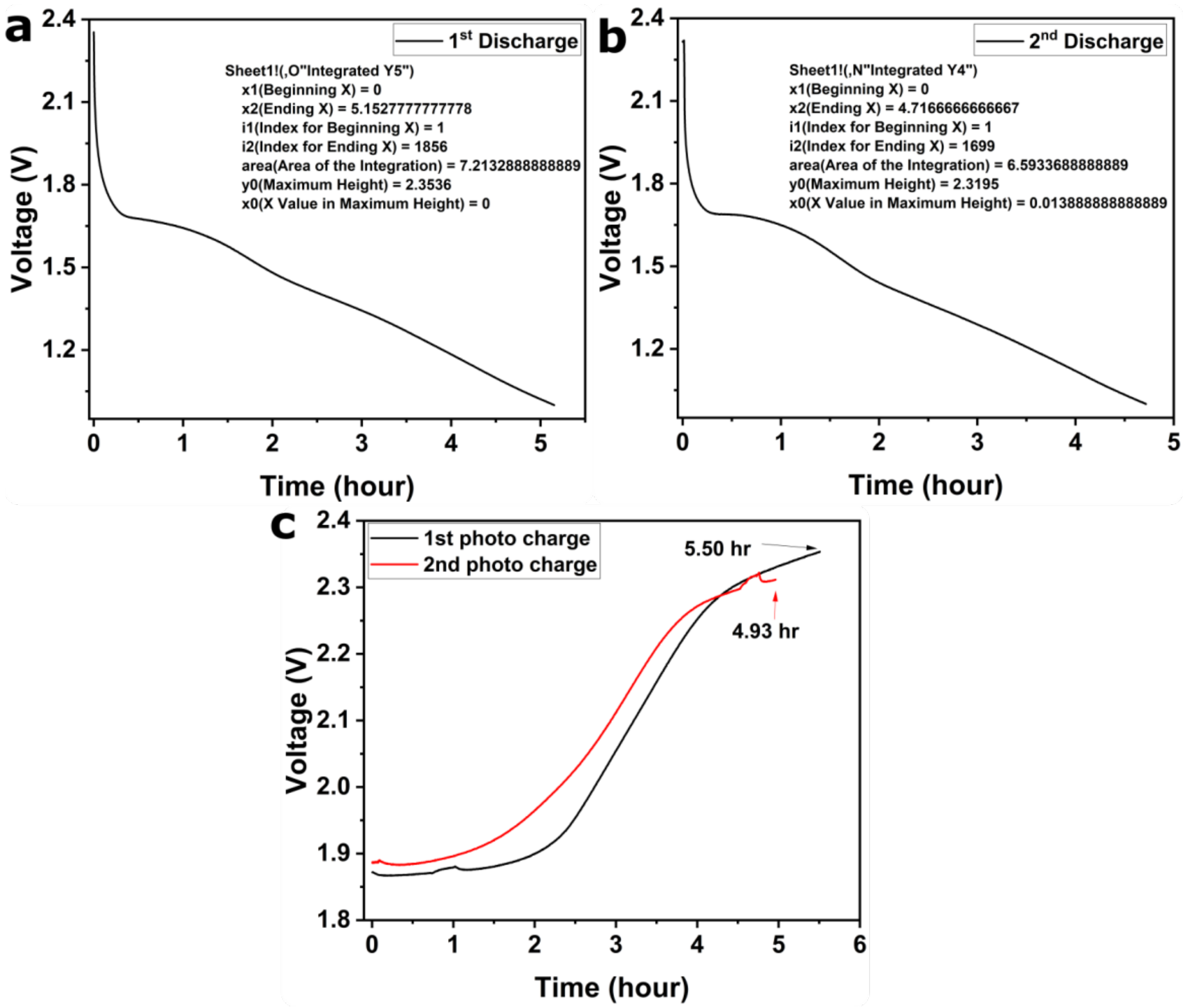

**Figure S13.** Discharge potential-time integration calculation for $TiS_2$-$TiO_2$ NSs: (a) 1[st]discharge after charge in white light and (b) 2[nd] discharge after charge in white light, and (c) photocharging duration in two cycles.

$$Photoefficiency = \frac{\text{Output energy (discharge)}}{\text{Input energy (solar energy)}}$$

$$\text{Photoefficiency for 1st Discharge} = \frac{\text{Integrated area for discharge curve (7.21 V.h)} \times \text{ current (0.12mA)}}{\text{Area (1 cm}^2) \times \text{power density (70 mWcm}^{-2}) \times \text{exposure time (5.5h)}}$$

$$\text{Photoefficiency \%} = 0.225\ \%$$

$$\text{Photoefficiency for 2nd Discharge} = \frac{\text{Integrated area for discharge curve (6.59 V.h)} \times \text{ current (0.12 mA)}}{\text{Area (1 cm}^2) \times \text{power density (70 mWcm}^{-2}) \times \text{exposure time (4.93h)}}$$

$$\text{Photoefficiency } \% = 0.229 \ \%$$

**Table S2**. Comparison of discharge specific capacity after light charge and photoefficiency of $TiS_2$-$TiO_2$ NSs based photoelectrode with other reported photo-rechargeable two electrode system

| Types of Photo-battery | Types of Metal ion | Photo-active Electrode | Specific Capacity ($mAhg^{-1}$) (after photo charge) | Photoefficiency (%) | References |
|---|---|---|---|---|---|
| **Two electrode** | Li ion | Hybrid $LiFePO_4$& N719 dye | 104 | 0.07 | Nat Commun 8, 14643 (2017) |
| | | | | | |
| **Two electrode** | Li ion | $V_2O_5$ | 31 | 0.22*) | Nano Lett. 2021, 21, 3527–3532 |
| | | | | | |
| **Two electrode** | Li ion | $MoS_2/MoO_x$ | 120 | 0.05#) | Small 2021, 17, 2105029 |
| | | | | | |
| **Two electrode** | Li ion | $TiS_2$-$TiO_2$ | 277 | 0.23#) | Current work |
| | | | | | |

(*) Composite electrode; #) Single component electrode)

**Section 6:**

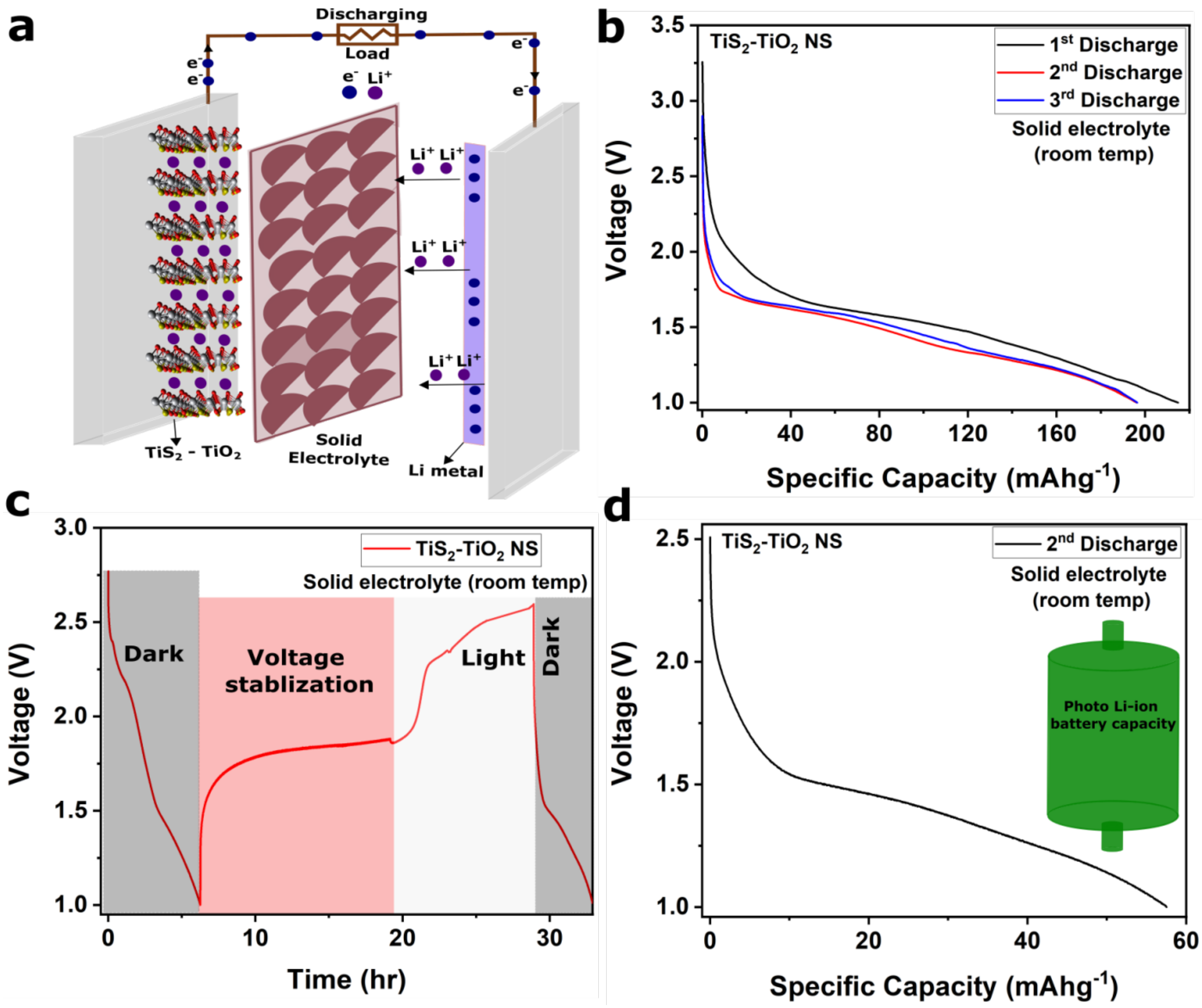

**Figure S14.** Solid state electrolyte (20% $LiClO_4$ in Polyethylene oxide) based conventional and photo battery performance: (a) Schematic illustrations for mechanism of lithiation by a single nanosheet structured type II semiconductor heterostructure of $TiS_2$-$TiO_2$ NSs with Li metal as the counter electrode in a two-electrode solid electrolyte (20 weight% $LiClO_4$ in polyethylene oxide) based battery. (b) Comparison of conventional battery specific capacity of 1$^{st}$ discharge, 2$^{nd}$ discharge and 3$^{rd}$ discharge of $TiS_2$-$TiO_2$ NSs electrodes (with Lithium metal as counter electrode). (c) photo battery cycle, discharging in dark by constant current load (black area) followed by voltage stablization (pink area), photo-charging (white area) and discharge after

photo charge (black area), for $TiS_2$-$TiO_2$ NSs based electrode with reported single material photo battery electrode material. (d) $2^{nd}$ discharge specific capacity after photo charge.

**Section 7: *In situ* and *ex situ* characterization of solar battery**

To understand Li ion intercalation and deintercalation in $TiS_2$-$TiO_2$ NSs electrodes by light irradiation as well as conventional electrochemical method, we performed *in situ* Raman characterisation. In the course of this experiment, we compared $A_{1g}$ and $E_g$ Raman modes of $TiS_2$ and $E_g$ Raman modes of $TiO_2$. First we assembled $TiS_2$-$TiO_2$ NSs electrode battery with lithium metal, where $TiS_2$-$TiO_2$ NSs was coated on ITO coated glass (**Figure S15 (a)**). We recorded Raman spectra of initial $TiS_2$-$TiO_2$ NSs from back side of ITO coated glass (**Figure S15 (b)**, black) (more detail schematics representation **Figure S16**). In this respect $TiS_2$ Raman modes observed at $A_{1g}$ - 332.02 $cm^{-1}$ (denoted as wine colour short dashed line in **Figure S15(b),** $E_g$ - 227.29 $cm^{-1}$ (denoted as deep yellow colour short dashed line in **Figure S15 (b)**) and $E_g$ - 147.08 $cm^{-1}$ for $TiO_2$ (denoted as deep purple colour short dashed line in (**Figure S15 (b)**)).[4,5] To understand the change in Raman modes after lithiation in $TiS_2$-$TiO_2$ NSs, we discharged the battery till 1 V and converted to $Li_xTiS_2$-$TiO_2$ NSs. Here, in electrochemically lithiated $Li_xTiS_2$-$TiO_2$ NSs, the Raman modes shifted to ($A_{1g}$) 352.09 $cm^{-1}$ and ($E_g$) 244.23 $cm^{-1}$ for $TiS_2$ and an extra peak is appeared at 177.66 $cm^{-1}$ (denoted as X, pink colour short dashed line in **Figure S15 (b)**, red). But the $E_g$ mode of $TiO_2$ was not observed after lithiation (**Figure S15 (b)**, red). This is matching with earlier report of lithiated $TiO_2$ Raman spectra.[10,11] Further, to know about Li ion deintercalation from $Li_xTiS_2$-$TiO_2$ NSs by light, we irradiated the electrode ($Li_xTiS_2$-$TiO_2$ NSs electrode) with white light and allow Li ion deintercalation from $Li_xTiS_2$-$TiO_2$ NSs. This time we noticed that $TiS_2$ Raman modes reappeared at the same positions as initial at, $A_{1g}$ - 337.6 $cm^{-1}$, $E_g$ - 227.49 $cm^{-1}$ but again the $E_g$ mode of $TiO_2$ was not observed and extra peak X peak - 170.39 cm$^{-1}$ was observed (**Figure S15 (b)**, blue). This is due to the fact that hole is only generated in $TiS_2$ which pushes Li ion for de-intercalation from $TiS_2$ only, in tune with the band structure. Hence $Li^+$ de-intercalation from $TiO_2$ is not possible by light irradiation. We discharged the battery to make $Li_xTiS_2$-$TiO_2$ NSs. Again the Raman modes shifted to $A_{1g}$ - 360.33 cm$^{-1}$, $E_g$ - 238.00 cm$^{-1}$ for $TiS_2$ with an extra X peak - 172.39 cm$^{-1}$ but $E_g$ mode of $TiO_2$ was not observed (**Figure S15 (b)**, dark green). Furthermore, to compare Li ion electrochemical de-intercalation and light assisted de-intercalation from $Li_xTiS_2$-$TiO_2$ NSs, we de-intercalate the Li ion by externally electrochemical charging. In this instance, we noticed that $TiS_2$ Raman modes appear at the same positions as, $A_{1g}$ - 332.0 cm$^{-1}$, $E_g$ - 236.07 cm$^{-1}$. Unlike light de-intercalation, due to electrochemically de-intercalation from $Li_xTiS_2$-$TiO_2$ NSs, Li ions got de-intercalated from $TiS_2$ as well as $TiO_2$. As a result, $E_g$ mode of $TiO_2$ is observed at 141.90 cm$^{-1}$ and no any extra peak was observed (**Figure S15 (b)**, pink). This is in tune with the earlier reports.[10,11] This establishes the hypothesis that light assisted de-intercalation helps de-intercalation only from $TiS_2$, due to the fact that hole is generated only in $TiS_2$ (due to the staggered band structure), after the absorption of light by the $TiS_2$-$TiO_2$ NSs.

To understand Li deposition on other electrode, we fabricated $Li_xTiS_2$-$TiO_2$ NSs after intercalation of $TiS_2$-$TiO_2$ NSs electrode from Li metal (discharged *half cell*). Further we assembled again a cell with graphite as other electrode (which was drop casted on ITO coated glass) and $Li_xTiS_2$-$TiO_2$ NSs as cathode (Schematic illustrations in **Figure S15 (c)**). During this experiment, we recorded Raman modes of graphite electrode from light window as shown in (**Figure S15 (c)**). We regularly irradiate light on the $Li_xTiS_2$-$TiO_2$ NSs for Li ion de-intercalation process to occur. Both D and G bands get broadened after Li ion intercalation (**Figure S15 (d)**).[12] This shows that Li ions were get intercalated in graphite electrode.[12] Further, we recorded powder X-ray diffraction of pristine graphite, electrochemically lithiated graphite (lithiation done in discharging mode of battery with Li metal as opposite electrode), and photo assisted lithiated graphite (lithiation done in battery by irradiated light on $Li_xTiS_2$-$TiO_2$ NSs in rest mode).

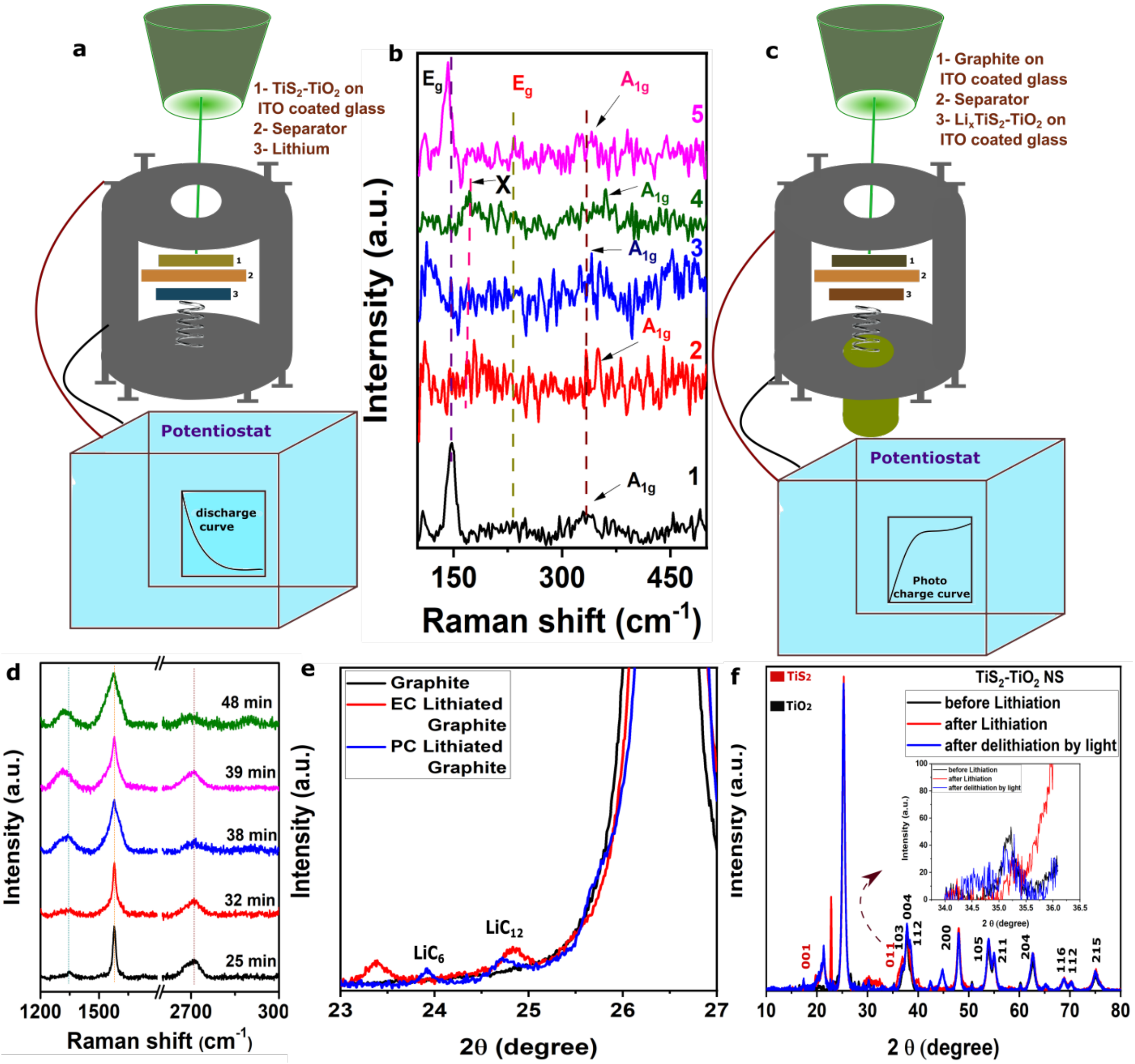

**Figure S15.** (a) Schematic illustrations of *in situ* Raman and *ex situ* powder X-ray diffraction characterisation and experiment of lithiation and delithiation in $TiS_2$-$TiO_2$ NSs electrodes (with Lithium metal as counter electrode) and Graphite electrodes (with $Li_xTiS_2$-$TiO_2$ NSs as counter electrode): (a) initial Raman spectra recorded of$TiS_2$-$TiO_2$ NSs electrodes (b) Raman spectra of 1- initial $TiS_2$-$TiO_2$ NSs, 2-$Li_xTiS_2$-$TiO_2$ NSs, 3- after delithiated $Li_xTiS_2$-$TiO_2$ NSs by light, 4-$Li_xTiS_2$-$TiO_2$ NSs, 5- after delithiated $Li_xTiS_2$-$TiO_2$ NSs by constant current. (c) *In situ* Raman spectra recorded of graphite electrodes. (d) *In situ* Raman spectra of Light assisted lithiation in graphite electrodes (other side $Li_xTiS_2$-$TiO_2$ NSs electrodes) at different time interval. (e) *Ex situ* powder X-ray diffraction spectra recorded of graphite electrodes (black), electrochemically lithiated graphite from Li metal as counter (red), Photo-assisted lithiated graphite from $Li_xTiS_2$-$TiO_2$ NSs electrodes (blue). (f) *Ex situ* powder X-ray diffraction spectra recorded of $TiS_2$-$TiO_2$ NSs electrodes (black), electrochemically converted $TiS_2$-$TiO_2$ NSs to Lithiated $Li_xTiS_2$-$TiO_2$ NSs from Li metal as counter (red), Photo assisted de-lithiated $TiS_2$-$TiO_2$ NSs from $Li_xTiS_2$-$TiO_2$ NSs electrodes (blue) [inset- zoom of $TiS_2$ (011 plan) change during lithiation and delithiation of $TiS_2$-$TiO_2$ NSs].

As shown in **Figure S15 (e)**, $Li^+$ appears as $LiC_6$ and $LiC_{12}$ in graphite in both electrochemically and light assisted intercalation method. In pristine graphite, 111 plane appears at $2\Theta \sim 26.44^o$ (**Figure S15 (e)**, black) indicating the interplanar distance of 3.3 Å, but for electrochemically lithiated graphite, apart from the peak at $2\Theta = 26.51^o$ additional XRD peaks of $LiC_6$ at $2\Theta = 23.38^o$ and $LiC_{12}$ at $2\Theta = 24.83^o$ (**Figure S15 (e)**, red) are appeared.[12] Furthermore, in light assisted intercalated graphite also, peaks at $2\Theta = 26.54^o$ and $LiC_6$ peak at $2\Theta = 23.91^o$ and $LiC_{12}$ peak at $2\Theta = 24.72^o$ (**Figure S15 (e)**, blue) are appeared. Both *in situ* Raman and *ex situ* XRD data confirms that Li ion deposited in other electrode. Furthermore we performed *ex situ* XRD for $TiS_2$-$TiO_2$ NSs, lithiated $Li_xTiS_2$-$TiO_2$ NSs from Li metal as counter and photo assisted de-lithiated $TiS_2$-$TiO_2$ NSs from $Li_xTiS_2$-$TiO_2$ NSs electrodes (**Figure S15 (f)**). We noticed that $TiS_2$ 011 plane is shifted after lithiation from $2\Theta = 35.15^o$ to $2\Theta = 35.35^o$ in $Li_xTiS_2$-$TiO_2$. This is due to the increased interplanar of 011 after intercalation. It was found that the de-intercalation is shifting this peak back to the original value indicating the photo assisted charging process.

But at the same time, changes in $TiO_2$ planes are not recoverable after light assisted de-lithiation of $Li_xTiS_2$-$TiO_2$ (**Figure S18)**. Hence *in situ* and *ex situ* data confirm that light irradiation helps in Li ion de-intercalation from $TiS_2$ phase of $Li_xTiS_2$-$TiO_2$ and the ions get intercalated at the graphite electrode forming different intercalated structures such as $LiC_6$ and $LiC_{12}$.

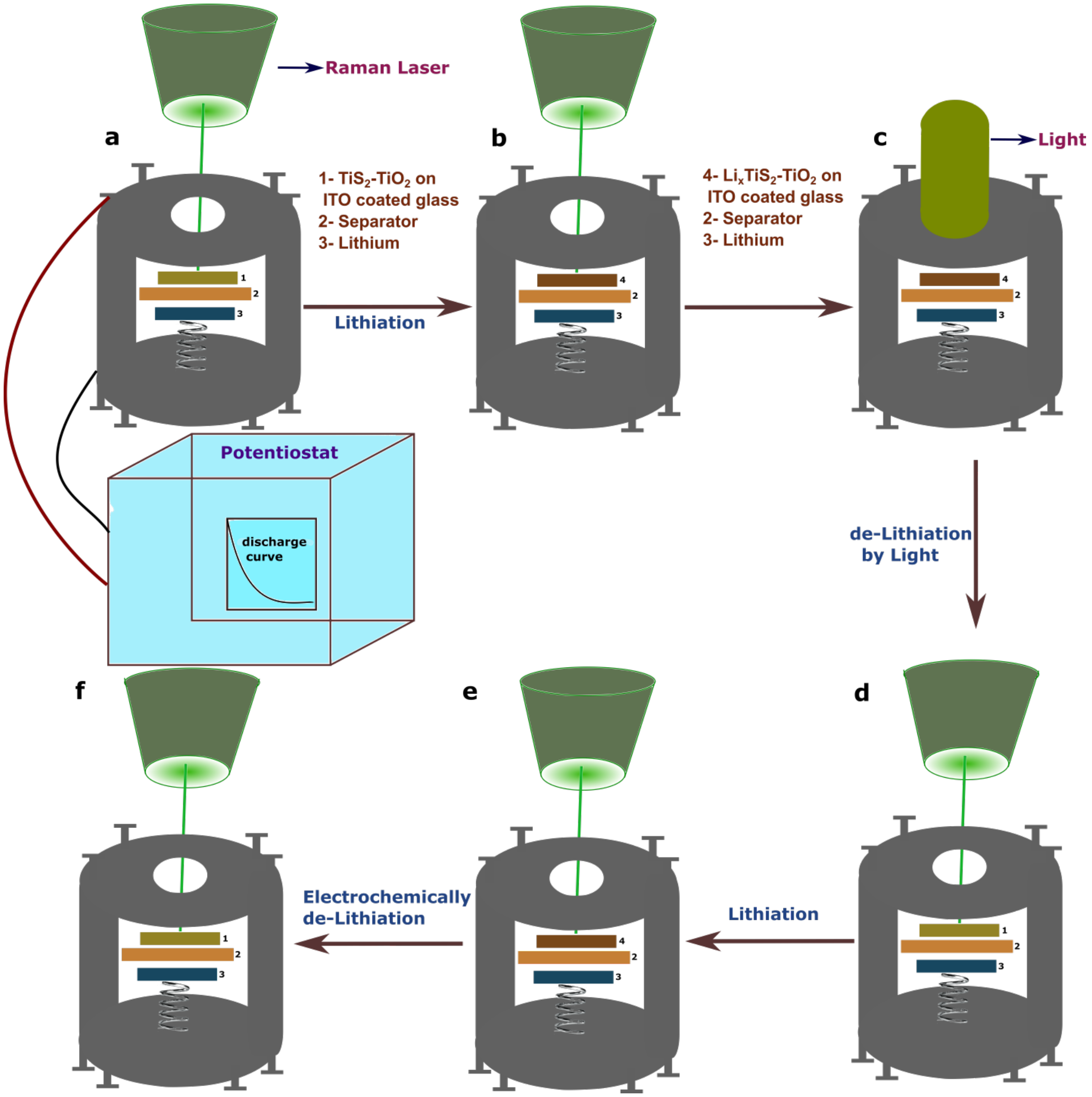

**Figure S16.** (a) Schematic illustrations of *in situ* Raman characterisation of lithiation and delithiation in $TiS_2$-$TiO_2$ NSs electrodes (with Lithium metal as counter electrode): (a) Initial Raman spectra recorded of $TiS_2$-$TiO_2$ NSs electrodes (b) after 1$^{st}$ discharge, Raman spectra recorded of $Li_xTiS_2$-$TiO_2$ NSs (c) Light assisted delithiation of $Li_xTiS_2$-$TiO_2$ NSs (d) after Light assisted delithiation of $Li_xTiS_2$-$TiO_2$ NSs Raman spectra recorded of $TiS_2$-$TiO_2$ (e) after 2$^{nd}$ discharge, Raman spectra recorded of $Li_xTiS_2$-$TiO_2$ NSs (f) after Electrochemically delithiation of $Li_xTiS_2$-$TiO_2$ NSs Raman spectra recorded of $TiS_2$-$TiO_2$

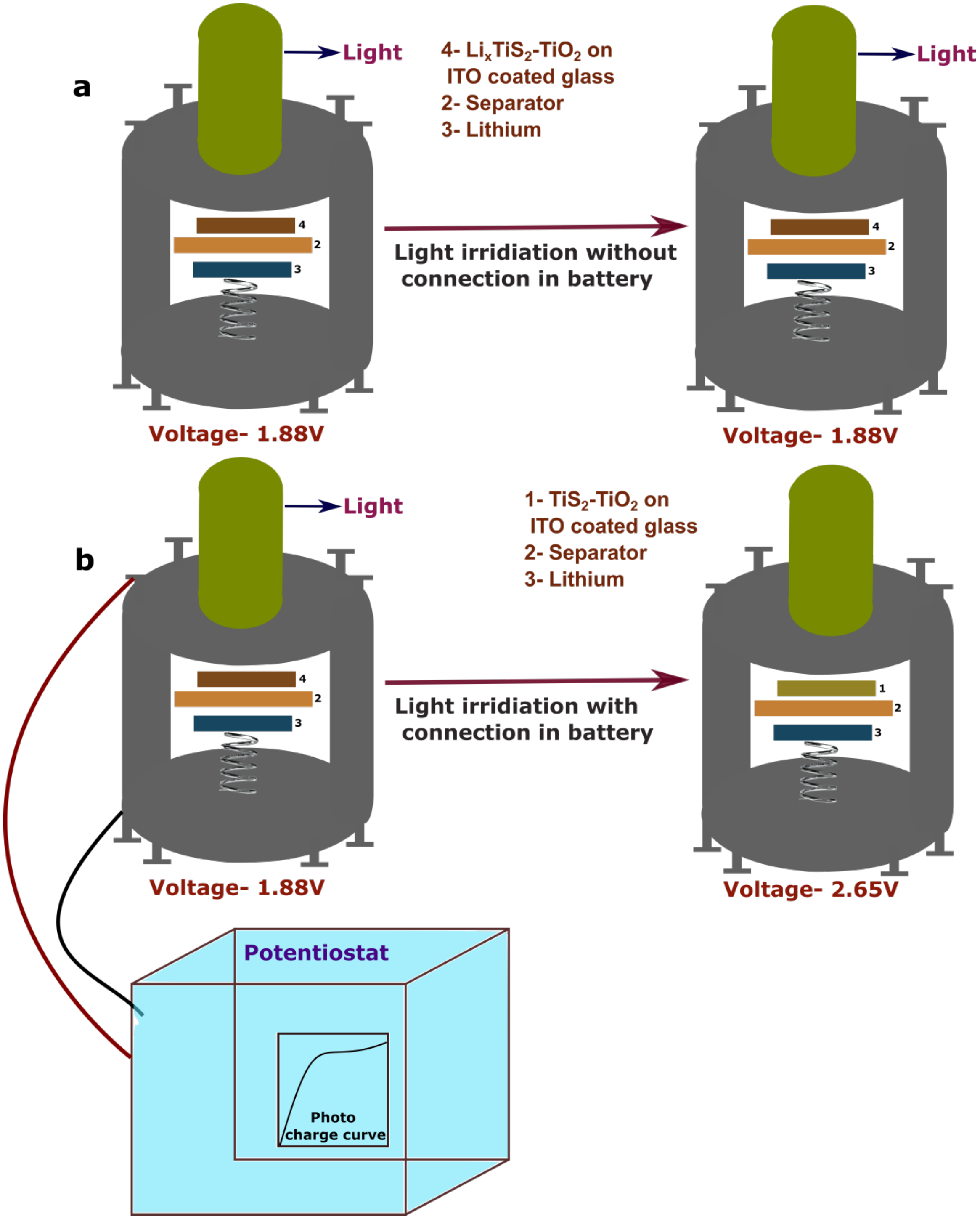

**Figure S17.** (a) Schematic illustrations of control experiment of de-lithiation with and without external connection for $Li_xTiS_2$-$TiO_2$ NSs electrodes (with Lithium metal as counter electrode)

(a) Light irradiation on $Li_xTiS_2$-$TiO_2$ NSs without external connection (b) Light irradiation on $Li_xTiS_2$-$TiO_2$ NSs with external connection by potentiostat.

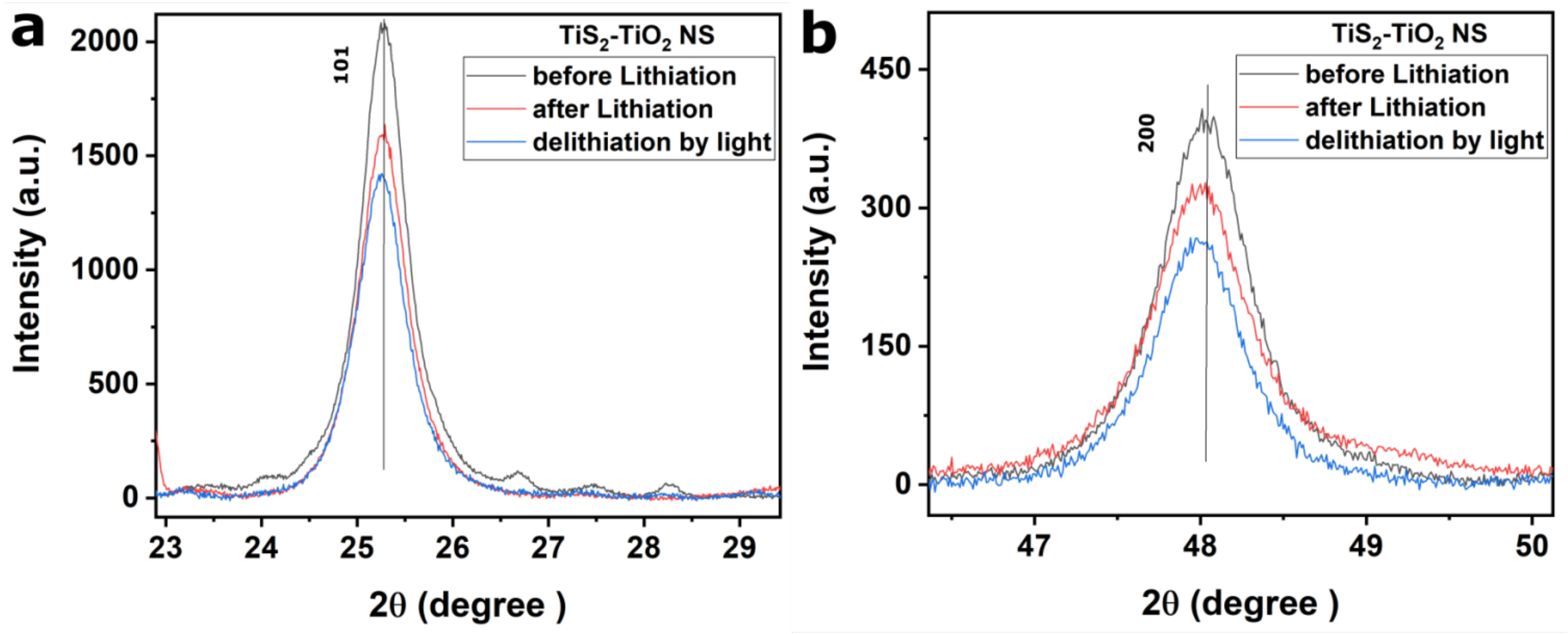

**Figure S18.** *Ex situ* powder X-ray diffraction spectra recorded of $TiS_2$-$TiO_2$ NSs electrodes(a) comparison of $TiO_2$ 101 plane (b) and $TiO_2$ 200 plane, before lithiation (black), electrochemically converted $TiS_2$-$TiO_2$ NSs to lithiated $Li_xTiS_2$-$TiO_2$ NSs from Li metal as counter (red), Photo assisted de-lithiated $TiS_2$-$TiO_2$ NSs from $Li_xTiS_2$-$TiO_2$ NSs electrodes (blue)

**Section 8:**

The mechanism of photo charging of the solar battery *full cell* can be now depicted as seen in **Figure S19**. When light irradiated on $Li_xTiS_2$-$TiO_2$ NSs, Li ions get de-intercalated from $Li_xTiS_2$-$TiO_2$ NSs and intercalated in graphite as shown in schematics **Figure S19**) (mechanism discussed in main text and proven *via in situ* Raman and *ex situ* XRD studies, as shown before). This mechanism is further verified theoretically.

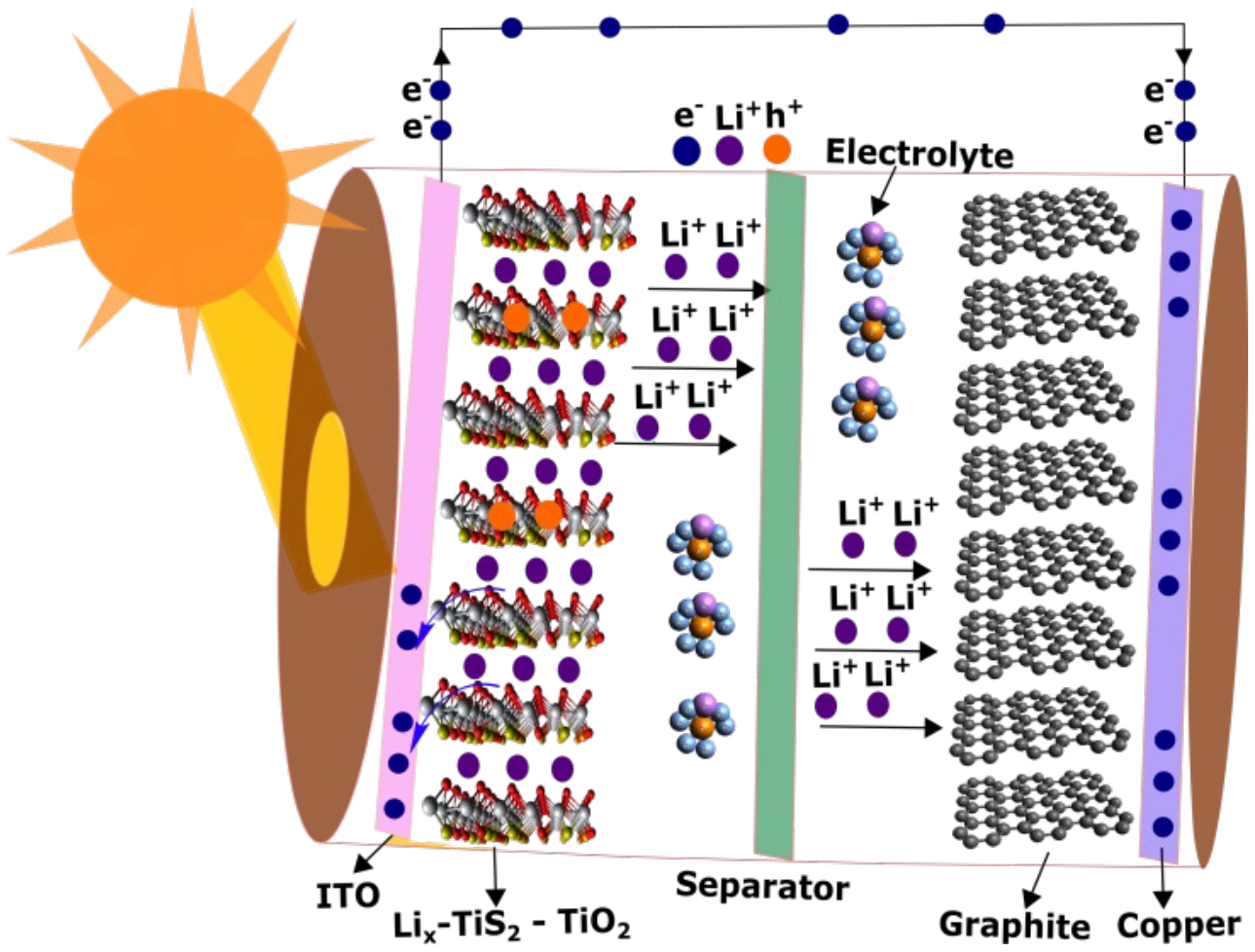

**Figure S19.** Schematic illustrations for mechanism of Light-assisted photo delithiation by a single nanosheet structured type II semiconductor heterostructure of $Li_xTiS_2$-$TiO_2$ NSs with graphite as the counter electrode in a two-electrode photo-rechargeable battery.

**Section 9:**

A. **Computational details**:

Our calculations are based on Density functional theory (DFT) which uses PBE form for the Generalized Gradient approximation(GGA).[13] These calculations were performed in Vienna ab-initio simulation package(VASP)[14] using a plane wave basis set with the projector augmented plane-wave method (PAW).[15] For all calculations the kinetic energy cutoff for plane-waves was set to be 550 eV. The energy and force convergence criteria were set to be $10^{-4}$ eV and 0.02 eV $Å^{-1}$ respectively. As pointed out by Zhenpeng Hu *et.al* accurate Titanium Oxide calculations require Hubbard corrections on top of DFT.[16] Therefore we have employed DFT+U approach introduced by Dudradev *et al.*[17] For our $TiO_2$ (anatase) calculations the Hubbard parameter was set to 3.5 eV,[18–20] while a Hubbard parameter of 2.1 eV was used for $TiS_2$.[21] A Monkhorst-Pack mesh of 12x12x5 and 9x9x3 was used for $TiS_2$ and $TiO_2$ respectively, to sample the Brillouin zone for band structure calculations.[22] For binding energy and work function calculations, a Monkhorst-Pack mesh of 5x5x1 is used for mono-layers of $TiS_2$ and $TiO_2$ with 48 atoms. For band offset calculations, we have used planar averaged Hartree potential to compute the offset between $TiS_2$ and $TiO_2$. Bulk calculations of $TiS_2$ and $TiO_2$ were performed using HSE06 hybrid functional. VBM (valence band maxima) for both systems were referenced to their respective macroscopic averages, which in turn are referenced to the vacuum level. For this calculation multi-layered slabs of $TiS_2$[001] and $TiO_2$[101] are taken along with vacuum. The lateral dimensions of materials are kept same as the bulk and the geometries were relaxed. Central layers in this system represent the bulk and using the macroscopic average of this region one can reference the VBM of the bulk to the vacuum.[23,24] We can obtain the valence band offset (VBO) between the two materials using this method. The simplified formula for the above method is given below

$$VBO = \left(VBM_{TiS2} - \bar{\bar{V}}_{TiS2} + \bar{\bar{V}}_{TiS2} - \bar{\bar{V}}_{vac}\right) - \left(VBM_{TiO2} - \bar{\bar{V}}_{TiO2} + \bar{\bar{V}}_{TiO2} - \bar{\bar{V}}_{vac}\right)$$
$$= \left(\widetilde{VBM}_{TiS2} + \bar{\bar{V}}_{TiS2-vac}\right) - \left(\widetilde{VBM}_{TiO2} + \bar{\bar{V}}_{TiO2-vac}\right) = \Delta\widetilde{VBM} + \Delta\bar{\bar{V}}_{x-vac}$$

where $\bar{\bar{V}}_x$ is the macroscopic average of the bulk, $\bar{\bar{V}}_{x-vac}$ is the macroscopic average of the bulk referenced to the vacuum, and $\widetilde{VBM}_x$ is the VBM with respect to the macroscopic average of the bulk. The conduction band offset is simply obtained by adding the band-gap to the VBM. However, PBE is known to underestimate the band gap, and therefore the experimental band gap can be combined with the above VBO for band alignment.[24,25] Another option is to compute the band structure (VBM referenced to the macroscopic average and also the band gap) using a hybrid functional, e.g. HSE06 , while the potential alignment ($\bar{\bar{\bar{V}}}_{x-vac}$) for the multi-layered slab is done using a GGA functional.[26] In this case we are making the approximation that $\Delta\bar{\bar{\bar{V}}}_{x-vac}(HSE06) \approx \Delta\bar{\bar{\bar{V}}}_{x-vac}(PBE + U + D3)$.

## B. $TiS_2$ Band-structure

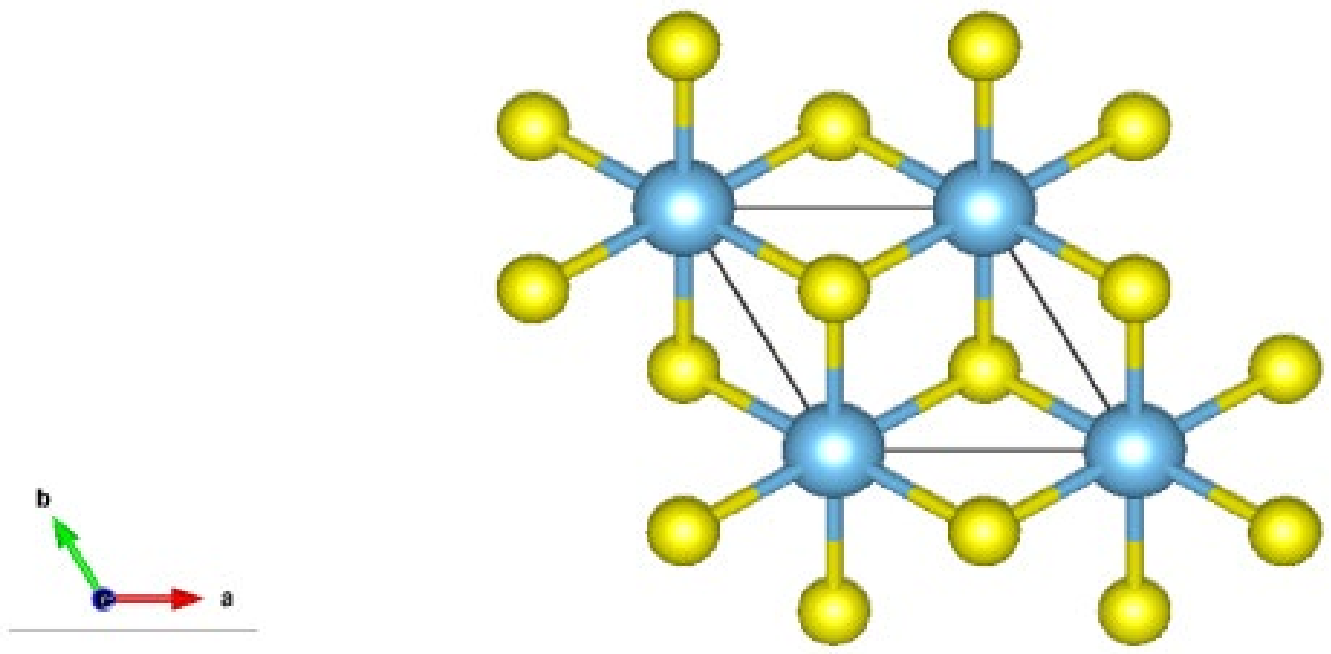

**Figure S20.** Structural unit of $TiS_2$

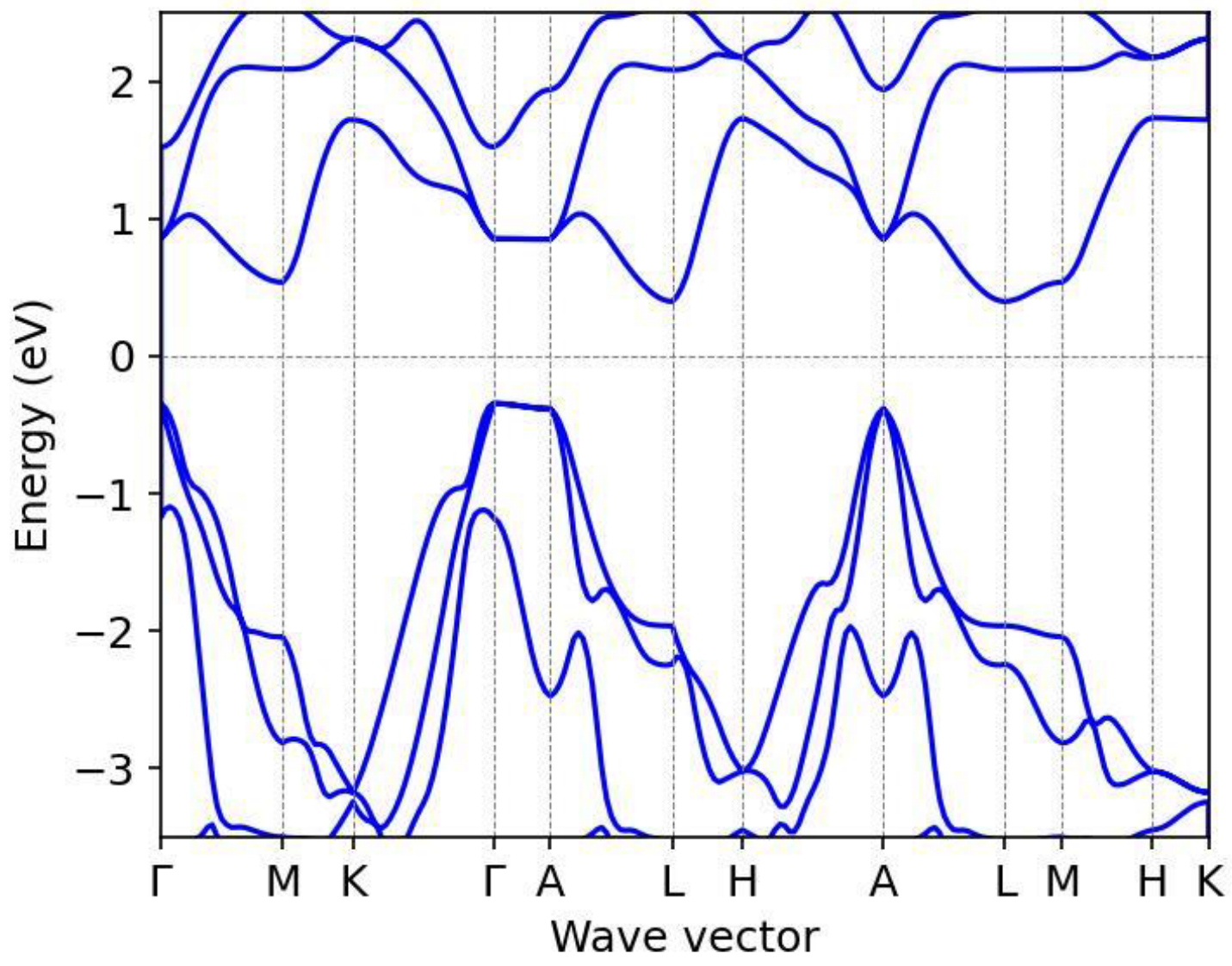

**Figure S21.** Band-structure of $TiS_2$ (bulk) computed using HSE06

Indirect gap of 0.47 eV is obtained from DFT calculations while the experimental gap is 0.5 eV.[21]

C. **$TiO_2$ Band-structure**

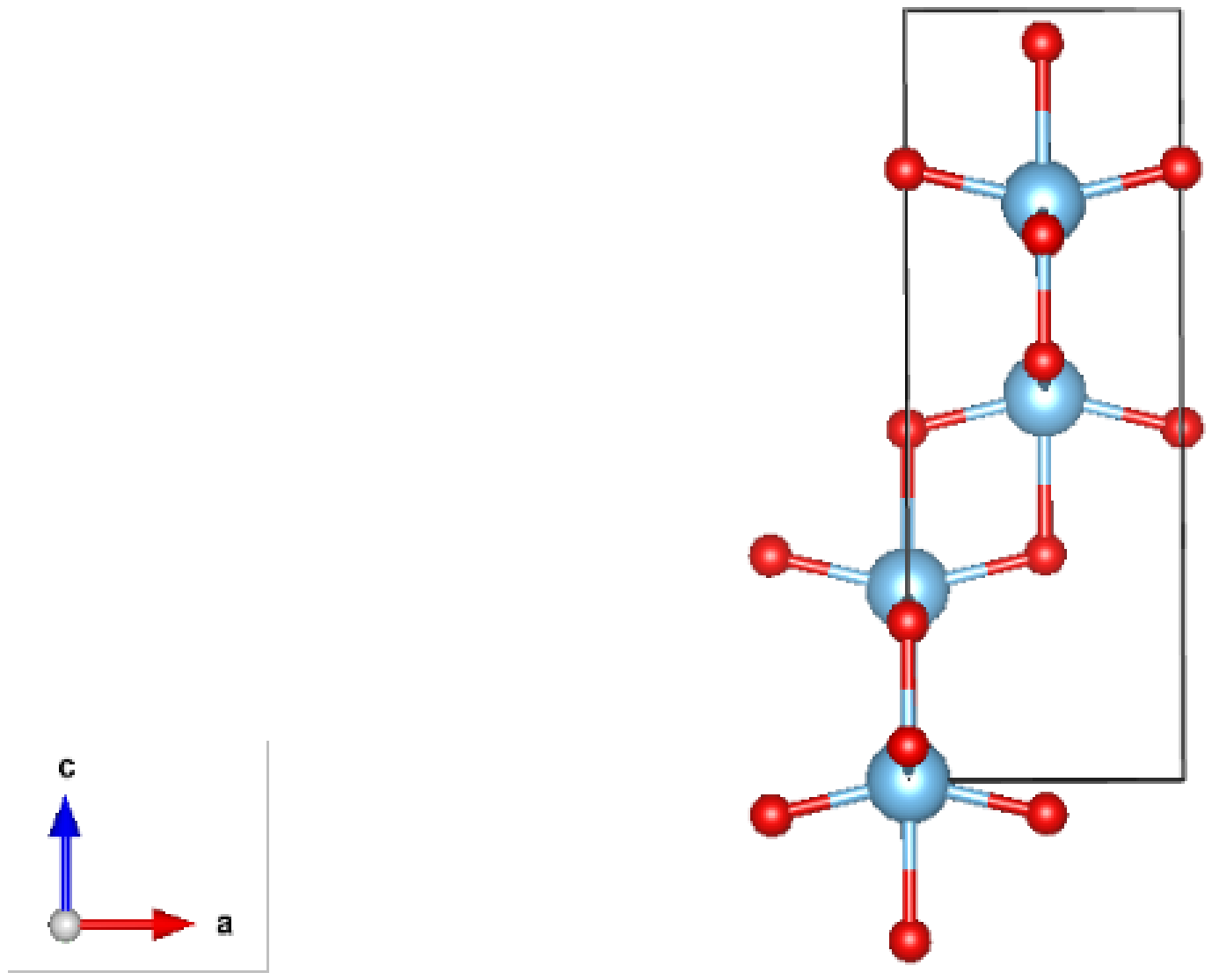

**Figure S22.** Structural unit of $TiO_2$

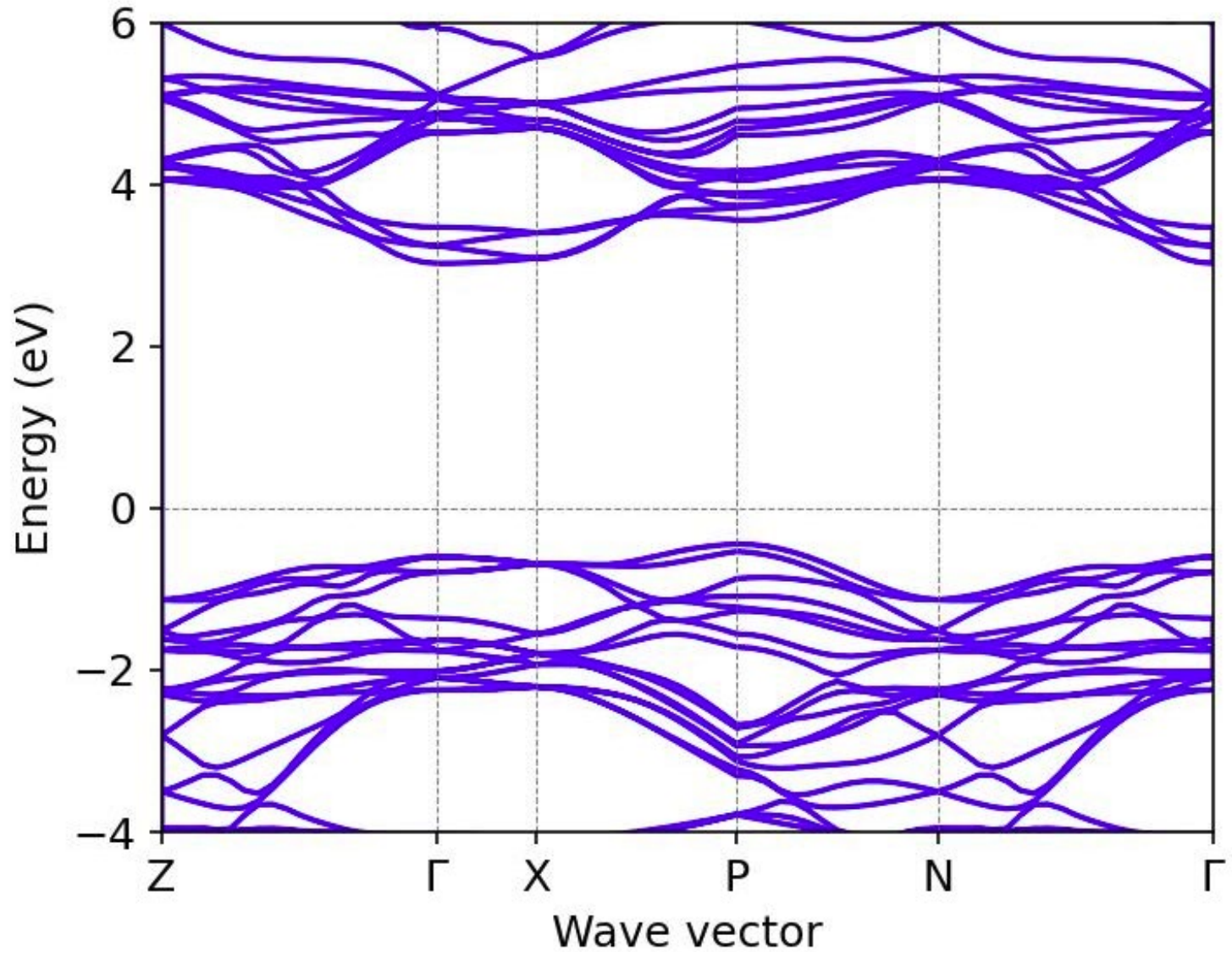

**Figure S23.** Band structure of $TiO_2$ (bulk) computed using HSE06

Indirect gap of 3.34 eV is obtained from DFT calculations while the experimental gap is 3.2 eV.[18]

## D. Work-Function Calculations [Mono-layer]

Work function ($E_{vac} - E_{Fermi}$) of $TiS_2$ [001]monolayer is **5.758 eV**

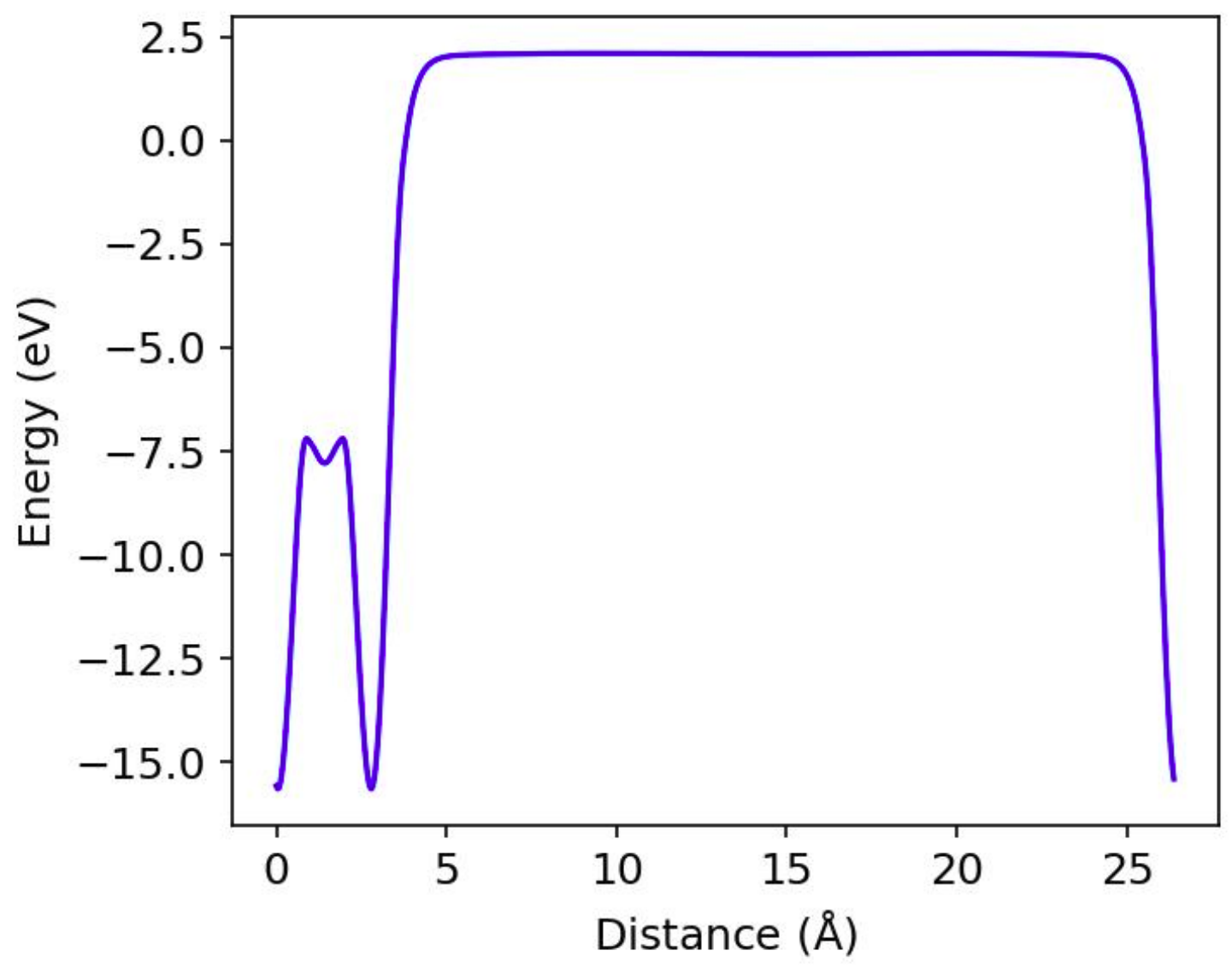

**Figure S24.** Planar averaged potential of $TiS_2$ [001] mono-layer

Work function of $TiO_2$ [101]mono-layer is **6.025 eV**

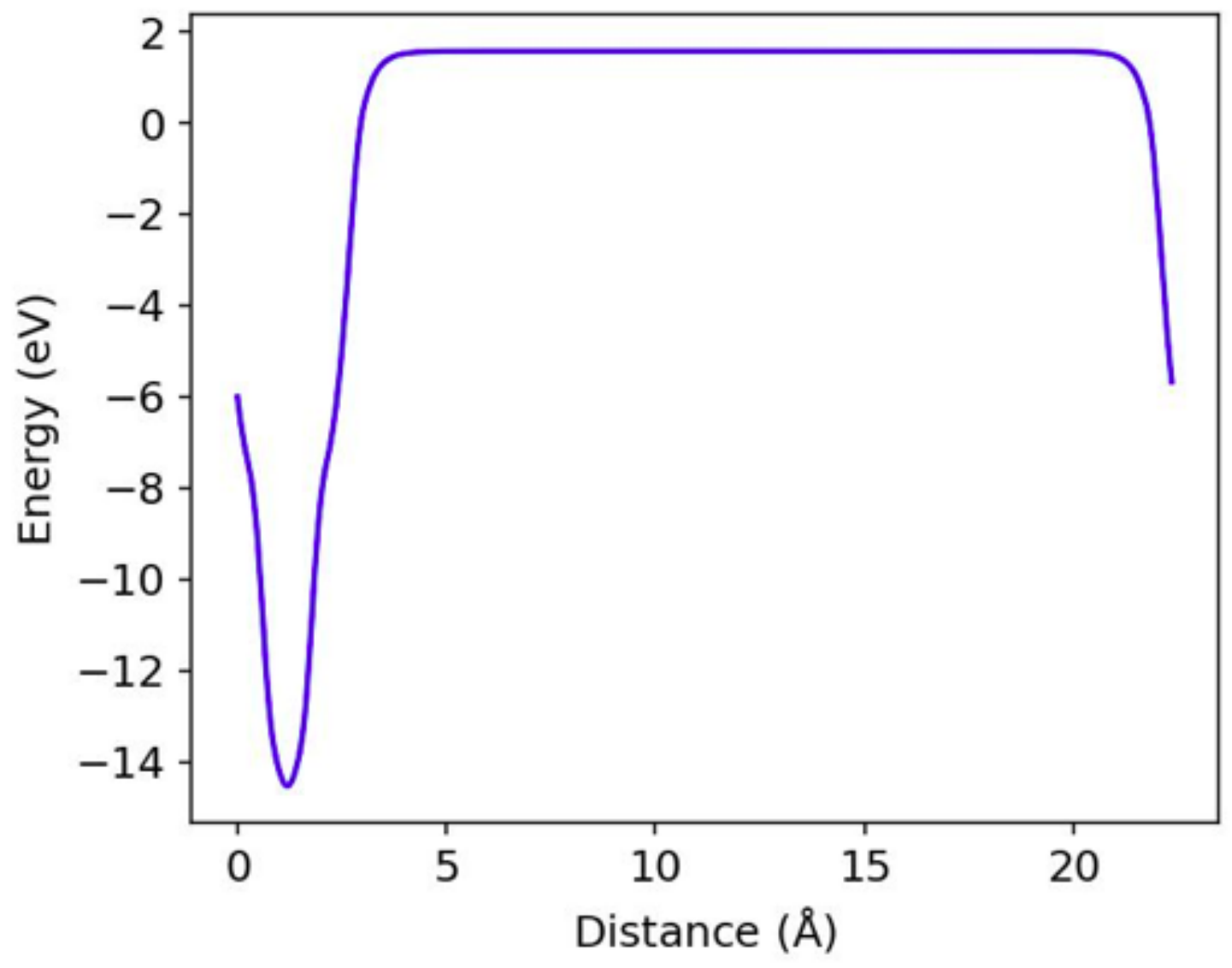

**Figure S25.** Planar averaged potential of $TiO_2$ [101] mono-layer.

## E. Work-Function Calculations [Multi-layer]

Unrelaxed calculations

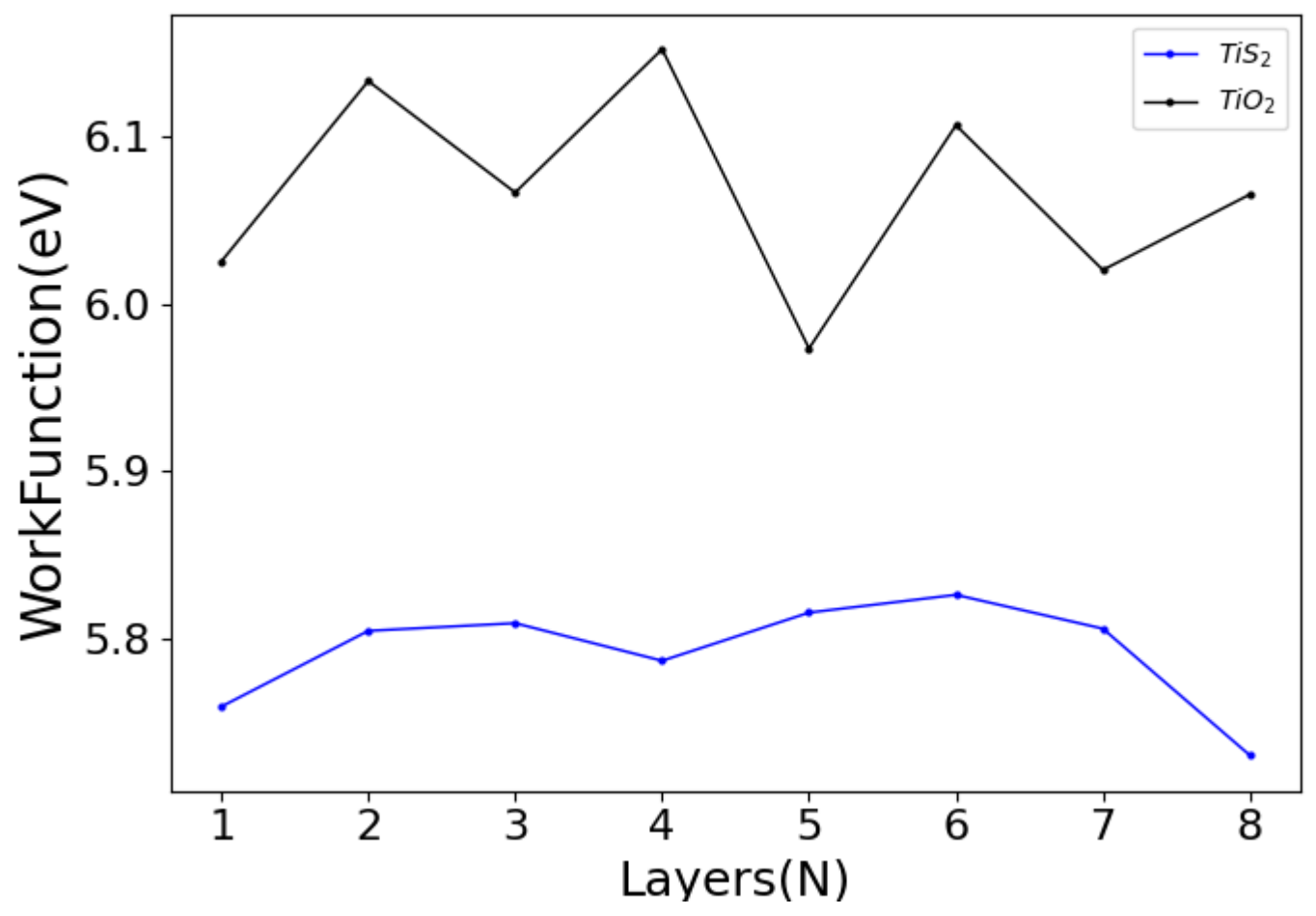

**Figure S26.** Work-Function of $TiS_2$ and $TiO_2$ as a function of number of Layers.

## F. Band Offset for $TiS_2$-$TiO_2$

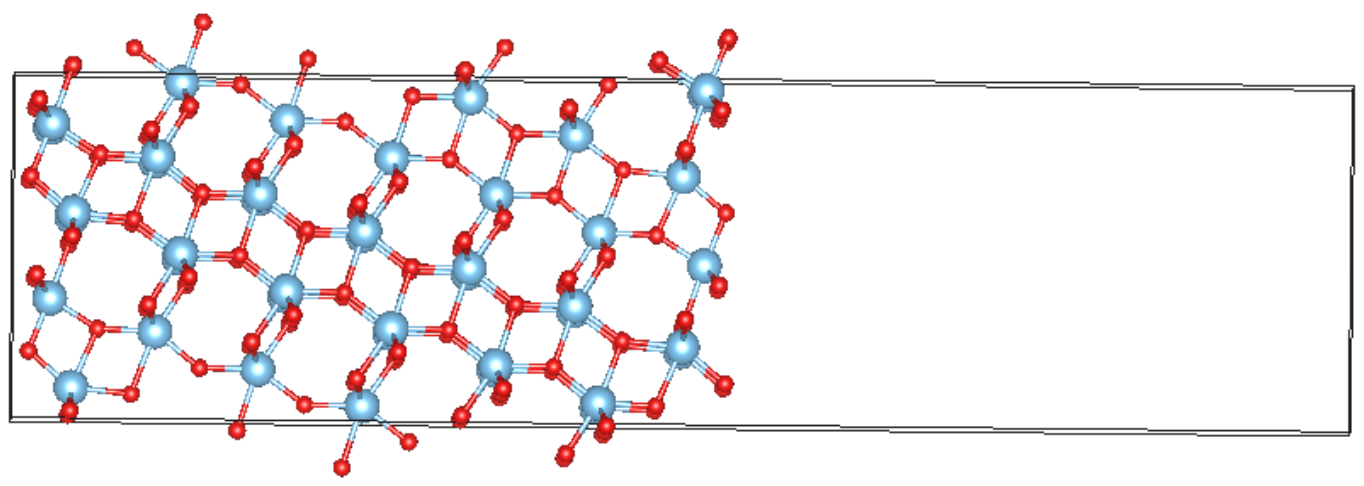

**Figure S27.** Multiple layers of $TiO_2$ [101] with vacuum.

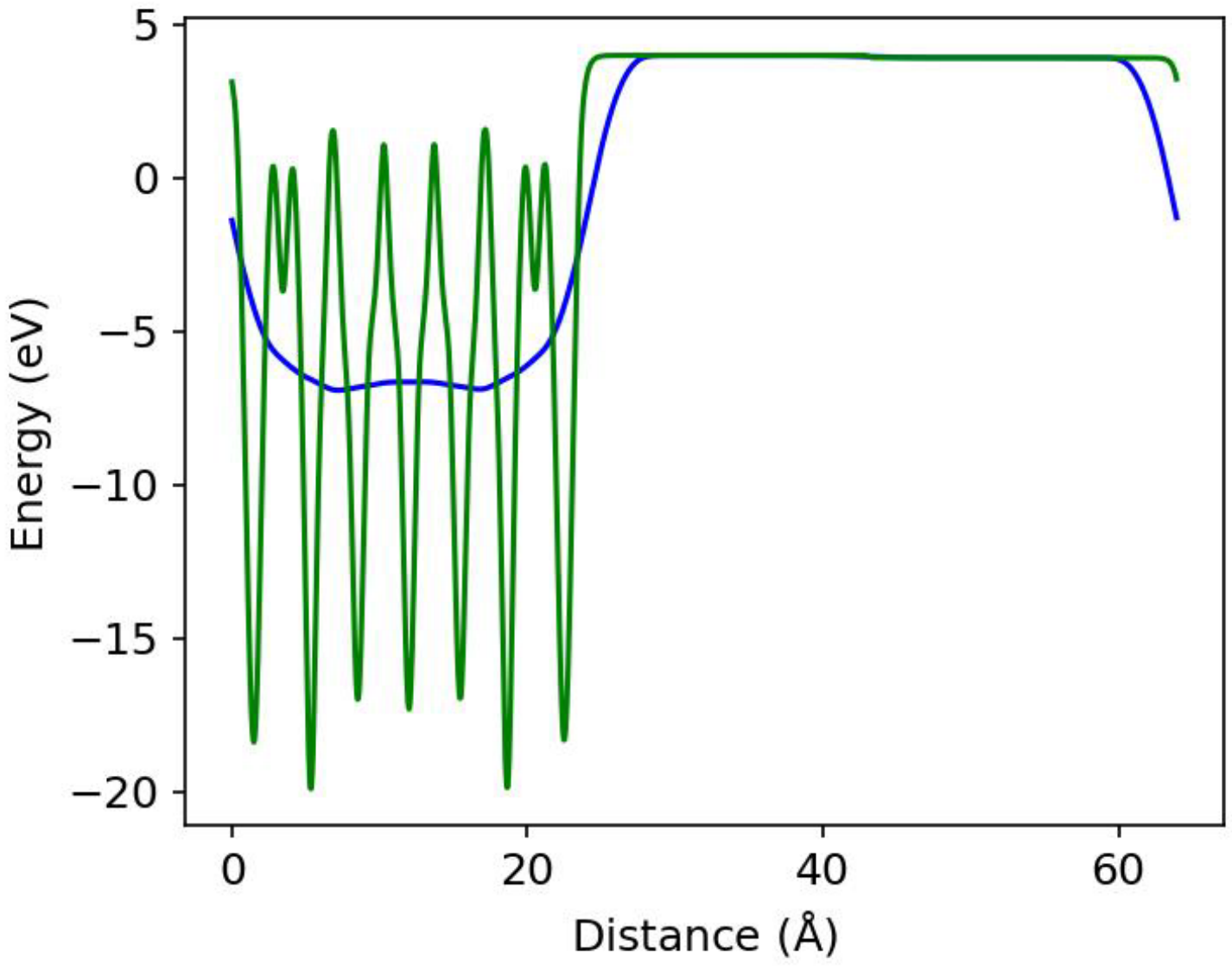

**Figure S28.** Planar averaged and macroscopic averaged potential of multiple layers of $TiO_2$ [101] with vacuum, represented by green and blue colors respectively.

**Figure S28** shows how the macroscopic average of the bulk is referenced to the vacuum. The macroscopic average of the central layer represents the macroscopic average of the bulk, this value is now compared to the vacuum level.

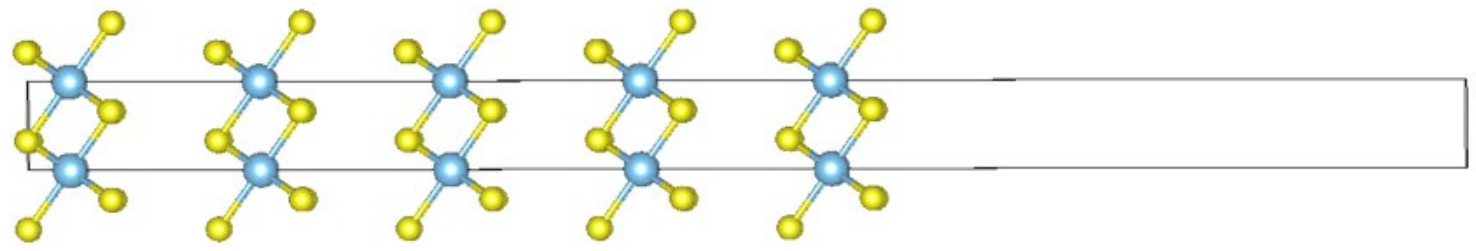

**Figure S29.** Multiple layers of $TiO_2$ [101] with Vacuum.

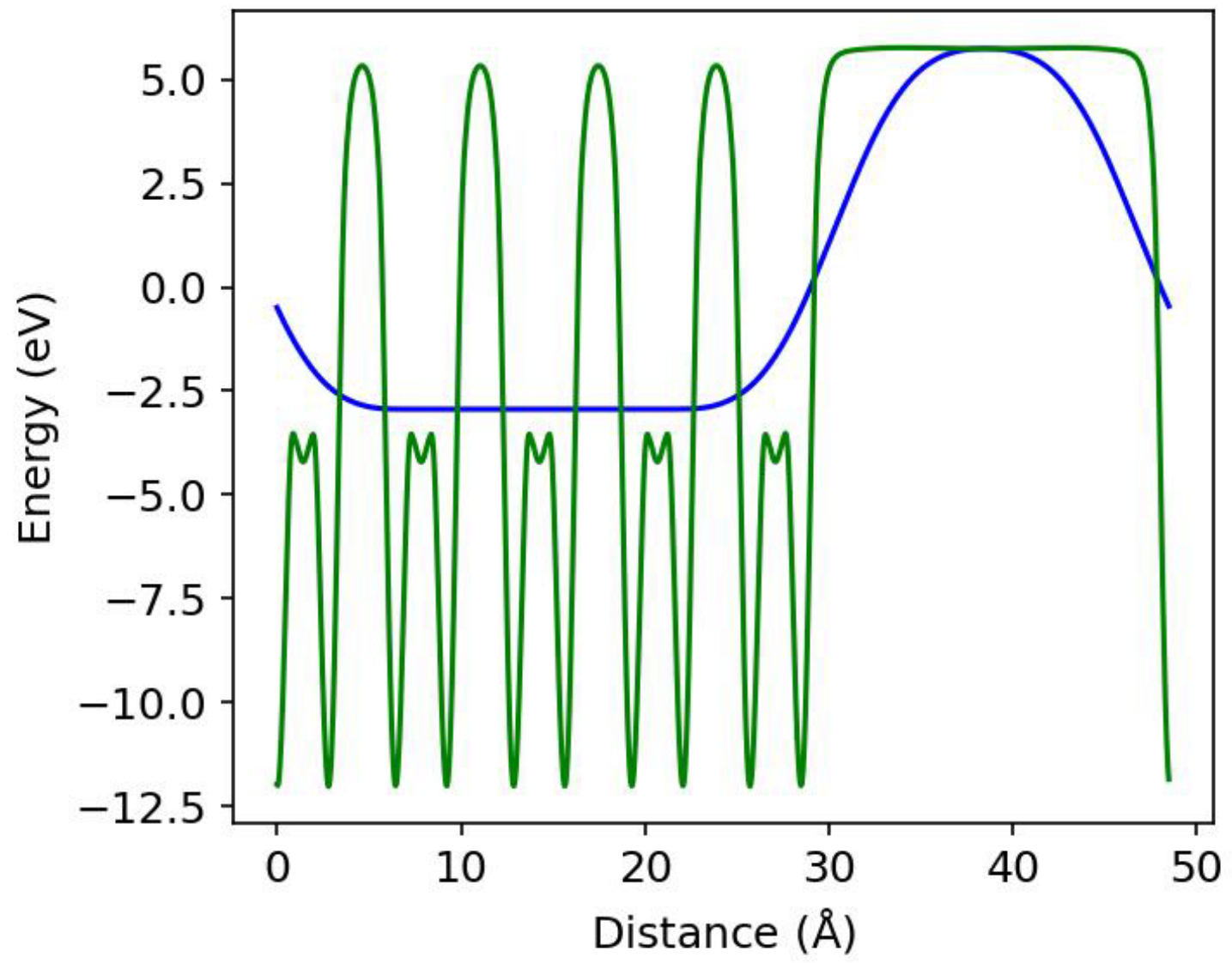

**Figure S30.** Planar averaged and macroscopic averaged potential of multiple layers of $TiS_2$ [001] with vacuum, represented by green and blue respectively.

## G. Band offset for $TiS_2$-$TiO_2$-$Li_1C_{12}$

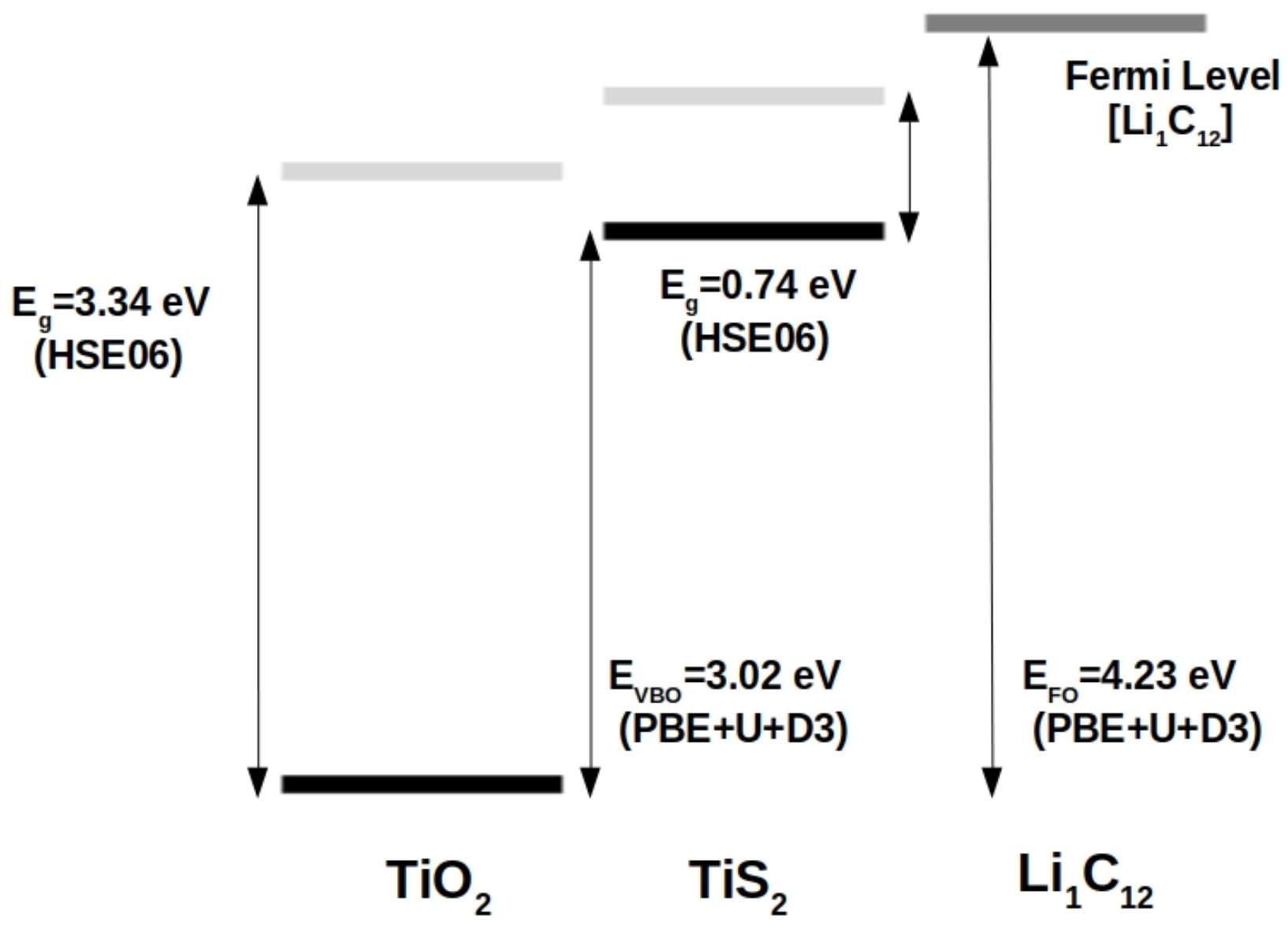

**Figure S31.** Band offset for $TiS_2$, $TiO_2$ and $Li_1C_{12}$. Black and light grey colors represent the valence and conduction band respectively while dark grey level represents the Fermi level of $Li_1C_{12}$.

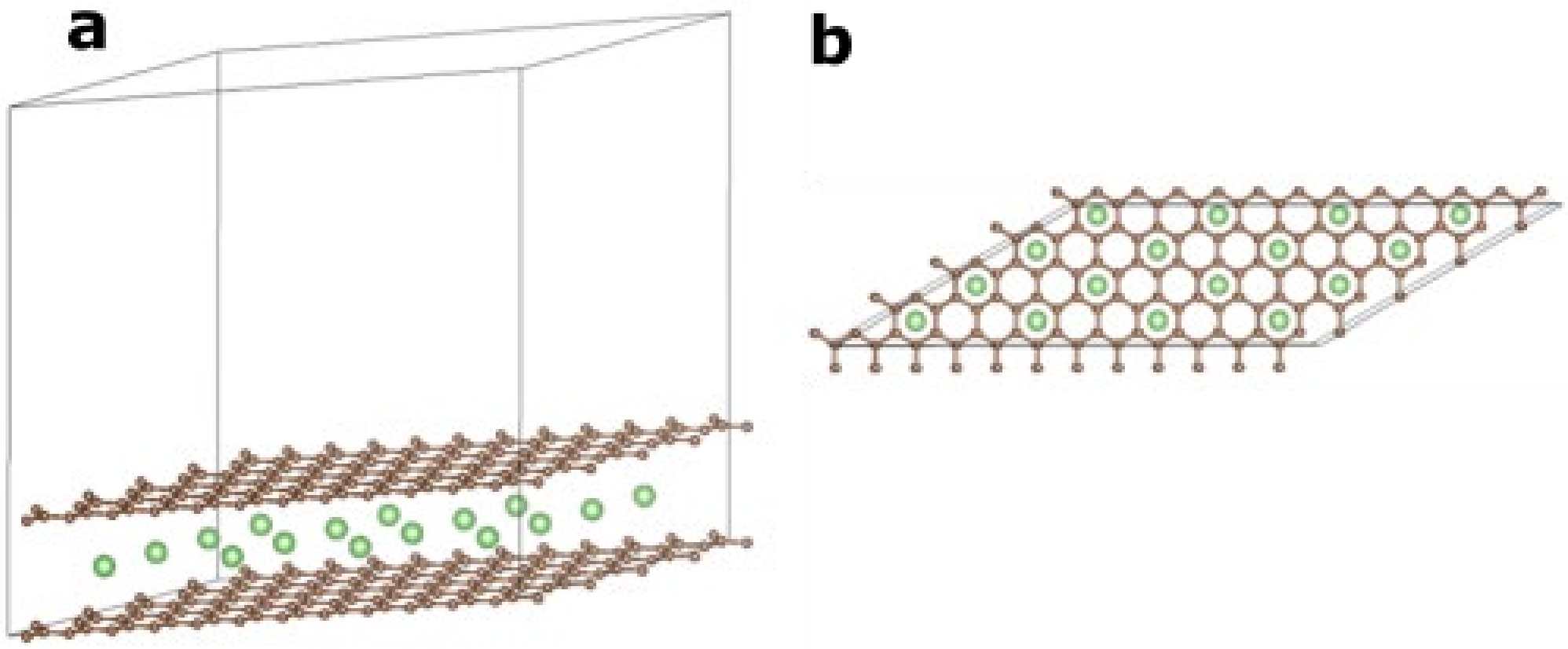

**Figure S32.** Structure of $C_6Li_1C_6$ with additional vacuum to decouple periodic images.(a) side view (b) top view

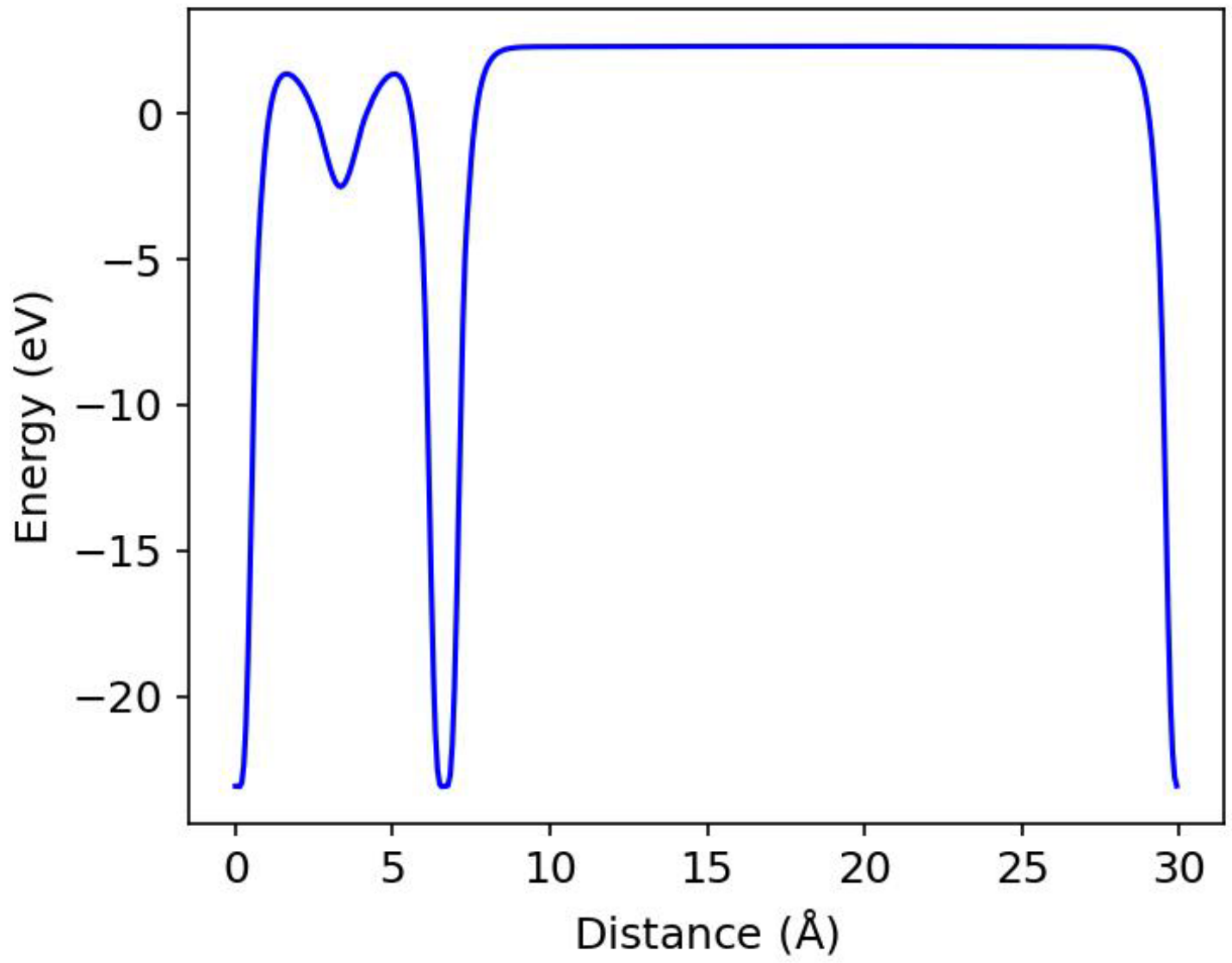

**Figure S33.** Planar averaged Hartree potential to compute the vacuum level for the system

## References:

(1) Bayhan, Z.; Huang, G.; Yin, J.; Xu, X.; Lei, Y.; Liu, Z.; Alshareef, H. N. Two-Dimensional TiO2/TiS2Hybrid Nanosheet Anodes for High-Rate Sodium-Ion Batteries. *ACS Appl. Energy Mater.* **2021**, *4* (9), 8721–8727. https://doi.org/10.1021/ACSAEM.1C01818/SUPPL_FILE/AE1C01818_SI_001.PDF.

(2) Kumar, A.; Thakur, P.; Sharma, R.; Puthirath, A. B.; Ajayan, P. M.; Narayanan, T. N. Photo Rechargeable Li-Ion Batteries Using Nanorod Heterostructure Electrodes. *Small* **2021**, *17* (51), 2105029. https://doi.org/10.1002/SMLL.202105029.

(3) Patra, S.; Thakur, P.; Soman, B.; Puthirath, A. B.; Ajayan, P. M.; Mogurampelly, S.; Karthik Chethan, V.; Narayanan, T. N. Mechanistic Insight into the Improved Li Ion Conductivity of Solid Polymer Electrolytes. *RSC Adv.* **2019**, *9* (66), 38646–38657. https://doi.org/10.1039/C9RA08003A.

(4) Frank, O.; Zukalova, M.; Laskova, B.; Kürti, J.; Koltai, J.; Kavan, L. Raman Spectra of Titanium Dioxide (Anatase, Rutile) with Identified Oxygen Isotopes (16, 17, 18). *Phys. Chem. Chem. Phys.* **2012**, *14* (42), 14567–14572. https://doi.org/10.1039/C2CP42763J.

(5) Sherrell, P. C.; Sharda, K.; Grotta, C.; Ranalli, J.; Sokolikova, M. S.; Pesci, F. M.; Palczynski, P.; Bemmer, V. L.; Mattevi, C. Thickness-Dependent Characterization of Chemically Exfoliated TiS2 Nanosheets. *ACS Omega* **2018**, *3* (8), 8655–8662. https://doi.org/10.1021/ACSOMEGA.8B00766/ASSET/IMAGES/MEDIUM/AO-2018-00766D_M002.GIF.

(6) Zhu, L.; Lu, Q.; Lv, L.; Wang, Y.; Hu, Y.; Deng, Z.; Lou, Z.; Hou, Y.; Teng, F. Ligand-Free Rutile and Anatase TiO2 Nanocrystals as Electron Extraction Layers for High Performance Inverted Polymer Solar Cells. *RSC Adv.* **2017**, *7* (33), 20084–20092. https://doi.org/10.1039/C7RA00134G.

(7) Etghani, S. A.; Ansari, E.; Mohajerzadeh, S. Evolution of Large Area TiS2-TiO2 Heterostructures and S-Doped TiO2 Nano-Sheets on Titanium Foils. *Sci. Reports 2019 91* **2019**, *9* (1), 1–14. https://doi.org/10.1038/s41598-019-53651-y.

(8) Han, J. H.; Lee, S.; Yoo, D.; Lee, J. H.; Jeong, S.; Kim, J. G.; Cheon, J. Unveiling Chemical Reactivity and Structural Transformation of Two-Dimensional Layered Nanocrystals. *J. Am. Chem. Soc.* **2013**, *135* (10), 3736–3739. https://doi.org/10.1021/JA309744C.

(9) Ahmad, S.; George, C.; Beesley, D. J.; Baumberg, J. J.; De Volder, M. Photo-Rechargeable Organo-Halide Perovskite Batteries. *Nano Lett.* **2018**, *18* (3), 1856–1862. https://doi.org/10.1021/ACS.NANOLETT.7B05153/SUPPL_FILE/NL7B05153_SI_001.PDF.

(10) Laskova, B.; Frank, O.; Zukalova, M.; Bousa, M.; Dracinsky, M.; Kavan, L. Lithium Insertion into Titanium Dioxide (Anatase): A Raman Study with 16/18O and 6/7Li Isotope Labeling. *Chem. Mater.* **2013**, *25* (18), 3710–3717. https://doi.org/10.1021/CM402056J/SUPPL_FILE/CM402056J_SI_001.PDF.

(11) Gentili, V.; Brutti, S.; Hardwick, L. J.; Armstrong, A. R.; Panero, S.; Bruce, P. G. Lithium Insertion into Anatase Nanotubes. *Chem. Mater.* **2012**, *24* (22), 4468–4476. https://doi.org/10.1021/CM302912F/SUPPL_FILE/CM302912F_SI_001.PDF.

(12) Haik, O.; Ganin, S.; Gershinsky, G.; Zinigrad, E.; Markovsky, B.; Aurbach, D.; Halalay, I. On the Thermal Behavior of Lithium Intercalated Graphites. *J. Electrochem. Soc.* **2011**, *158* (8), A913. https://doi.org/10.1149/1.3598173/XML.

(13) Perdew, J. P.; Burke, K.; Ernzerhof, M. Generalized Gradient Approximation Made Simple. *Phys. Rev. Lett.* **1996**, *77* (18), 3865. https://doi.org/10.1103/PhysRevLett.77.3865.
(14) Kresse, G.; Furthmüller, J. Efficient Iterative Schemes for Ab Initio Total-Energy Calculations Using a Plane-Wave Basis Set. *Phys. Rev. B* **1996**, *54* (16), 11169. https://doi.org/10.1103/PhysRevB.54.11169.
(15) Kresse, G.; Joubert, D. From Ultrasoft Pseudopotentials to the Projector Augmented-Wave Method. *Phys. Rev. B* **1999**, *59* (3), 1758. https://doi.org/10.1103/PhysRevB.59.1758.
(16) Hu, Z.; Metiu, H. Choice of U for DFT+ U Calculations for Titanium Oxides. *J. Phys. Chem. C* **2011**, *115* (13), 5841–5845. https://doi.org/10.1021/JP111350U/ASSET/IMAGES/LARGE/JP-2010-11350U_0002.JPEG.
(17) Dudarev, S. L.; Botton, G. A.; Savrasov, S. Y.; Humphreys, C. J.; Sutton, A. P. Electron-Energy-Loss Spectra and the Structural Stability of Nickel Oxide: An LSDA+U Study. *Phys. Rev. B* **1998**, *57* (3), 1505. https://doi.org/10.1103/PhysRevB.57.1505.
(18) Dette, C.; Pérez-Osorio, M. A.; Kley, C. S.; Punke, P.; Patrick, C. E.; Jacobson, P.; Giustino, F.; Jung, S. J.; Kern, K. TiO2 Anatase with a Bandgap in the Visible Region. *Nano Lett.* **2014**, *14* (11), 6533–6538. https://doi.org/10.1021/NL503131S/SUPPL_FILE/NL503131S_SI_001.PDF.
(19) Mattioli, G.; Filippone, F.; Alippi, P.; Amore Bonapasta, A. Ab Initio Study of the Electronic States Induced by Oxygen Vacancies in Rutile and Anatase TiO2. *Phys. Rev. B - Condens. Matter Mater. Phys.* **2008**, *78* (24), 241201. https://doi.org/10.1103/PHYSREVB.78.241201/FIGURES/5/MEDIUM.
(20) Morgan, B. J.; Watson, G. W. A DFT + U Description of Oxygen Vacancies at the TiO2 Rutile (1 1 0) Surface. *Surf. Sci.* **2007**, *601* (21), 5034–5041. https://doi.org/10.1016/J.SUSC.2007.08.025.
(21) Wang, H.; Qiu, Z.; Xia, W.; Ming, C.; Han, Y.; Cao, L.; Lu, J.; Zhang, P.; Zhang, S.; Xu, H.; Sun, Y. Y. Semimetal or Semiconductor: The Nature of High Intrinsic Electrical Conductivity in TiS2. *J. Phys. Chem. Lett.* **2019**, *10* (22), 6996–7001. https://doi.org/10.1021/ACS.JPCLETT.9B02710/ASSET/IMAGES/LARGE/JZ9B02710_0004.JPEG.
(22) Monkhorst, H. J.; Pack, J. D. Special Points for Brillouin-Zone Integrations. *Phys. Rev. B* **1976**, *13* (12), 5188. https://doi.org/10.1103/PhysRevB.13.5188.
(23) Conesa, J. C. Computing with DFT Band Offsets at Semiconductor Interfaces: A Comparison of Two Methods. *Nanomater. 2021, Vol. 11, Page 1581* **2021**, *11* (6), 1581. https://doi.org/10.3390/NANO11061581.
(24) Fu, H.; Goodrich, J. C.; Tansu, N. Band Alignment of ScAlN/GaN Heterojunction. *Appl. Phys. Lett.* **2020**, *117* (23), 231105. https://doi.org/10.1063/5.0029488.
(25) Yang, W.; Wen, Y.; Zeng, D.; Wang, Q.; Chen, R.; Wang, W.; Shan, B. Interfacial Charge Transfer and Enhanced Photocatalytic Performance for the Heterojunction WO3/BiOCl: First-Principles Study. *J. Mater. Chem. A* **2014**, *2* (48), 20770–20775. https://doi.org/10.1039/C4TA04327H.
(26) Weston, L.; Tailor, H.; Krishnaswamy, K.; Bjaalie, L.; Van de Walle, C. G. Accurate and Efficient Band-Offset Calculations from Density Functional Theory. *Comput. Mater. Sci.* **2018**, *151*, 174–180. https://doi.org/10.1016/J.COMMATSCI.2018.05.002.